\documentclass[epj,epsf,psfig]{svjour}
\pdfoutput=1
\usepackage{amsmath}

\newcommand{\beq}{\begin{equation}}
\newcommand{\eeq}{\end{equation}}

\newcommand{\di}{\displaystyle}

\newcommand{\si}{\sigma}

\newcommand{\nub}{\bar{\nu}}

\newcommand{\GeV}{{\rm{GeV}}}
\newcommand{\cm}{{\rm{cm}}}

\usepackage{array, graphicx, subfigure}
\usepackage{graphics}
\begin{document}
\title{Methods to Determine Neutrino Flux at Low Energies: }
\subtitle{Investigation of the Low $\nu$ Method}

\author{A. Bodek\inst{1}, U. Sarica\inst{1}, D. Naples\inst{2} and L. Ren\inst{2} }
\institute{Department of Physics and Astronomy, University of
Rochester, Rochester, NY  14627-0171 USA
\and University of Pittsburgh, Pittsburgh, PA  15260  }

\date{Received: date / Revised version: date  March 8, 2012}
\abstract{
We investigate the ``low-$\nu$''  method (developed by the CCFR/NUTEV collaborations) to determine the  neutrino flux in a wide
band neutrino beam at very low energies, a region of interest to neutrino oscillations experiments.
Events with low  hadronic final state energy $\nu<\nu_{cut}$ (of 1, 2 and 5 GeV) were used by the MINOS collaboration to determine the neutrino flux in their
measurements of neutrino ($\nu_\mu$) and antineutrino ($\nub_\mu$) total cross sections. 
The lowest $\nu_\mu$  energy for which the method was used in MINOS
is 3.5 GeV,  and the lowest $\nub_\mu$ energy is 6 GeV. At these energies,  the cross sections  are dominated
by inelastic processes. 
We investigate the application of the method to determine the
neutrino flux  for $\nu_\mu$, $\nub_\mu$ energies as low as 0.7 GeV where the cross sections are dominated by quasielastic scattering and $\Delta$(1232) resonance production.   We find that the method can be extended to low energies by using $\nu_{cut}$ values of  0.25 and 0.50 GeV, which is feasible in  fully active neutrino detectors such as MINERvA.
} 
\maketitle
%

\section{Introduction}
\label{intro}

A detailed understanding of neutrino ($\nu_\mu$)  and antineutrino ($\nub_\mu$) interaction cross sections for various final
states is required for the next generation neutrino oscillations experiments.
The relevant neutrino energy region of interest
for the large neutrino detectors  such as T2K\cite{T2K},  MINOS\cite{MINOS,MINOS2}, 
and  NOVA\cite{NOVA}  is  $0.5<E_{\nu}<3$ GeV.

The MINERvA\cite{minerva} experiment at the NUMI wide band
beam at Fermilab uses a fine grain fully
active scintillator target-detector to investigate  neutrino ($\nu_\mu$) and antineutrino
($\nub_\mu$) cross sections for
energies above 0.5 GeV. These measurements require a reliable  determination
of the flux as a function of $\nu_\mu$, $\nub_\mu$ energy.

 Previous neutrino experiments in wide band beams
 used five methods for the determination of flux as
 a function of energy. 

\begin{enumerate}
\item Modeling the distribution of pions and kaons produced by 
incident proton beam in the target.  Then,  tracking the pions and kaons though the Horn
focussing  magnetic
fields,  and  modeling the decays of  pions and kaons  in the decay pipe.
\item Measuring the muon flux that exits  the decay pipe and relating it to the
neutrino flux.
\item Monitoring Inverse muon decay   events ( $\nu_\mu+  e \to \mu^- + \nu_e$) in the detector.
\item Monitoring neutrino-electron scattering events  ($\nu_\mu+  e \to \nu_\mu+ e$) in the detector.
\item The ``low-$\nu$'' method for the determination of the energy dependence of
the relative neutrino and antineutrino flux.
\end{enumerate}

Here,  ``low-$\nu$''~\cite{mishra,seligman} refers to events with  low energy transfer to the target nucleon in the scattering
processes  $ \nu_\mu + N \rightarrow \mu^- + X$  and $ \nub_\mu + N \rightarrow \mu^+ + X$.
This energy transfer manifests itself as the energy ($\nu=E_{had}$)  of the final state hadrons (X)  in the laboratory frame. 

There are inherent difficulties in the each of those techniques: 
\begin{enumerate}
\item In method 1, the differential cross sections for the production of pions  and kaons  by protons incident on a thick nuclear target must be known very well. In addition, the magnetic field of the horn focussing magnets  must be modeled reliably. 
 \item  In method 2, the response of the muon detectors at the end of the decay pipe must be very well understood (for absolute calibration of the neutrino flux). The response of the muon detectors is sensitive to $\delta$ rays.  In addition, since the energy of the muons is not measured, it is difficult to  determine the energy dependence of the neutrino flux. 
 \item In method 3, the threshold for
 the reaction    $\nu_\mu+  e \to \mu^- + \nu_e$  is
 about 12 GeV. Therefore, this  method can only be used at higher energies.
  Unfortunately, this method cannot be used for the determination of the
 flux for antineutrinos.
  Inverse muon decay 
 was used by NOMAD to constrain their neutrino flux at high energies.  
 \item In method 4,  only the sum of the fluxes for
 neutrinos and antineutrinos can be measured.
 This is because
calorimetric detectors such as MINERvA cannot determine
the charge of final state electron in $\nu_\mu+  e \to \nu_\mu+  e$ events.
\end{enumerate} 

Both methods 4 and 5 are statistically limited. In addition,  
in both  methods,  the total final state energy in the  events is not fully 
reconstructed since there is a neutrino in the final state. This places a limitation
on the determination of  the energy dependence
of the neutrino and antineutrino fluxes.  Despite of these limitations, these
two additional methods are valuable as important
consistency checks.

Consequently, having another independent technique such as the ``low-$\nu$'' method is extremely valuable. 

The ``low-$\nu$'' method was initially  developed
by the CCFR/NUTEV~\cite{mishra,seligman} collaboration. At high energy, the method was used to determine the relative neutrino flux as a function of neutrino energy ($E_{\nu}$). The method relies on
the observation that the  charged current 
differential cross section, $\frac{d \sigma^{\nu,\nub}}{d\nu}$ in the  limit $\nu\rightarrow 0$,
only depends on the structure function  ${\cal F}_2$, and therefore is independent of energy. 

The ``low-$\nu$'' method was used by the CCFR/NuTeV collaborations  to measure the energy dependence  of 
 $\sigma_{\nu}/E$ and $\sigma_{\nub}/E$   for charged current interactions
 for   energies higher than 30 GeV for an iron target. 
The absolute level of the  charged cross sections is normalized to previous  measurements of  $\sigma_{\nu}/E$ in a  high energy narrow band neutrino beam.

 Most recently,  the method was extended to lower energies by the MINOS~\cite{MINOS2} collaboration. The lowest neutrino energy for which this method was used in MINOS is 3.5 GeV for neutrinos and 6 GeV for antineutrinos.  
 
 The absolute normalization used by MINOS is to the world average  value
of   charged current   $\sigma_{\nu}/E$  measurements for an isoscalar target  for neutrino energies between 30 to 50 GeV.  The average value used by MINOS is
   $$\langle {\sigma_{\nu}/E} \rangle_{30-50}= 0.675 \times 10^{-38}\cm^2/\GeV$$ per nucleon. The antineutrino sample is not independently normalized
but is related to the neutrinos by using the same normalization factor.
  
 In this communication we investigate 
 the application of the technique to  much lower neutrino energies ($E_\nu >0.5$ GeV). Neutrino interactions in this energy range
are currently being studied at MINERvA.

\section{The ``low-$\nu$'' method at high energies \label{lnu_high}}

If we neglect terms which are proportional to the muon mass, 
the differential cross section  $\frac{d^2\sigma^{\nu,\nub}}{dxdy}$ for charged current scattering of $\nu_{\mu}$ ($\nub_{\mu}$)
 with an incident energy $E_{\nu}$, muon final energy
$E_{\mu}$ and scattering angle $\theta$ can be written in terms of
the structure functions ${\cal F}_1= M{\cal W}_1 (x,Q^2)$,  ${\cal F}_2=\nu {\cal W}_2 (x,Q^2)$
and   ${\cal F}_3=\nu {\cal W}_3 (x,Q^2)$:

\begin{eqnarray}
\label{cross1}
\frac{d^2\sigma^{\nu(\overline{\nu})}}{dxdy} &=& \frac{G_F^2 M
 E_{\nu}}{\pi} {\Big(}\Big[1-y(1+\frac{Mx}{2E_{\nu}}) \nonumber\\
&+&\frac{y^2}{2}\Big(\frac{1+(\frac{2Mx}{Q})^2}{1+R}\Big)\Big] {\cal F}_2 
\pm \Big[y-\frac{y^2}{2}\Big]x{\cal F}_3{\Big)},
\end{eqnarray}
where $G_F$ is the Fermi weak coupling constant, $M$ is the
proton mass,  $y=\nu/E_\nu$, $\nu=E_{had}=E_{\nu}-E_{\mu}$,
$Q^2=4E_{\nu}E_{\mu}\sin^2(\theta/2)$ 
is the square of  four momentum transfer,  and $x=Q^2/(2M\nu)$ is  the  Bjorken scaling variable.
The plus sign in front of the $xF_3$ term is for neutrinos and the minus is for
antineutrinos.

Here, $R(x,Q^2)$ is defined as the ratio of the longitudinal
and transverse structure functions ($\sigma_L/\sigma_T$). It  is related to the other structure functions by,
\begin{equation}
 R(x,Q^2)
   = \frac {\sigma_L }{ \sigma_T}
   = \frac{{\cal F}_2 }{ 2x{\cal F}_1}(1+\frac{4M^2x^2 }{Q^2})-1
   = \frac{{\cal F}_L }{ 2x{\cal F}_1},
\end{equation}
where ${\cal F}_L$ is called the longitudinal structure function,  
\begin{eqnarray}
 {\cal F}_L(x,Q^2) &=& {\cal F}_2 \left(1 + \frac{4 M^2 x^2 }{ Q^2}\right) - 2x{\cal F}_1.
\end{eqnarray}
Other useful relations are: 
\begin{eqnarray}
2x{\cal F}_1 (x,Q^{2}) &=& {\cal F}_2 (x,Q^{2})
\frac{1+4M^2x^2/Q^2}{1+R(x,Q^{2})},  
\end{eqnarray}
\begin{eqnarray}
{\cal W}_1 (x,Q^{2}) &=& {\cal W}_2(x,Q^{2})
\frac{1+\nu^2/Q^2}{1+R(x,Q^{2})}.  \nonumber 
\end{eqnarray}
The three  structure functions ${\cal F}_2(x,Q^2)$, ${\cal F}_1(x,Q^2)$ and  $x{\cal F}_3(x,Q^2)$
depend on $x$ and $Q^2$.

Integrating over $x$, the differential dependence on $\nu$ can be written in the
simplified form
\begin{equation}
\frac{d\sigma^{\nu,\nub}}{d\nu}=A\left(1+\frac{B}{A}\frac{\nu}{E_{\nu}}-\frac{C}{A}\frac{\nu^2}{2E_{\nu}^2}\right) \label{eq:dsigmadnu}.
\end{equation}
The coefficients $A,$ $B$, and $C$ depend on integrals over
structure functions, where
\begin{eqnarray}
\label{Efcminos}
A &=& \frac{G^2_FM}{\pi} \displaystyle\int_{0}^{1}{\cal F}_2(x) dx,\\
\nonumber
B &=& -\frac{G^2_FM}{\pi} \displaystyle\int_{0}^{1} \Big( {\cal F}_2(x)  \mp x{\cal F}_3(x)\Big) dx,\\
\nonumber
C &=& B - \frac{G^2_FM}{\pi} \displaystyle\int_{0}^{1} {\cal F}_2(x) ~ \tilde{R}~ dx,
\nonumber
\end{eqnarray}
and
\begin{eqnarray*}
\tilde{R} &=& \left(\frac{1+\frac{2Mx}{\nu}}{1+ R(x,Q^2)}-\frac{Mx}{\nu}-1\right).
\end{eqnarray*}
In the  limit $\nu/E_{\nu} \rightarrow 0$,  the $A$ term dominates
and the $B$ and $C$ terms are very small. 
The MINOS collaboration used the number of ``low-$\nu$'' events  (with $\nu<\nu_{cut}$) in the detector
to determine the relative flux of neutrinos and antineutrinos as a function of  $E_{\nu}$.

In the MINOS analysis, the relative flux is determined using events with  $\nu<1$ GeV for
$\nu_\mu$ energies in the range  $3<E_{\nu}<9$ GeV,  and for
$\nub_\mu$ in the range $5<E_{\nu}<9$ GeV.  Events with   $\nu<2$ GeV are used
for $\nu_\mu$ and $\nub_\mu$ events in the range $9<E_{\nu}<18$ GeV,  and events with 
$\nu<5$ GeV  are used for $E_{\nu}>18$ GeV.

MINOS divides the number of ``low-$\nu$'' events with $y<y_{cut}=\nu_{cut}/E_{\nu}$
by correction term $f_C$ to account for the energy dependence from
the $B$ and $C$ terms.  Here 
\begin{equation}
f_C (E_\nu)= 1+  \displaystyle\int_{0}^{y_{cut}} \frac{B}{A}\frac{\nu}{E_{\nu}} dy-\displaystyle\int_{0}^{y_{cut}}\frac{C}{A}\frac{\nu^2}{2E_{\nu}^2}dy.
\end{equation}
As seen in equation~\ref{Efcminos}, the negative contribution of  ${\cal F}_2(x)$ in $B$ partially
cancels the positive contribution of  $x{\cal F}_3(x) $ for $\nu_\mu$'s.  For $\nub_\mu$'s  both
contributions are negative.  There are additional small corrections that are applied to equation \ref{Efcminos} to correct for differences in ${\cal F}_2$
between neutrinos and antineutrinos.

\begin{figure}
\includegraphics[width=3.2in,height=2.2in]{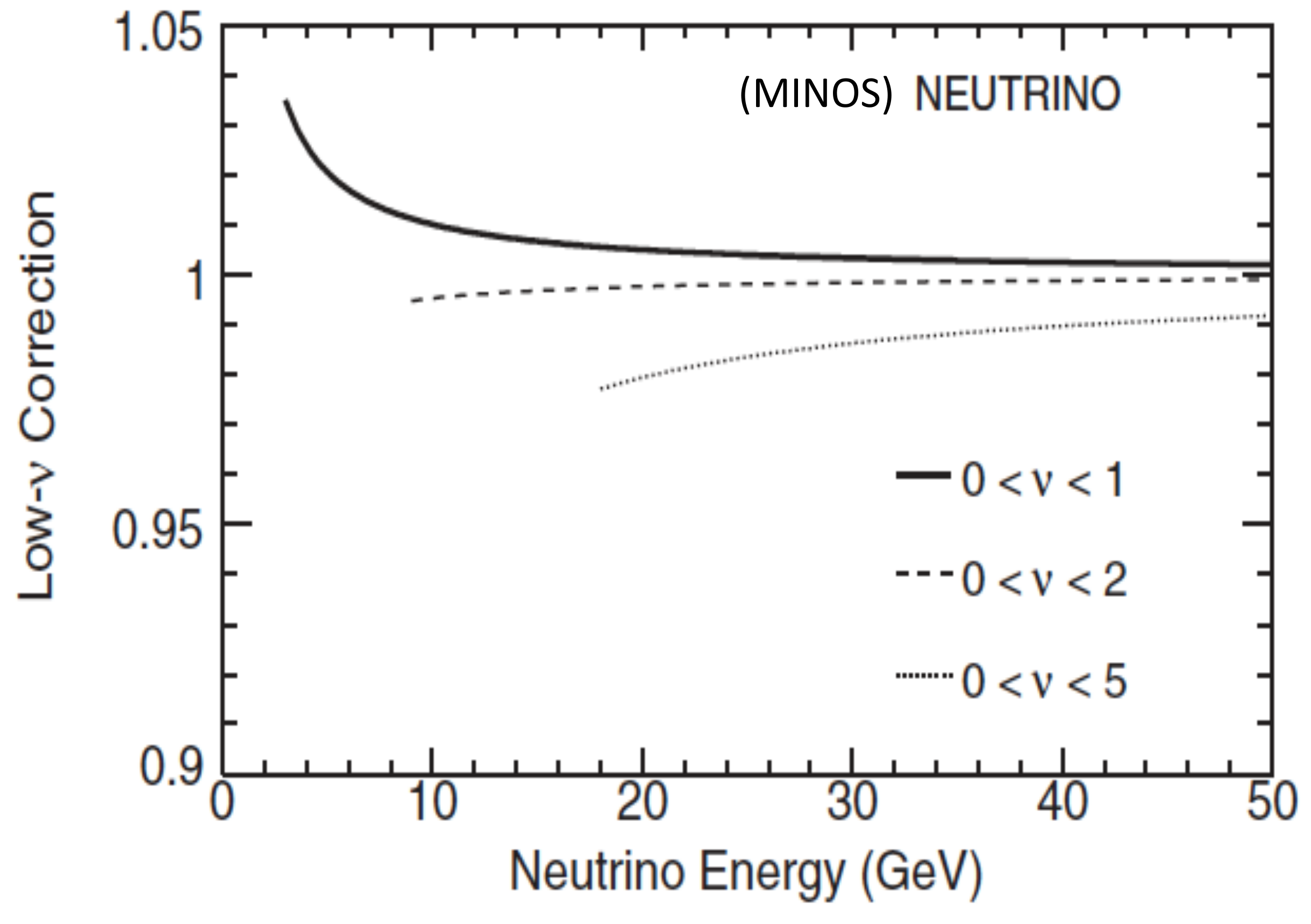}
\includegraphics[width=3.2in,height=2.2in]{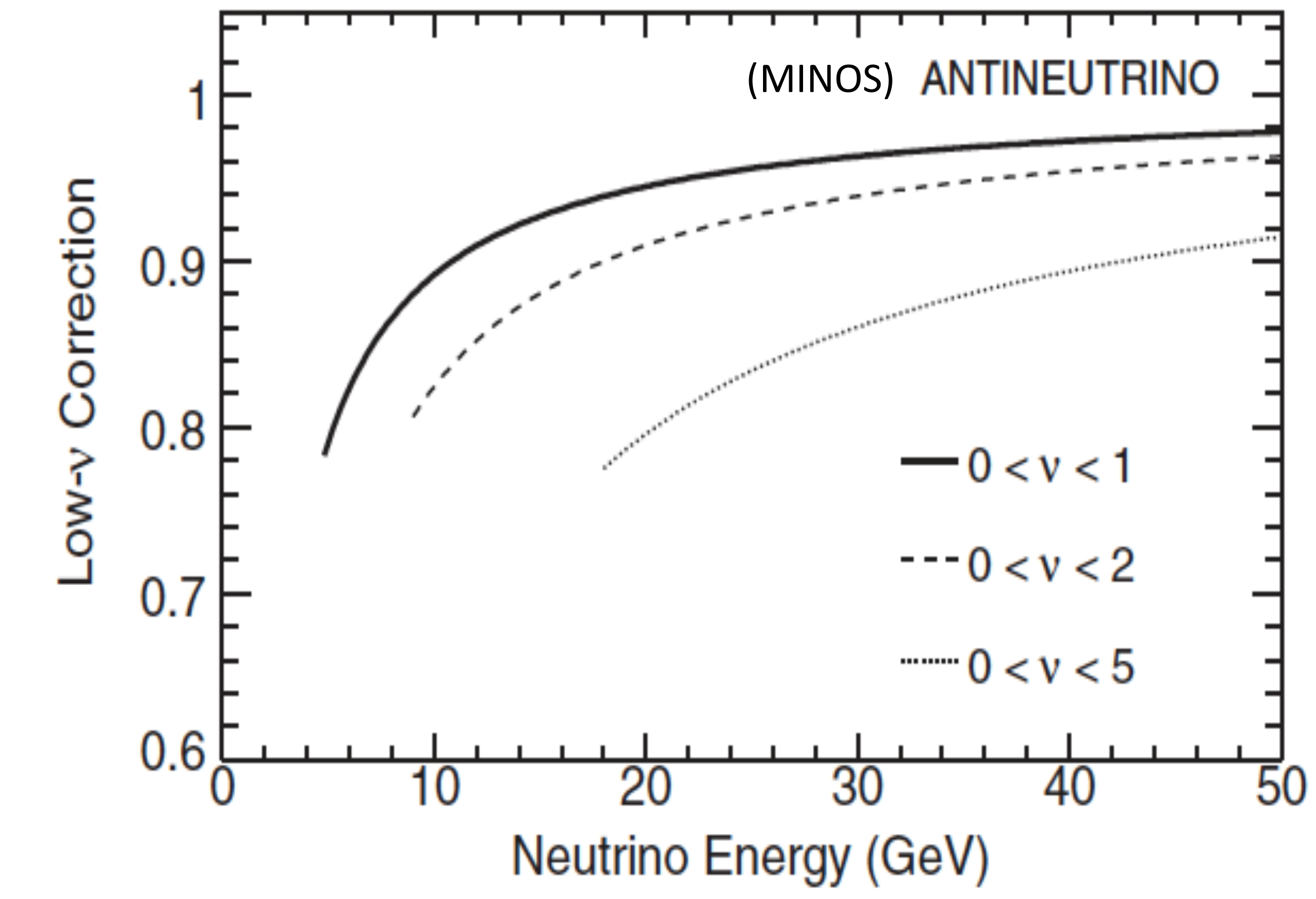}
\caption{The  ``low-$\nu$'' correction factors $f_C$ used by MINOS for neutrinos (shown in the top panel) and
antineutrinos (shown in the bottom panel).}
 \label{minosfcnu}
\end{figure}

%
In practice, a neutrino interaction generator model~\cite{neugen3} is used to compute $f_C$ from
\begin{equation}
f_C (E_\nu)= \frac {  \sigma (\nu<\nu_{cut},E_\nu)} {  \sigma (\nu<\nu_{cut},E_\nu=\infty)}.
\label{eq:minosfc}
\end{equation}
%
The corrections factors $f_C$ used by MINOS for $\nu_\mu$ and $\nub_\mu$
 are shown in Fig.~\ref{minosfcnu}.
%

The measured ``low-$\nu$'' sample is corrected for detector smearing and acceptance by multiplying the number of observed ``low-$\nu$'' events in the data
in each energy bin by
\begin{eqnarray}
\nonumber
R^{MC}(E_\nu) &= & \frac { N^{GEN}(E_{reconstructed},\nu <\nu_{cut})   }  { N^{REC}(E_{reconstructed},\nu <\nu_{cut})},
\end{eqnarray}
which is obtained from a Monte Carlo detector simulation.  $N^{GEN}$ and   $N^{REC}$
are the number of generated and reconstructed events below $\nu_{cut}$ in each reconstructed energy bin, respectively.
In the first pass the initial input flux from a beam model is used. It is then
replaced by the extracted ``low-$\nu$'' flux and the procedure is reiterated to account for the effect of the flux model on 
the acceptance corrections. (The change in the extracted flux is found to be negligible). The ``low-$\nu$'' sample 
is further corrected for radiative effects using Ref.~\cite{bardin}. 
The absolute level of the flux is set by normalizing the cross section in data
to a nominal world average charged current cross section at some high energy. As mentioned earlier, in  MINOS the normalization is set to the average of previous
$\sigma_{total} /E$ measurements for neutrino energies between 30 and 50 GeV. 

 There are three criteria for the  effectiveness of the ``low-$\nu$'' method. 
\begin{enumerate}
\item The  number of  ``low-$\nu$'' events 
that are used in the determination of the flux should not be a large fraction 
  of the total number of neutrino events in  each energy bin.
 \item The systematic uncertainty in the energy dependent  correction factor $f_C$ should be small. 
  \item The  number of  ``low-$\nu$'' events 
that are used in the determination of the flux should be sufficiently large
to have flux sample with small statistical errors.
\end{enumerate}
The first two criteria require a $\nu_{cut}$ which is as low as possible. The third
requires a $\nu_{cut}$  which is as large as possible.

 The MINOS collaboration uses the criteria that the fractional contribution of  
events with  $\nu<$ $\nu_{cut}$  to the total charged current cross section  should  be less than 60$\%$. 
 MINOS uses events with $\nu<1$ GeV for determination of the flux at their lowest
$\nu_\mu, \nub_\mu$ energies. 

  The  fraction of events with $\nu<1$ GeV  is less than  60$\%$  for $\nu_\mu$ interactions with  $E_{\nu}>3$ GeV and for  $\nub_\mu$ interactions with $E_{\nub}>5$ GeV.
Therefore, to determine the flux for $E_{\nu}< $ 3 GeV and  $E_{\nub}< 5$ GeV  we need to use a $\nu_{cut}$ which is smaller than 1 GeV.

We investigate $\nu_{cut}=0.25$ GeV  to be used for   $ E_{\nu,\nub}>0.7$ GeV,
and $\nu_{cut}=0.5$ GeV to be used for  $ E_{\nu,\nub}>1.4$ GeV.   These  samples  can be cross calibrated   against  the   $\nu_{cut}=1$ GeV sample
in the range  $ E_{\nu}>3$ GeV for neutrinos and $ E_{\nub}>5$ GeV for antineutrinos.  Similarly, they can be calibrated
against  the $\nu_{cut}=2$ GeV  and $\nu_{cut}=5$ GeV  samples in the range  $ E_{\nu,\nub}>9$ GeV  and $ E_{\nu,\nub}>18$ GeV, respectively.

\begin{figure}[t]
\includegraphics[width=3.5in,height=3.5in]{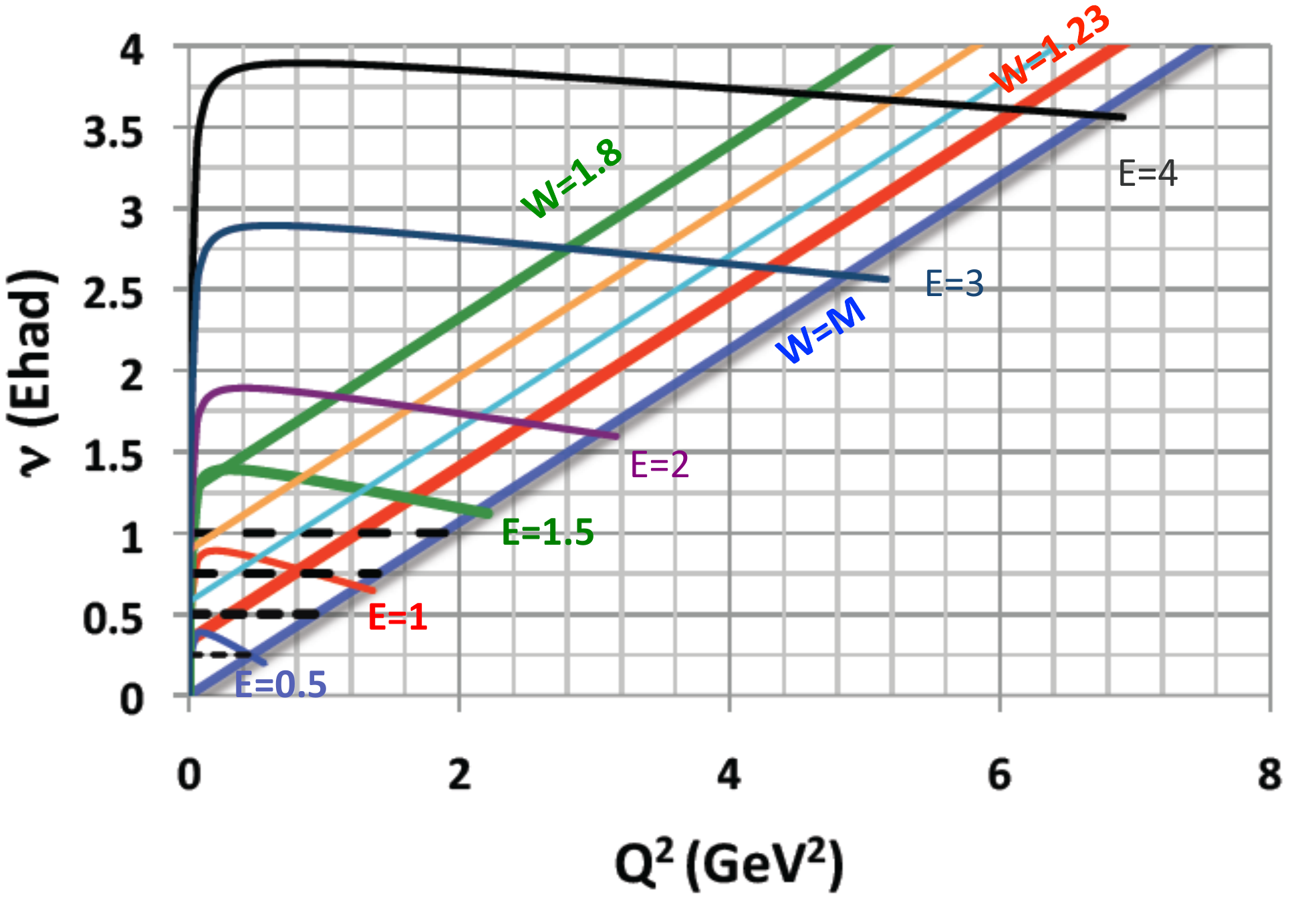}
\caption{The kinematic region in  the  $Q^2$ (in GeV$^2$),   $\nu=E_{had}$ (in GeV) plane for
 $\nu_{\mu}$ ($\nub_{\mu}$) energies less than 4 GeV. The shaded area is $\nu<$ 0.25 GeV (color online).  } 
\label{nuvsq24}
\vspace{-0.05in}
\end{figure}
     \vspace{-0.1in} 
%
\begin{figure}[t]
\includegraphics[width=3.5in,height=3.1in]{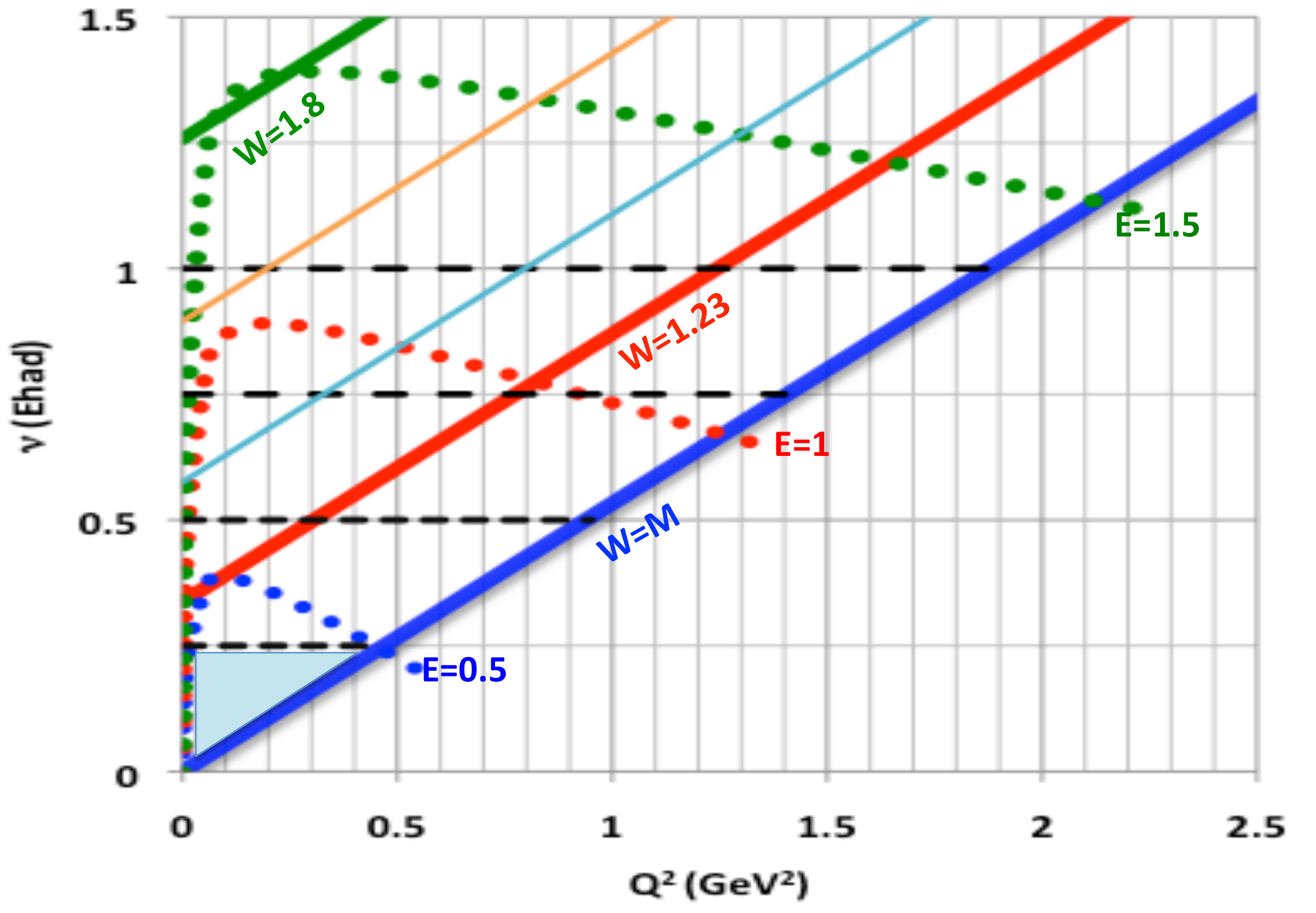}
\caption{ The kinematic region in  the  $Q^2$ (in GeV$^2$),   $\nu=E_{had}$ (in GeV) plane for
 $\nu_{\mu}$ ($\nub_{\mu}$) energies less than 1.5 GeV.  The shaded area is $\nu<$ 0.25 GeV (color online).  } 
\label{nuvsq22}
\vspace{-0.08in}
\end{figure}
%
\section{The ``low-$\nu$'' method at low energies \label{lnu}}
In the few GeV region, there are several types
of neutrino interaction processes as defined by 
the final state invariant mass $W$.   These 
 include quasielastic (QE) reactions ($W<1.07$ GeV),  production of
 the  $\Delta$(1232)  resonance ($1.1<W<1.4$ GeV), coherent pion production, production of higher mass
  resonances  ($1.4<W<2.0$ GeV)
 and the  inelastic continuum ($W> 2.0$ GeV ).
 Fig.~\ref{nuvsq24}  shows the  kinematic region in  $Q^2$ (in GeV$^2$ and $\nu=E_{had}$ (in GeV) for   $E_{\nu}<$ 4 GeV. 
 
 Fig.~ \ref{nuvsq22} shows
the kinematic region in  $Q^2$ and $\nu$ for $E_{\nu}< $  1.5 GeV.  
In this paper we focus on   $\nu<$0.25 GeV region  (shaded area in Fig.~ \ref{nuvsq24} and Fig.~\ref{nuvsq22})
for the lowest neutrino energies.  In addition, we  investigate the  $\nu<$0.50 GeV
region as an additional check.

 For  $E_{\nu}$= 3 GeV, about 1/3 of the total charged current  cross section
 originates from QE scattering, 1/3 from resonance production and
 1/3 from inelastic scattering.  
  
 As seen in Fig.~\ref{nuvsq22} the  $\nu<0.25$ GeV sample is dominated almost
 entirely by QE  events with $Q^2<0.45$  GeV$^2$.
 
 The  $\nu<0.5$ GeV   sample  includes  both  QE  events with $Q^2<0.95$ GeV$^2$ and  
also  $\Delta$(1232) resonance events with $Q^2<0.3$ GeV$^2$.  Both samples include a very small fraction of  events originating from coherent pion production (as discussed in Appendix II). 
  
 In the very ``low-$\nu$''  region it is more convenient to write the expression
 for the  charged current  differential cross sections as follows\cite{paschos,pyu}:
\begin{eqnarray} \di
\frac{d^2\si}{dQ^2 d\nu}=  S_{cos}  
\frac{1}{2E^2}{\cal W}_1 \left[Q^2+m_\mu^2\right] \nonumber \\
+ S_{cos}{\cal W}_2  \left[ (1-\frac {\nu}{E} )  - \frac {(Q^2+m_\mu^2)}{4E^2}  \right]  \nonumber\\  
+S_{cos}{\cal W}_3 \left[ \frac{Q^2}{2ME} -  \frac{\nu}{4E}  \frac { Q^2+m_\mu^2}{ME}  \right] \nonumber\\  
+ S_{cos} {\cal W}_4 \left[ m_\mu^2 \frac{(Q^2+m_\mu^2)}{4M^2E^2}\right] \nonumber \\ 
 - S_{cos}{\cal W}_5 \left[ \frac {m_\mu^2} {ME} \right],  
\label{cross-small}
\end{eqnarray}
where  $S_{cos}=\frac{G^2}{2\pi}\cos^2\theta_C = 80\times 10^{-40}~$\cm$^2$/GeV$^2$.
In the scattering process, there are additional small contributions from 
strangeness and charm
non-conserving processes. In the discussion below we do not
show these terms explicitly, but  
 charm and strangeness changing contributions are assumed to be  included in the analysis. 
(The strangeness changing  valence quark contributions are proportional to $\frac{G^2}{2\pi}\sin^2\theta_C$).

Each of the structure functions  has a vector and axial component (except for ${\cal W}_3$ which originates from  axial-vector interference).
The  vector part of ${\cal W}_4$  and ${\cal W}_5$ are well known since they are related to the
vector part of  ${\cal W}_2$ and  ${\cal W}_1$ by the following expressions\cite{paschos}:
\begin{eqnarray} 
{\cal W}_4^{vector}&=&{\cal W}_2^{vector}\frac{ M^2\nu^2 }{Q^4}-{\cal W}_1^{vector}\frac{ M^2 }{Q^2}.
\nonumber \\
{\cal W}^{vector}_5&=&{\cal W}_2^{vector}\frac{ M\nu }{Q^2}. \nonumber  
\end{eqnarray}

At `low-$\nu$' and  very high energy the charged current cross section is
only a function of ${\cal W}_2$.  If we integrate the
cross section  from $\nu_{min}\approx 0$ up to  $\nu$= $\nu_{cut}$ (where $\nu_{cut}$ is small), we can write the expression for
the cross section  in terms of  ${\cal W}_2$ only,  and energy dependent corrections
ratios to  the ${\cal W}_2$ component:
\begin{eqnarray}
 \sigma_{\nu cut}(E)  &=& \displaystyle\int_{\nu_{min (E)}}^{\nu_{cut}}   \di\frac{d^2\si}{dQ^2 d\nu}dQ^2 d\nu 
\\  \nonumber
&=& \sigma_{{W}_2} + \sigma_{2}+\sigma_{1}+\sigma_{3}+\sigma_{4}+\sigma_{5}, \nonumber
\label{cross-w123}
\end{eqnarray}
Here,
$\sigma_{{W}_2} \approx  \sigma_{{W}_2} (\infty)$, where
\begin{eqnarray} 
\sigma_{{W}_2}  &=& S_{cos}  \displaystyle\int_{\nu_{min (E)}}^{\nu_{cut}}  {\cal W}_2 ~d\nu.  \\
\sigma_{{W}_2} (\infty) &=& S_{cos}  \displaystyle\int_{\nu_{min (E=\infty)}}^{\nu_{cut}}  {\cal W}_2 ~d\nu.  
\end{eqnarray}
and the small corrections to the  QE cross section are:
\begin{eqnarray} 
\sigma_{2} &=&S_{cos} \displaystyle\int_{\nu_{min (E)}}^{\nu_{cut}} 
  \left[ -\frac{\nu}{E} -\frac{Q^2+m_\mu^2}{4E^2}  \right]{\cal W}_2 ~d\nu.  
\\
\sigma_{1} &=&S_{cos} \displaystyle\int_{\nu_{min (E)}}^{\nu_{cut}} 
 -  \left[ \frac{(Q^2+m_\mu^2)}{2E^2}   \right]{\cal W}_1 ~d\nu. \nonumber \\
\sigma_{3} &=&S_{cos} \displaystyle\int_{\nu_{min (E)}}^{\nu_{cut}} 
\left[ \frac{Q^2}{2ME} -  \frac{\nu}{4E}  \frac { Q^2+m_\mu^2}{ME}  \right]   {\cal W}_3 ~d\nu .\nonumber\\  
\sigma_{4} &=&S_{cos} \displaystyle\int_{\nu_{min (E)}}^{\nu_{cut}} 
 \left[ m_\mu^2 \frac{(Q^2+m_\mu^2)}{4M^2E^2} \right] {\cal W}_4 ~d\nu. \nonumber \\ 
\sigma_{5} &=&S_{cos} \displaystyle\int_{\nu_{min (E)}}^{\nu_{cut}}  \nonumber 
  \left[ \frac {-m_\mu^2} {ME} \right]  {\cal W}_5 ~d\nu. 
\label{w123}
\end{eqnarray}
The above can be written in terms of fractional corrections:
\begin{eqnarray} 
\sigma_{\nu cut}(E)&=&   \sigma_{{W}_2}({\infty}) \left[f_{C} \right],  \\ 
f_{C}&=&\left[ f_{W2}+ f_{2} + f_{1} + f_{3}+f_{4}+f_{5} \right],\nonumber \\ 
f_{W2} &=&\frac {\sigma_{W2} }{\sigma_{{W}_2}(\infty)}  \approx 1, \nonumber \\
f_{i} &=&\frac {\sigma_{i} }{\sigma_{{W}_2}(\infty)}.\nonumber 
\label{fw123}
\end{eqnarray}

 The energy dependent corrections  $f_{W2}$, 
 $f_{1}$,  $f_{2}$,  $f_{3}$,  $f_{3}$, and  $f_{4}$ and $f_{5}$  can be calculated within a specific models. 
 The theoretical uncertainty in $f_C$  determines the systematic
  uncertainty in the relative flux which can extracted from the ``low-$\nu$'' events.
 
 \begin{enumerate}
\item $f_{W2}
 =\frac {\sigma_{W2}  }{\sigma_{{W}_2} (\infty)} 
 \approx 1$ is well known and does not contribute to the uncertainty in $f_C$.
 \item The energy dependent correction $f_{2}$ is explicit and
therefore does not contribute to the uncertainty in $f_C$.
\item The contributions of $f_{4}$ and $f_{5}$ are small since they are proportional to the square of the  muon mass,
and therefore have a negligible contribution to the uncertainty in $f_C$. (Note that  the vector parts of  $f_{4}$ and $f_{5}$ are known very well since they can be expressed in terms of the vector parts of  ${\cal W}_1$ and ${\cal W}_2$).
\item The only  non-negligible uncertainty originates from the modeling of the  contributions of   $f_{1}$ and  $f_{3}$ (primarily from  $f_{3}$). 
 \end {enumerate}

   \begin{figure}
\includegraphics[width=3.7in,height=3.0in]{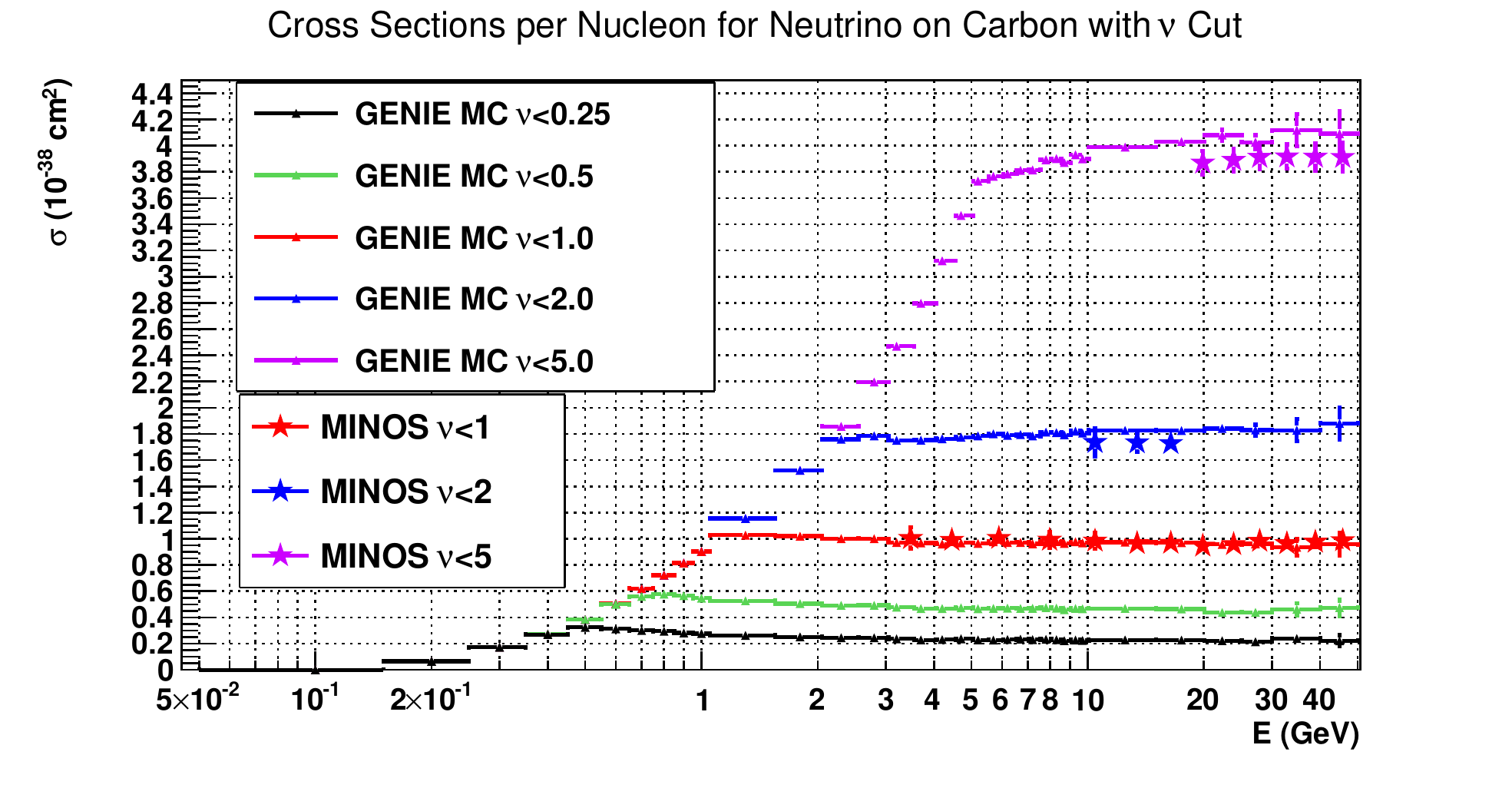}
     
     \vspace{-0.2in} 
     
\includegraphics[width=3.7in,height=3.0in]{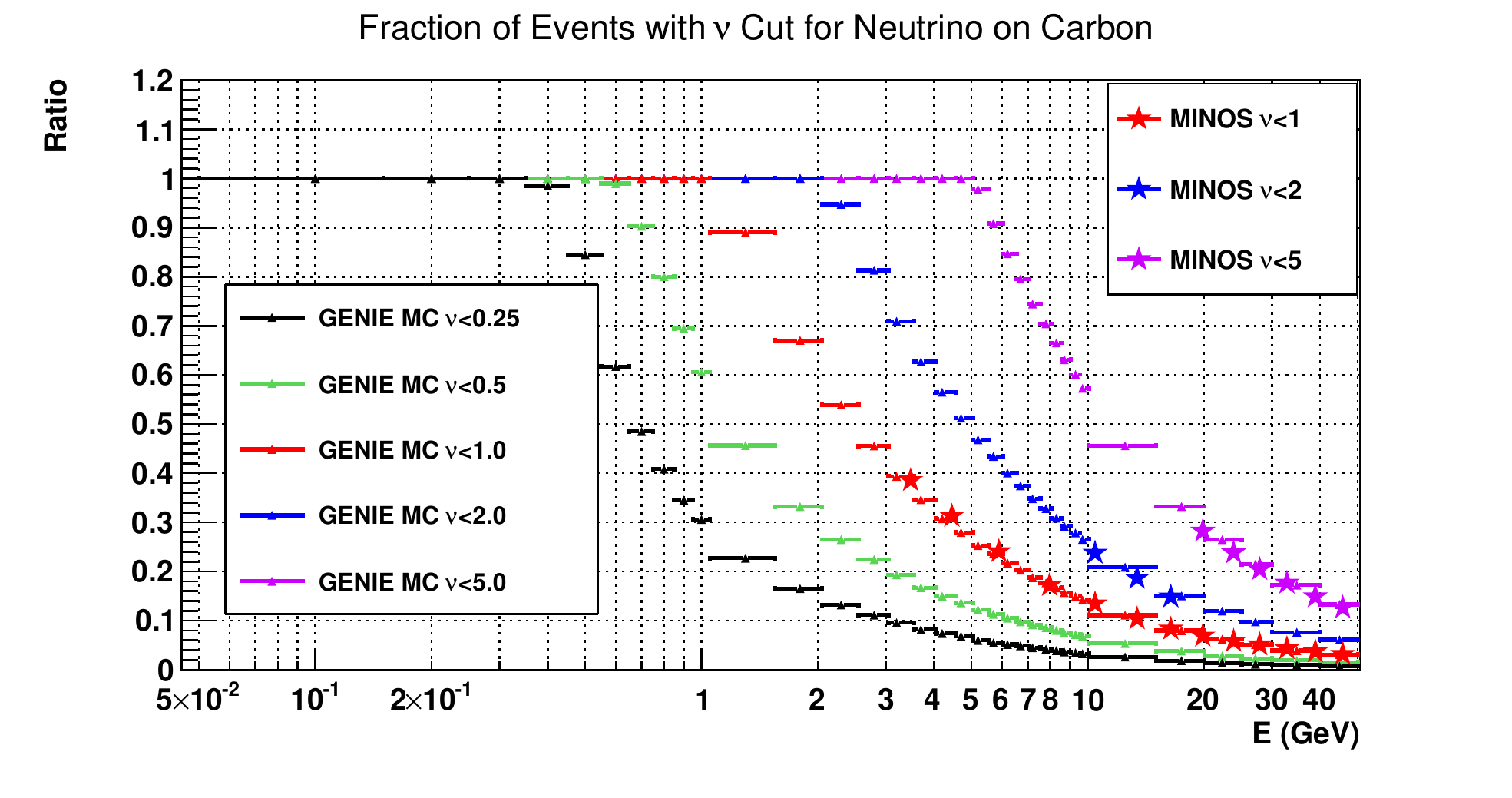}
\vspace{-0.3in}
\caption{Top panel: Neutrino partial charged cross sections per nucleon for  ``low-$\nu$'' events 
 determined from the GENIE Monte Carlo\cite{GENIE}.
 Also shown
are the  measurements of MINOS on iron (per nucleon corrected for the excess number of neutrons). 
 Bottom panel:  The  fraction of  ``low-$\nu$''  neutrino events  
in the GENIE\cite{GENIE} Monte Carlo as
compared with the measurements of MINOS (color online).  } 
 \label{nufrac}
\end{figure}

  \begin{figure}
\includegraphics[width=3.7in,height=3.0in]{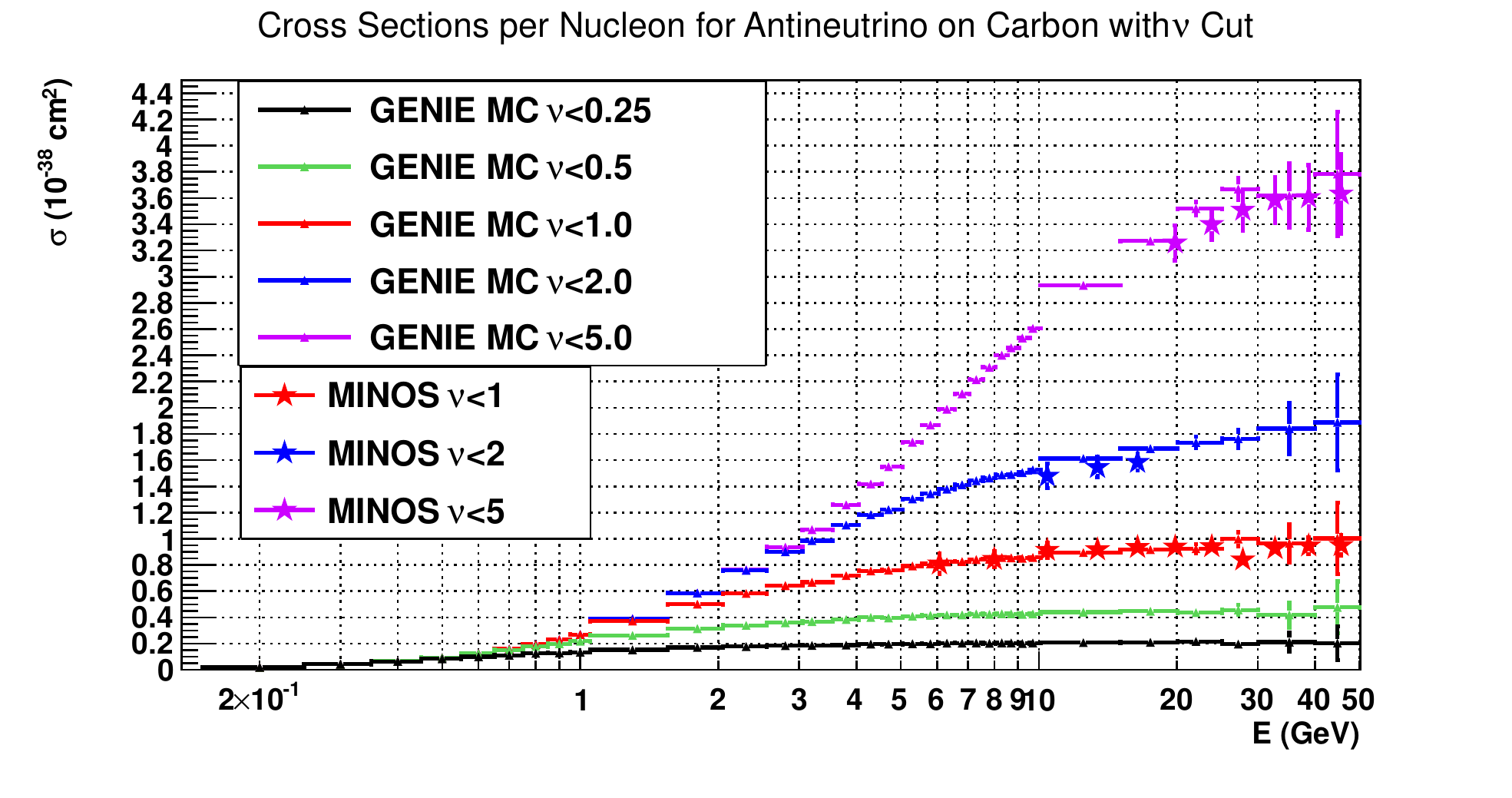}
     
     \vspace{-0.2in} 
     
\includegraphics[width=3.7in,height=3.0in]{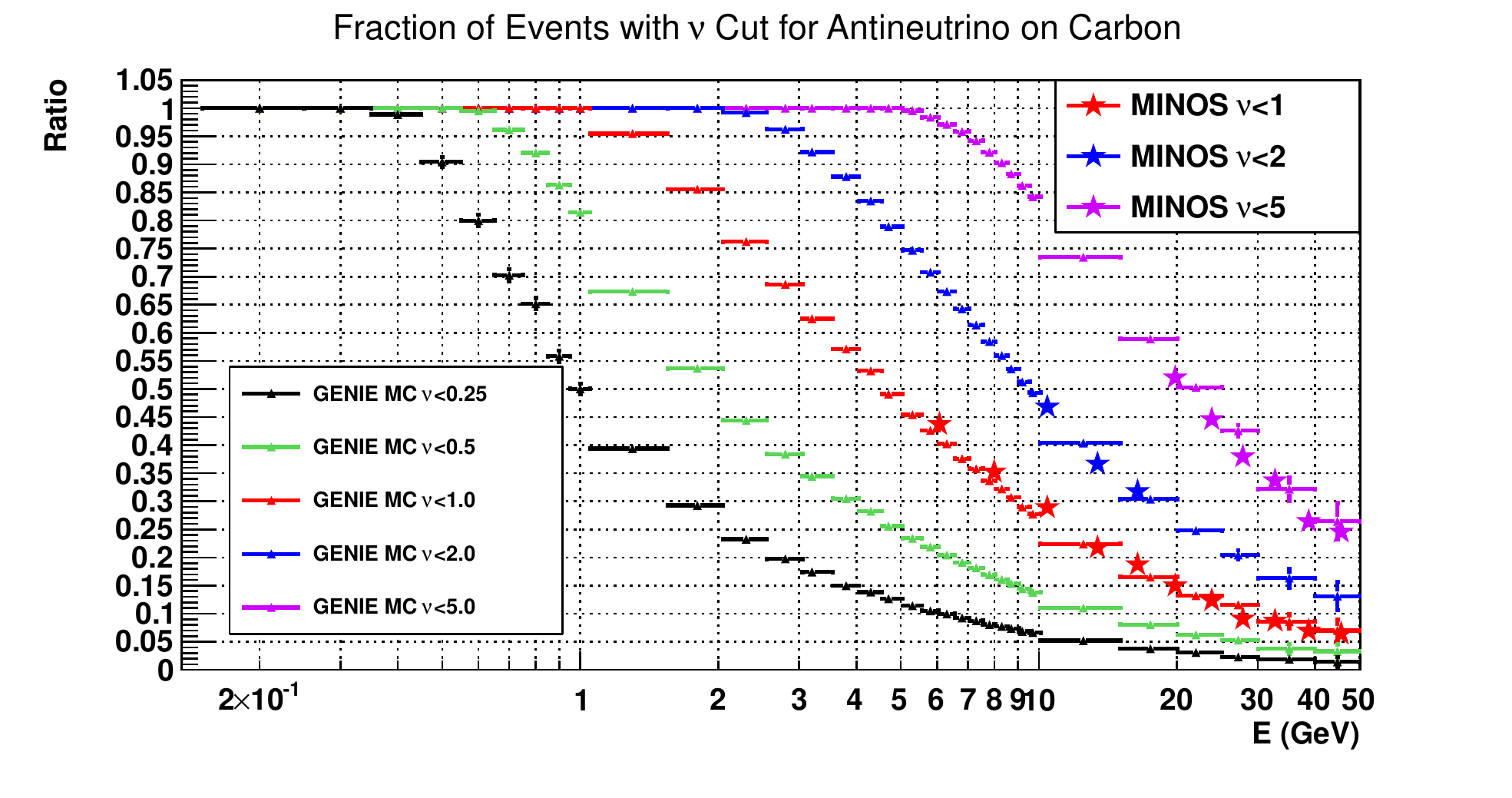}
\vspace{-0.3in}
\caption{  Same as  Fig.  \ref{nufrac} for the  case of antineutrinos.
(color online). 
} 
 \label{nufrac2}
\end{figure}

The technique does not depend on  the modeling of ${\cal W}_2$ because  the $\sigma_{{W}_2}$  cross section is the same at all energies.  All energy dependent corrections
 are expressed in terms of ratios to $\sigma_{{W}_2}$.   In quark parton language,
 the uncertainty in  $f_1$ is related  to the uncertainty in the longitudinal structure function at low $Q^2$
  and the uncertainty in  $f_3$ is related to the uncertainty in  level of antiquarks in the nucleon at low $Q^2$. 
   For  QE   scattering and resonance production  the structure functions are expressed in terms of form factors.

\subsection{ Partial charged current  cross sections}
 
The top panel of Fig. \ref{nufrac} shows 
the partial neutrino charged current cross section per nucleon for  ``low-$\nu$'' events (for $\nu$ cuts of 0.25, 0.5, 1, 2 and 5 GeV)  as a function of energy as determined by the  GENIE\cite{GENIE}  Monte Carlo for a carbon target.  
 The top panel of   Fig. \ref{nufrac2}  shows the corresponding  partial charged current cross sections for antineutrinos.

 Also shown
are the  measurements of the partial charged current  cross sections on iron from the MINOS collaboration (for $\nu$ cuts of  1, 2 and 5 GeV).  
The MINOS cross sections for iron have been corrected for the excess number of neutrons in iron.  Note that the
nuclear corrections to the structure functions in iron nucleus are larger than in carbon. 
Therefore,  the partial cross sections on carbon and on iron may not be the same.

 At high energies (as shown in Fig. \ref{nufrac} and \ref{nufrac2})  the partial cross sections for a fixed  $\nu_{cut}$ are independent of energy and are approximately equal for neutrinos and antineutrino.  The fact that these partial charged current cross section are relatively independent of energy is the basis for the ``low-$\nu$'' method. 

The bottom panels of figures  \ref{nufrac} and \ref{nufrac2} show  the  fraction of  ``low-$\nu$''  events predicted by the GENIE Monte Carlo as
compared with the measurements in MINOS.
In order to use the technique at low energies the fractions must be smaller than 0.6. Therefore,
at the lowest energies  we  must use $\nu$ cuts of 0.25 and 0.50 GeV.
  
  MINOS is a sampling
target calorimeter which  has poor resolution at low hadron energy. Therefore,  ``low-$\nu$'' samples with 
  $\nu< 0.25~$GeV and $\nu< 0.5~$GeV cannot be  defined reliably.
  On the other hand, since the  MINERvA detector is a  fully active  target  calorimeter,   ``low-$\nu$'' samples with   $\nu< 0.25~$GeV and $\nu< 0.5~$GeV can be used.
 
 \subsection{ Absolute normalization}
  
Since the neutrino energy range for MINERvA is limited to lower energies,
we propose  that the MINERvA charged current cross section
measurements  be normalized to the   cross section in the energy range
between 10 to 20 GeV (e.g. at a mean energy of  15.1 GeV).
  The absolute level of the charged current cross section
at this energy range has been measured by both the MINOS and NOMAD collaborations.

  The MINOS total cross section measurement
for an isoscalar iron target at a neutrino energy of 15.1 GEV is
 $$\sigma^{MINOS}_\nu  /E  =  0.708\pm0.020\times 10^{-38}\cm^2/\GeV $$
  per nucleon in iron. Here the total error of 0.02 is the combined 
 statistical, systematic and normalization errors of  $0.008\pm0.012\pm0.015$,
 respectively.

The NOMAD cross section  measurement 
for an isoscalar carbon target at a neutrino energy of 15.1 GEV is
 $$\sigma^{NOMAD}_\nu/ E=  0.698\pm0.025\times 10^{-38}\cm^2/\GeV  $$
  per nucleon in carbon.
  
   The MINOS total cross section measurement
for an isoscalar iron target at an antineutrino energy of 15.1 GEV is
 $$\sigma^{MINOS}_{\nub}  /E  =  0.304\pm0.012\times 10^{-38}\cm^2/\GeV $$
  per nucleon in iron. Here,  the total error of 0.012 is the combined 
 statistical, systematic and normalization errors of  $0.007\pm0.007\pm0.006$,
 respectively.
 
 Alternatively, it may be possible for MINERvA  to normalize to the partial cross sections measured by
  MINOS  for  $\nu< 1~$GeV  and $\nu< 2~$GeV  at  15.1 GeV.   These partial cross sections (which were used by MINOS to determine their relative flux) 
 are relatively constant between 10 and 20 GeV.   However,  the MINOS partial  cross sections are measured on iron. The MINERvA target
 is solid scintillator (i.e. carbon), and the partial cross sections for iron and carbon can be different.
  For a  neutrino 
 energy of 15.1  GeV MINOS 
  measured  the following  isoscalar partial cross sections on iron (per nucleon):
 $$\sigma^{MINOS}_{\nu} (15.1)=1.729\pm 0.049 \times 10^{-38}\cm^2 ( \nu<2~\GeV) $$
  $$\sigma^{MINOS}_{\nu}(15.1) = 0.968\pm 0.027\times 10^{-38}\cm^2 ( \nu<1~\GeV) .$$
  
  For an antineutrino 
 energy of 15.1 GeV MINOS 
 has measured  the following   isoscalar partial cross sections:
 $$\sigma^{MINOS}_{\nub} (15.1)= 1.585\pm 0.063 \times 10^{-38}\cm^2 ( \nu<2~\GeV) $$
  $$\sigma^{MINOS}_{\nub}(15.1) = 0.939\pm 0.039 \times 10^{-38}\cm^2 ( \nu<1~\GeV). $$

            \begin{figure}
\includegraphics[width=3.7in,height=2.6in]{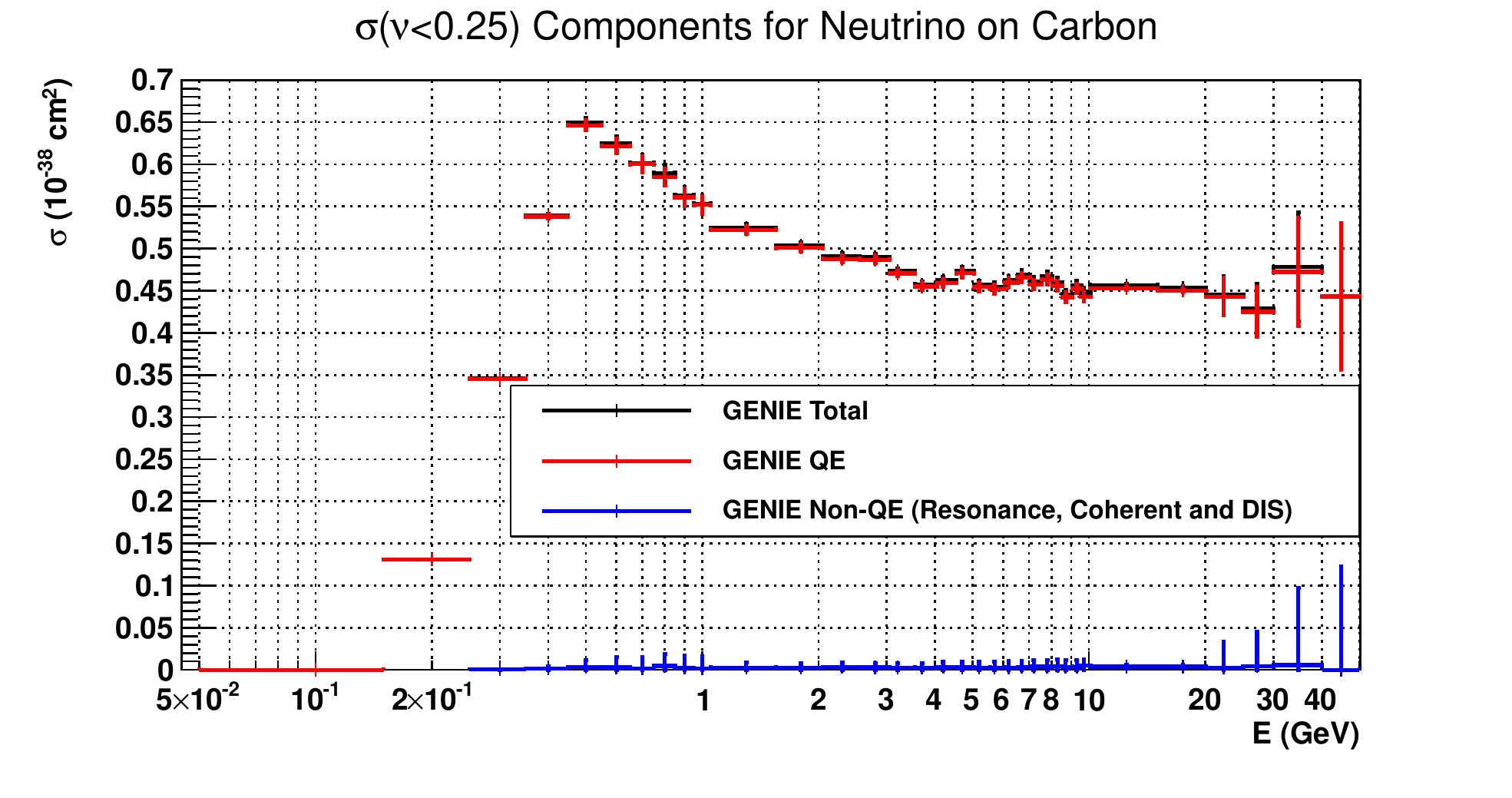}
     
     \vspace{-0.15in} 
     
\includegraphics[width=3.7in,height=2.6in]{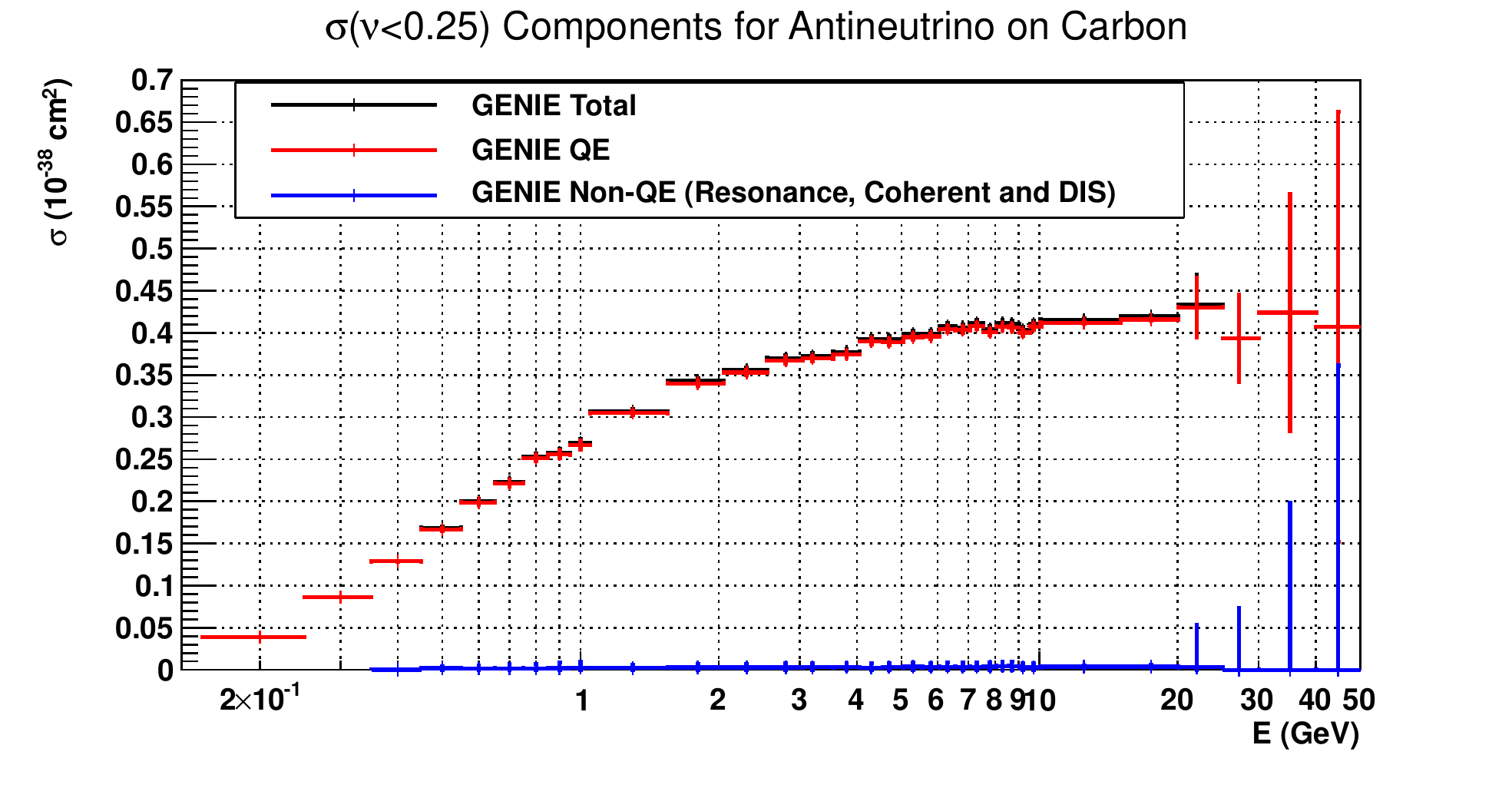}    
\vspace{-0.2in}       
\caption{The $\nu<0.25$ GeV partial charged current cross sections (per nucleon)  as a function of energy from the GENIE Monte Carlo
(for  carbon target). 
 Shown are the QE contribution, the contribution from pion production process (e.g.  $\Delta$, inelastic and coherent pion) and
 the total.  The $\nu<0.25$ GeV cross sections for $\nu_\mu$
 are shown on the top panel, and the $\nu<0.25$ GeV cross sections for $\nub_\mu$
are shown on the bottom panel. 
  (color online).}
\label{QEvsDelta25}
\end{figure}

\section{Using  ``low-$\nu$'' events with $ \nu<0.25~$GeV }
 As seen in Fig.~\ref{nuvsq22} the $\nu<0.25$ GeV region is dominated by QE  events.   
 This is illustrated in Fig.~\ref{QEvsDelta25} which shows the relative contributions
 of QE and non-QE processes to    $\nu<0.25~$GeV cross section as a function of energy  (as determined from the GENIE Monte Carlo).  The $\nu<0.25~$GeV cross sections for $\nu_\mu$
are shown on the top panel, and the $\nu<0.25~$GeV cross sections for $\nub_\mu$
are shown on the bottom panel.
The QE contribution is shown in red, the contribution from pion production process (e.g.  $\Delta$, inelastic and coherent pion)  is shown in blue and  the total is shown in black.
   Most of the   events are QE and the contribution from pion production processes is
   negligible.

  As mentioned earlier, the  technique does not rely on the modeling of ${\cal W}_2$,
 or the modeling of nuclear effects (e.g.  Fermi motion smearing)   on ${\cal W}_2$. This is because  the cross section   $\sigma_{{W}_2}$  (including nuclear effects)   is the same at all neutrino energies.
 
The uncertainty in the flux extracted from the  event sample with $ \nu<0.25~$GeV  is determined by how well we can model  the relative contributions of  ${\cal W}_1$ and ${\cal W}_3$ for the case of  QE  scattering on bound nucleons, or equivalently  the relative contributions of  $f_{1}$ and  $f_{3}$ to $f_C$.    Here  $f_{1}$ and  $f_{3}$   are  proportional to  the ratios  $\frac{{\cal W}_1}{{\cal W}_2}$ and $\frac{{\cal W}_3}{{\cal W}_2}$. Since the  ratios $\frac{{\cal W}_1}{{\cal W}_2}$ and $\frac{{\cal W}_3}{{\cal W}_2}$  for QE  scattering on  free nucleons are very well known,  the uncertainty in $f_C$  originates primarily
from  modeling  the nuclear corrections to $\frac{{\cal W}_1}{{\cal W}_2}$ and $\frac{{\cal W}_3}{{\cal W}_2}$ for nucleons bound in a nuclear target. 
%
%
 %
 	 \begin{figure}[ht]
\includegraphics[width=3.3in,height=2.5in]{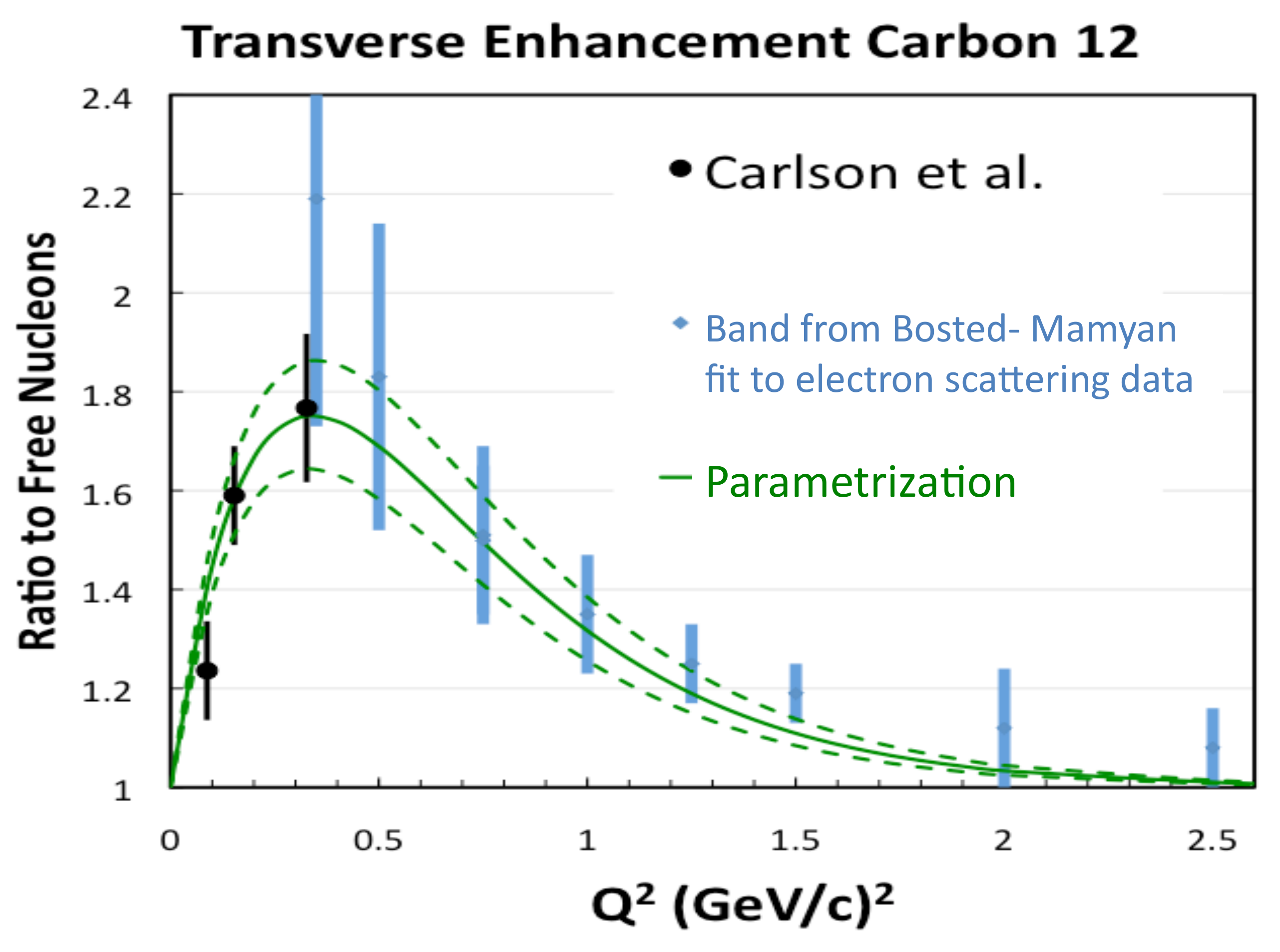}
\caption{ The transverse enhancement ratio\cite{transverse} (${\cal R}_{T}$) as a function of $Q^2$. Here, ${\cal R}_{T}$ is  ratio of the  integrated transverse response function
for QE  electron scattering on nucleons bound in  carbon divided by 
the integrated response function for independent nucleons. 
 The black points are extracted from Carlson $et~al$\cite{MEC4}, and
the blue bands are extracted from a fit\cite{vahe-thesis} to QE data from the JUPITER\cite{JUPITER} experiment (Jlab experiment E04-001). The curve is a fit to the data of the form  ${\cal R}_{T}=1+AQ^2e^{-Q^2/B}$. The dashed lines are the upper and lower error bands (color online).
}
\label{GMPN}
\end{figure} 

  \begin{figure}
\includegraphics[width=3.5 in,height=2.9in]{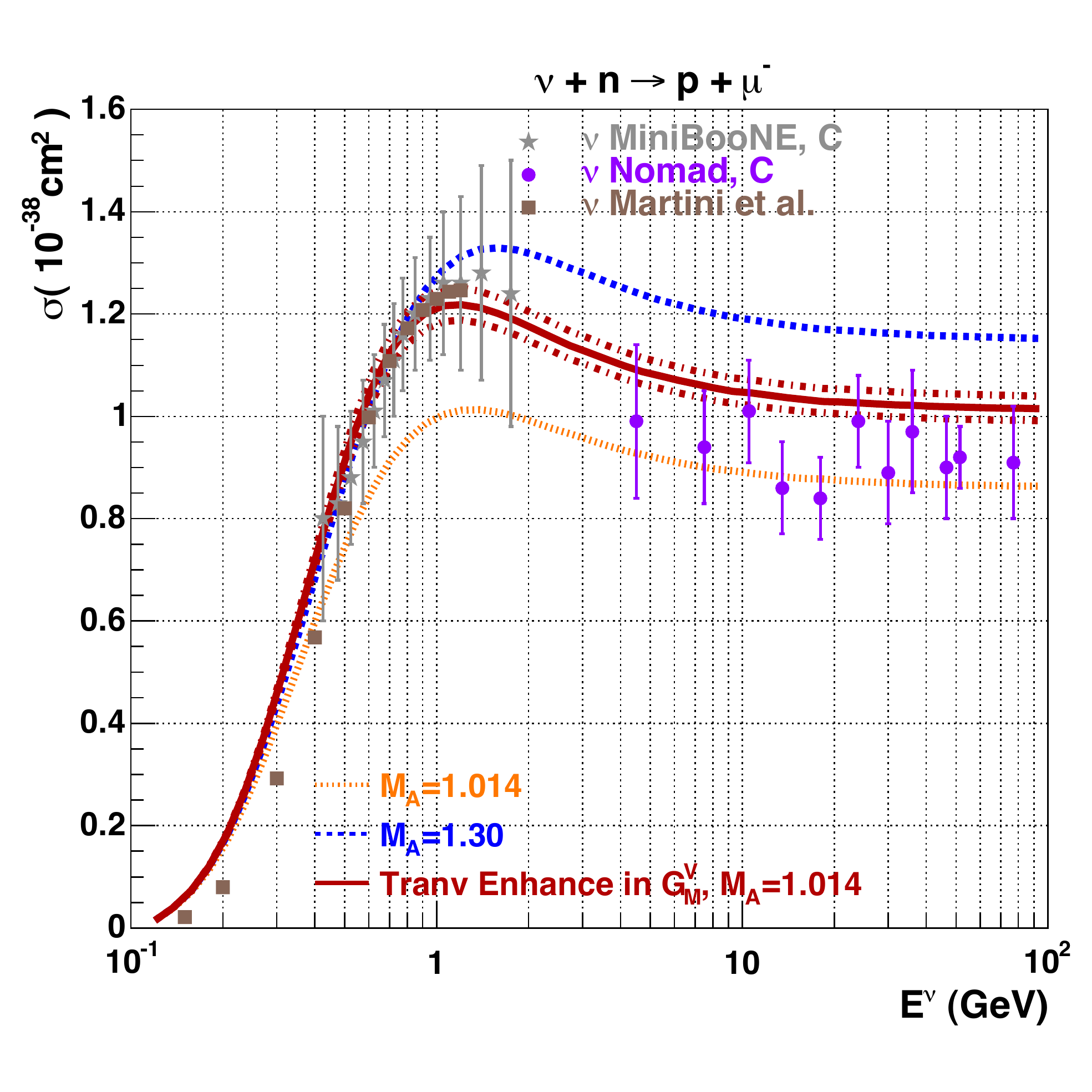}
     
     \vspace{-0.2in} 
     
\includegraphics[width=3.5 in,height=2.9in]{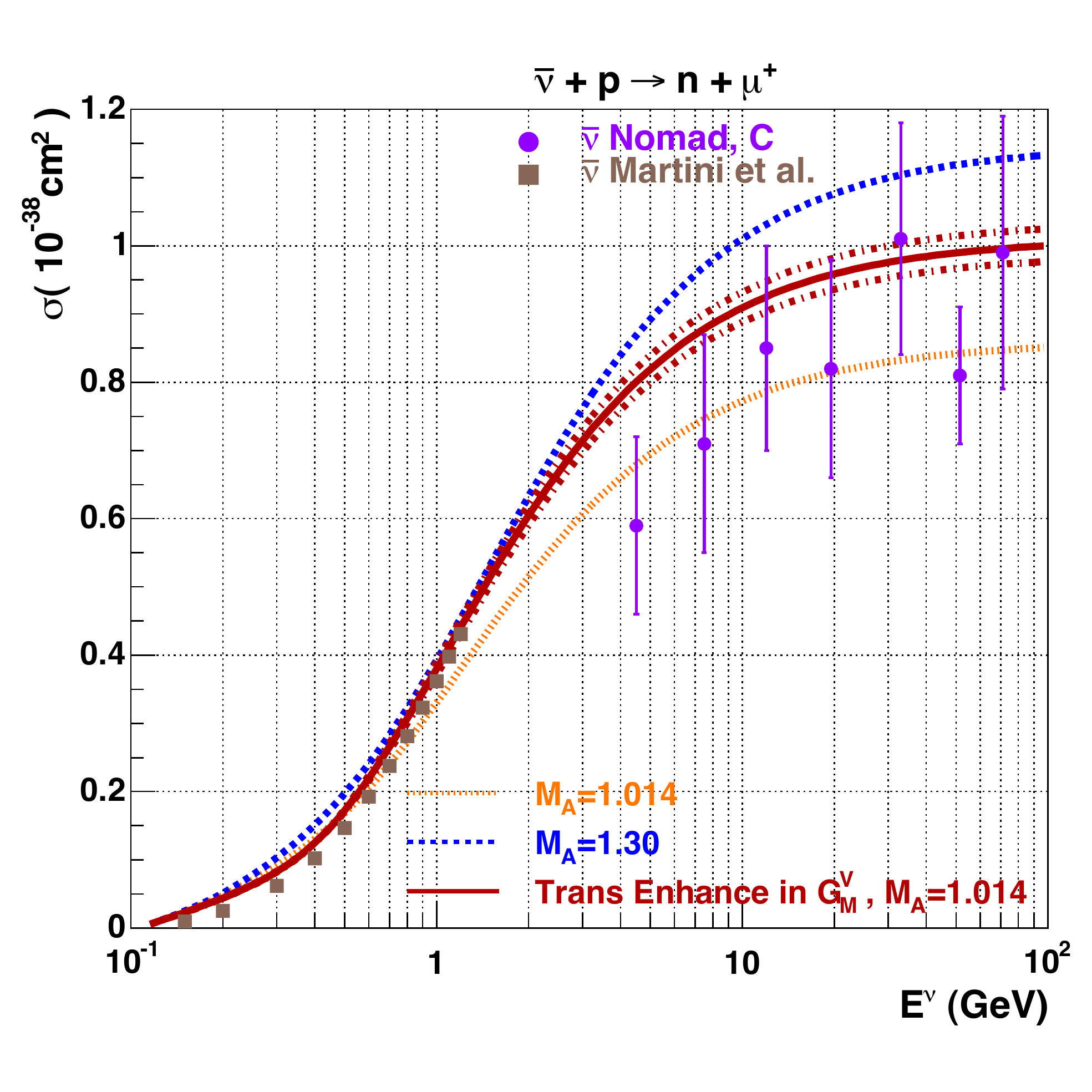}
\vspace{-0.3in}
\caption{Comparison of predictions for the   $\nu_{\mu}$, $\bar{\nu}_\mu$  total QE cross section sections from  the nominal  TE model,
the "Independent Nucleon (MA=1.014)"  model, the 
"Larger $M_A$  ($M_A$=1.3) model",   and  the 
  "QE+np-nh RPA"  MEC model of Martini et al.\cite{MEC5} 
  The data points are the   measurements of MiniBooNE\cite{MiniBooNE} (gray stars)  and NOMAD\cite{NOMAD} (purple circles) (color online).}
\label{totalMEClog}
\end{figure}
%
  \begin{figure}
\includegraphics[width=3.7 in,height=2.9in]{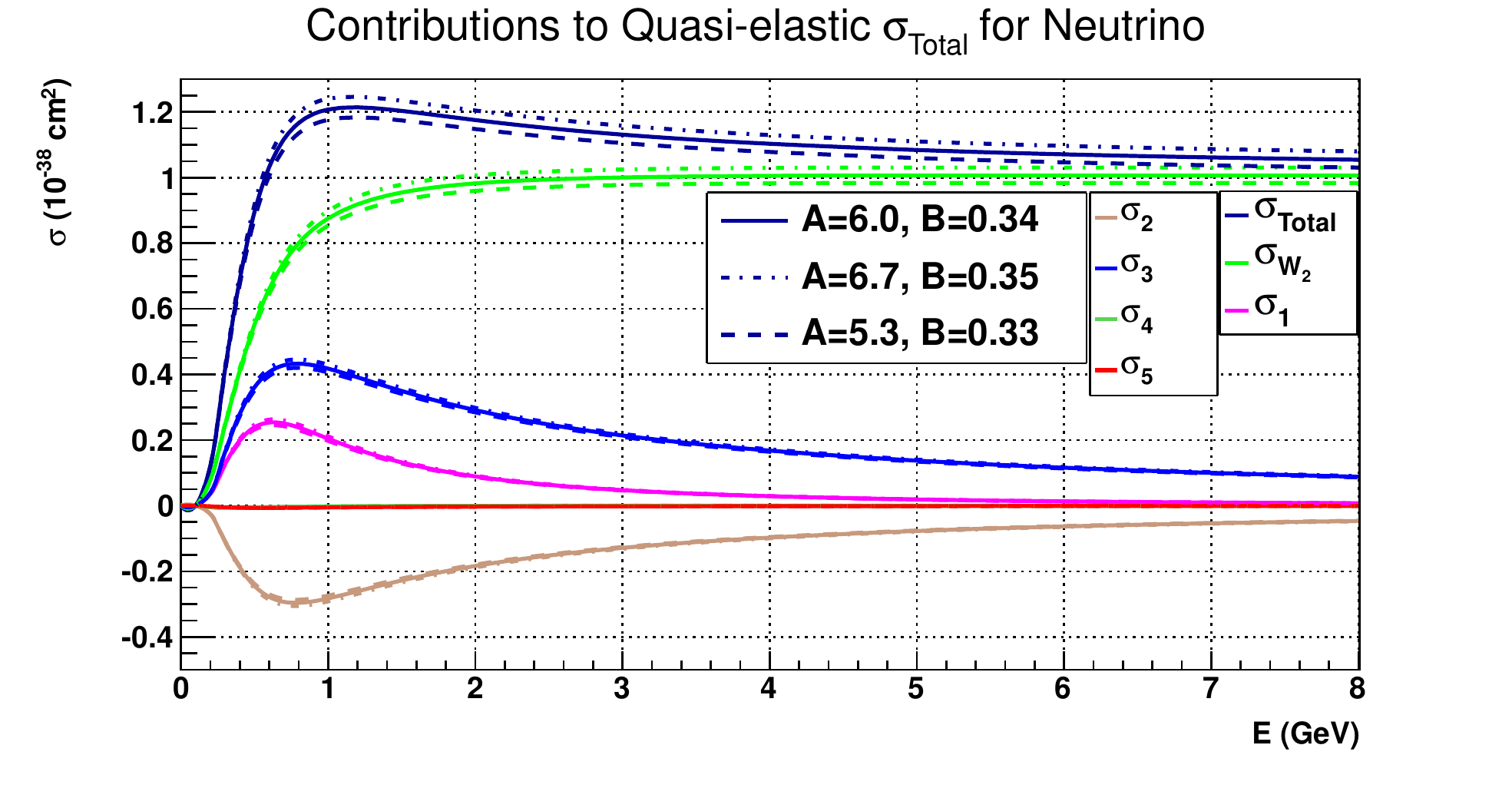}
     
     \vspace{-0.2in} 
     
\includegraphics[width=3.7 in,height=2.9in]{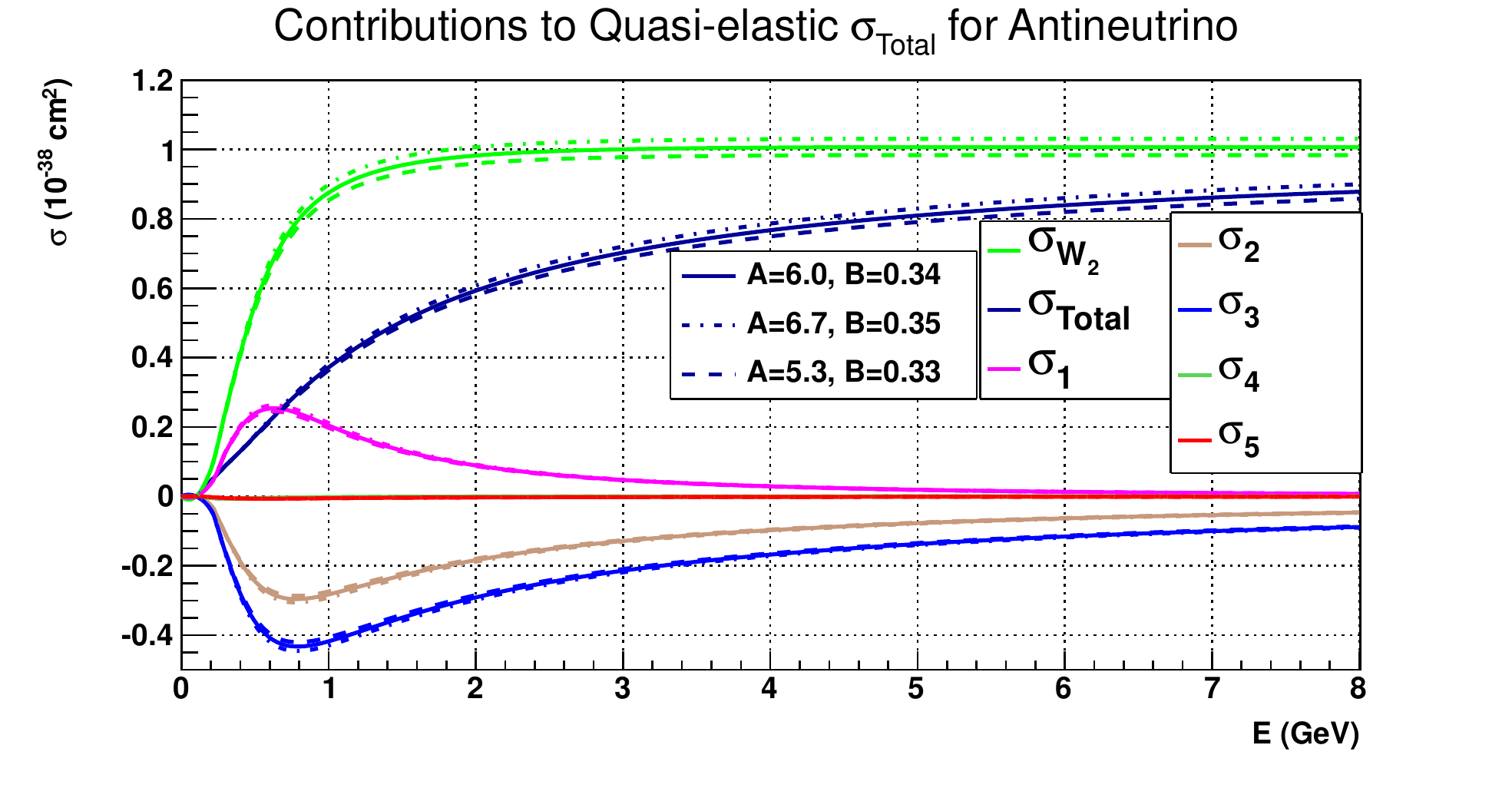}
\vspace{-0.2in}
\caption{Contribution of the various components
($\sigma_{W_2}$, $\sigma_{2}$,  $\sigma_{1}$, $\sigma_{3}$,  $\sigma_{4}$, $\sigma_{5}$) 
  to the total QE cross section (as predicted by the  TE  model).  Top panel: Neutrinos. Bottom panel: Antineutrinos (color online).
  }
\label{totalMECcomp}
\end{figure}

\subsection{Quasielastic   $\nu_{\mu},\bar{\nu}_\mu$   scattering}
The relationship between the structure functions and form factors for $\nu_\mu,\bar{\nu}_\mu$ 
QE scattering\cite{Lle72}  on free nucleons is given by\cite{transverse,steffens}:

$$W^{\nu-vector}_{1-Qelastic} =\delta(\nu-\frac{Q^2}{2M})\tau |{\cal G}_M^V (Q^2)|^2$$
$$W^{\nu-axial}_{1-Qelastic} = \delta(\nu-\frac{Q^2}{2M})(1+\tau)|{\cal F}_A (Q^2)|^2$$
$$W^{\nu-vector}_{2-Qelastic} =
 \delta(\nu-\frac{Q^2}{2M})|{
 \cal F}_V (Q^2)|^2$$
$$W^{\nu-axial}_{2-Qelastic} =
\delta(\nu-\frac{Q^2}{2M})|{\cal F}_A (Q^2)|^2$$
$$W^{\nu}_{3-Qelastic} = 
\delta(\nu-\frac{Q^2}{2M})|2 {\cal G}_M^V(Q^2) {\cal F}_A (Q^2)|$$
 $$W^{\nu-vector}_{4-Qelastic} =
 \delta(\nu-\frac{Q^2}{2M})\frac{1}{4}(|{\cal F}_V (Q^2)|^2 - |{\cal G}_M^V (Q^2)|^2)$$
  $$W^{\nu-axial}_{4-Qelastic} =
 \delta(\nu-\frac{Q^2}{2M})\times \frac{1}{4} \times $$
$$ \left[ {\cal F}_A^2(Q^2)+(\frac{Q^2}{M^2}+4)|{\cal F}_p(Q^2)|^2 -({\cal F}_A(Q^2) +2{\cal F}_P(Q^2))^2 \right]$$
 $$W^{\nu-vector}_{5-Qelastic} =
 \delta(\nu-\frac{Q^2}{2M})\frac{1}{2}|{\cal F}_V (Q^2)|^2$$
  $$W^{\nu-axial}_{5-Qelastic} =
 \delta(\nu-\frac{Q^2}{2M})\frac{1}{2}|{\cal F}_A (Q^2)|^2$$
 where
$$ {\cal G}_E^V(Q^2)=G_E^p(Q^2)-G_E^n(Q^2), $$
$$
{\cal G}_M^V(Q^2)=G_M^p(Q^2)-G_M^n(Q^2). 
$$
and
  $$ |  {\cal F}_V(Q^2)|^{2}=
\frac{[{\cal G}_E^V(Q^2)]^2+ \tau [{\cal G}_M^V(Q^2)]^2}{1+\tau}.$$
Here, $G_E^p(Q^2)$, $G_E^n(Q^2)$, $G_M^p(Q^2)$ and $G_M^n(Q^2)$
are the electric and magnetic nucleon form factors, which are measured
in electron scattering experiments. 
Note that:
 $$\sigma_T^{vector}\propto  \tau |{\cal G}^V_M (Q^2)|^2;~~\sigma_T^{axial}\propto (1+\tau)|{\cal F}_A (Q^2)|^2$$
 $$\sigma_L^{vector}\propto {({\cal G}_E^V(Q^2))^2}; ~~~\sigma_L^{axial}= 0$$
 
 Therefore, for QE  $\nu_{\mu},\bar{\nu}_\mu$  scattering  only ${\cal G}_M^V$ contributes
 to the   transverse  virtual
 boson absorption cross section.
 %
 \subsection{Transverse enhancement QE scattering from nuclei}
 Studies of QE  electron  scattering on nuclear targets\cite{MEC4} indicate
that only the longitudinal part of the QE cross section  can be described in terms of a universal response function of 
independent nucleons bound in a nuclear potential (and free nucleon form factors).  In contrast, a significant additional enhancement with respect to the model is observed in the transverse part of the QE cross section.

The enhancement  in the transverse QE cross section has been attributed\cite{MEC4} to meson exchange currents (MEC) in a nucleus. Within models of meson exchange currents\cite{MEC4} the enhancement is primarily in the transverse part of the QE cross section, while the  enhancement in the longitudinal QE cross section is small (in agreement with the electron scattering  experimental data).

 The conserved vector current hypothesis (CVC) implies that  the  corresponding  vector structure function for  the QE cross section in $\nu_{\mu},\bar{\nu}_\mu$ scattering can be expressed in terms of the  structure functions measured in electron scattering on nuclear targets. Therefore, there should also be a transverse enhancement in neutrino scattering. In models of  meson exchange currents the enhancement in the 
axial part of $\nu_{\mu},\bar{\nu}_\mu$ QE cross section on nuclear targets is also expected small.
 
The transverse enhancement observed in electron scattering is a function of both $Q^2$ and
$\nu$.  However, a simple way to account for the integrated transverse enhancement\cite{transverse}  from nuclear
effects is to assume that  $G_M^p(Q^2)$ and $G_M^n(Q^2)$ are enhanced 
in a nuclear targets by  factor
$\sqrt{R_{TL}}$.  

 Bodek, Budd and Christy\cite{transverse} have used
electron scattering data\cite{MEC4,JUPITER,vahe-thesis} to parametrize  ${R_{TL}}$ as follows: 
$$R_{TL}=1+AQ^2e^{-Q^2/B}$$
   with $A=6.0$ and $B=0.34$ GeV$^2$.  The electron scattering data indicates that the transverse enhancement is maximal near $Q^2$=0.3 GeV$^2$ and is small for $Q^2$ greater
than 1.5 GeV$^2$. 
The upper error band is given by  $A=6.7$ and $B=0.35$ GeV$^2$, and
the lower error band is given by $A=5.3$ and $B=0.33$ GeV$^2$.
 
 In modeling  $\nu_{\mu},\bar{\nu}_\mu$  QE scattering on nuclear targets we use   $BBBA2007_{25}$ parameterization\cite{quasi}
 of the free nucleon electromagnetic form factors $G_E^p(Q^2)$, $G_E^n(Q^2)$,
$G_M^p(Q^2)$ and $G_M^n(Q^2)$  (with $M_V^2=0.71$ GEV$^2$), and  a dipole axial form factor with  $M_A=1.014~$GeV.    We apply the transverse enhancement correction to  $G_M^p(Q^2)$ and $G_M^n(Q^2)$. We also apply  Pauli blocking corrections to the differential QE cross section as parametrized by Paschos and Yu\cite{pyu}. 
We refer to this model as the Transverse Enhancement (TE)  model. This is the nominal model that is used
in this paper. 

We also compare calculations based on the nominal TE model  to two other
models. The first model is  the  independent nucleon model with Pauli blocking with  $M_A=1.014~$GeV, without transverse enhancement.  We refer to this model as the "Independent Nucleon (MA=1.014)"  model.  This model,
which is used by the NOMAD \cite{NOMAD} collaboration,  is very close to the model which is currently implemented in the GENIE Monte Carlo (the GENIE default value is  $M_A=0.99~$GeV). 
The second model is the  independent nucleon model with Pauli blocking, $M_A=1.3$ GeV, without transverse enhancement.  This model is used by the MiniBooNE Collaboration\cite{MiniBooNE}.   We refer to this model as the "Larger $M_A$  ($M_A$=1.3) model". We use the difference between the three models  as a conservative systematic error on the flux extracted from the $\nu$ samples.

Fig. \ref{totalMEClog}  shows  a comparison of predictions of various  model predictions for the  $\nu_{\mu}$, $\bar{\nu}_\mu$  total QE cross section sections  to experimental data on nuclear targets. 
 Shown are  "Independent Nucleon (MA=1.014)"  model, the 
"Larger $M_A$  ($M_A$=1.3) model",  and  the TE model 
(with upper and lower error bands).
Also shown are the predictions of  the   "QE+np-nh RPA"  MEC model of Martini et al.\cite{MEC5}   The data points are the  QE cross section measurements  of MiniBooNE\cite{MiniBooNE} (gray stars)  and NOMAD\cite{NOMAD} (purple circles). Note that there is an overall $\approx10\%$  systematic error in the experimental QE cross sections because of uncertainties in the determination of the neutrino and antineutrino fluxes in each of the two experiments. 

In this paper we use the error band in the transverse enhancement parameters as a lower limit on  systematic error in the modeling.   We use the  "Independent Nucleon (MA=1.014)"   and   the "Larger $M_A$  ($M_A$=1.3) model" as conservative  upper limits on the errors in the modeling. 
 
Fig. \ref{totalMECcomp}  shows the contribution of the various components
($\sigma_{W_2}$, $\sigma_{2}$, $\sigma_{1}$, $\sigma_{3}$,  $\sigma_{4}$, $\sigma_{5}$)  to the total QE cross section (as defined by Eq. \ref {w123}) as a function
of incident energy.  These contributions are calculated using the TE model.  The top panel shows the contribution of the various 
components for the  neutrino QE cross section,  and the bottom panel shows
the contribution of the various components for the  antineutrino QE cross section. 
%
%
\begin{figure}
\includegraphics[width=3.5in,height=3.5in]{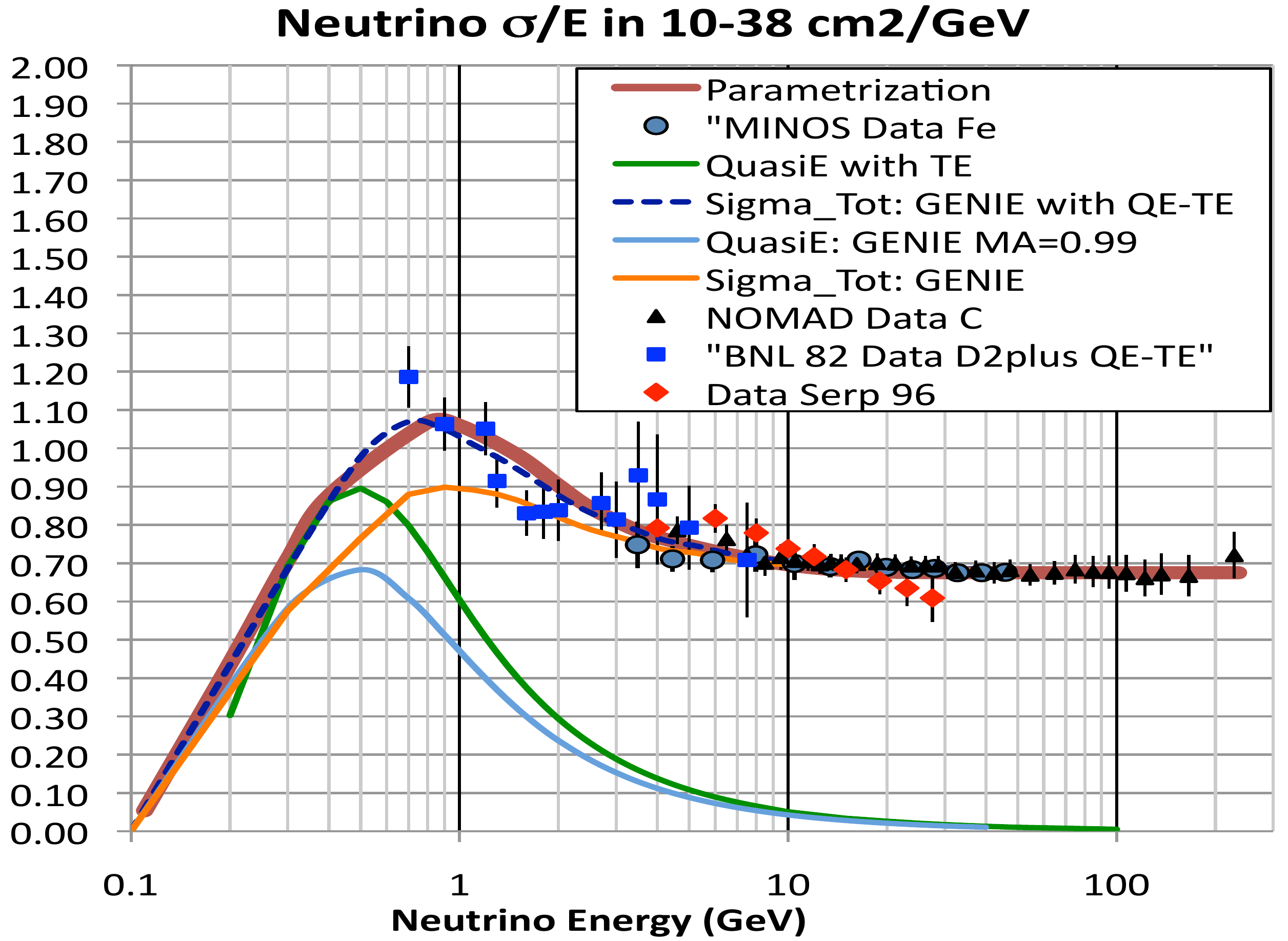}
\caption{The  MINOS\cite{MINOS2},  NOMAD\cite{NOMAD}, Serp96\cite{Serp96},  
and BNL82\cite{BNL82} (corrected) measurements of $\sigma_{total}/E$ per nucleon on isoscalar nuclear targets  for
$\nu$   in units of $10^{-38}\cm^2/$GeV. The orange line shows the predictions of the  unmodified GENIE Monte Carlo. The QE cross section in the GENIE MC is shown as the blue line.  The QE contribution calculated with the TE model
is shown as a green line.  The dashed blue line shows the prediction of the modified GENIE MC (using the TE model QE cross section instead).
The thick brown line is a parameterization described in the text (color online). } 
\label{cross}
\end{figure}
%
\begin{figure}
\includegraphics[width=3.5in,height=2.7in]{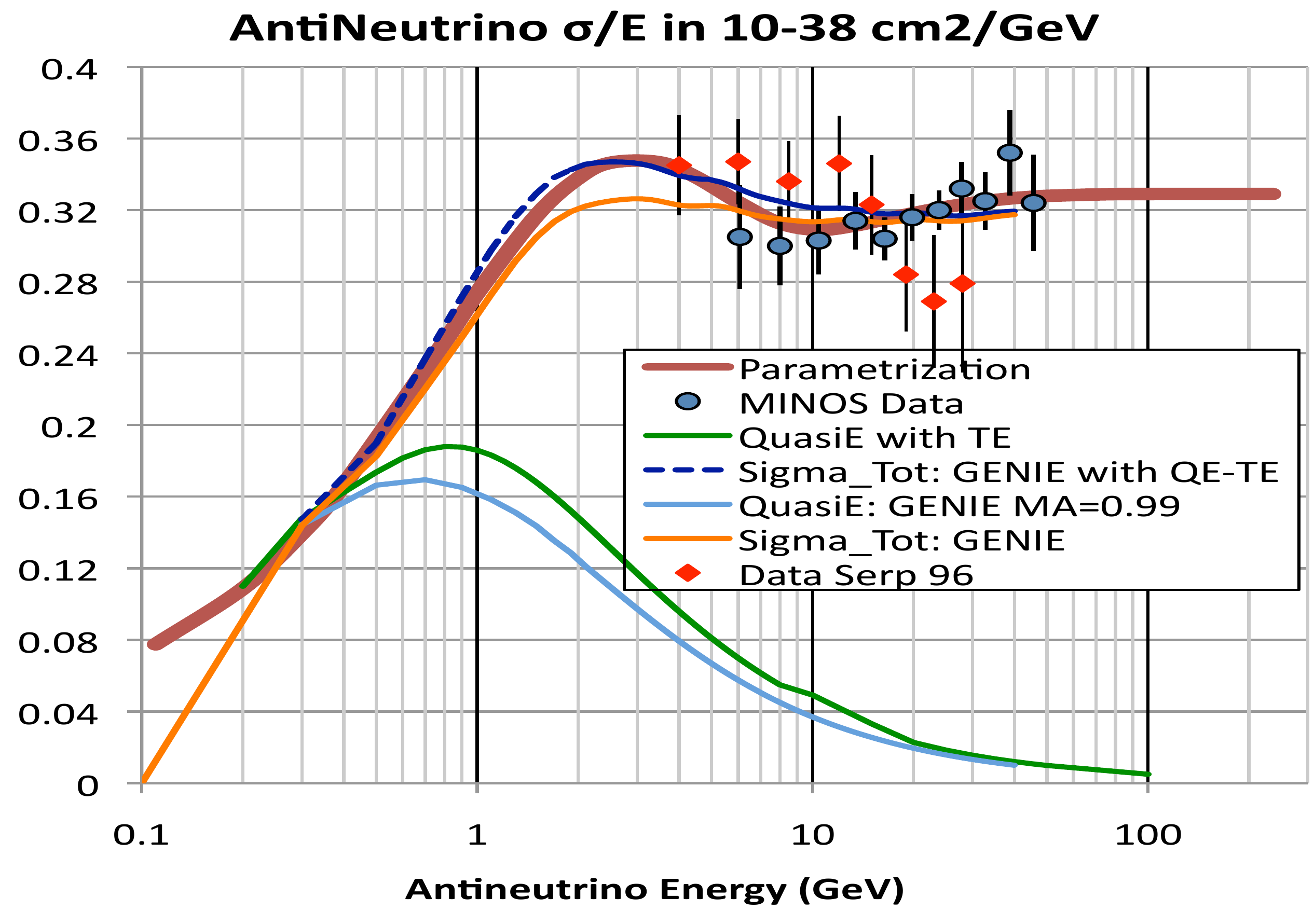}
\includegraphics[width=3.5in,height=2.7in]{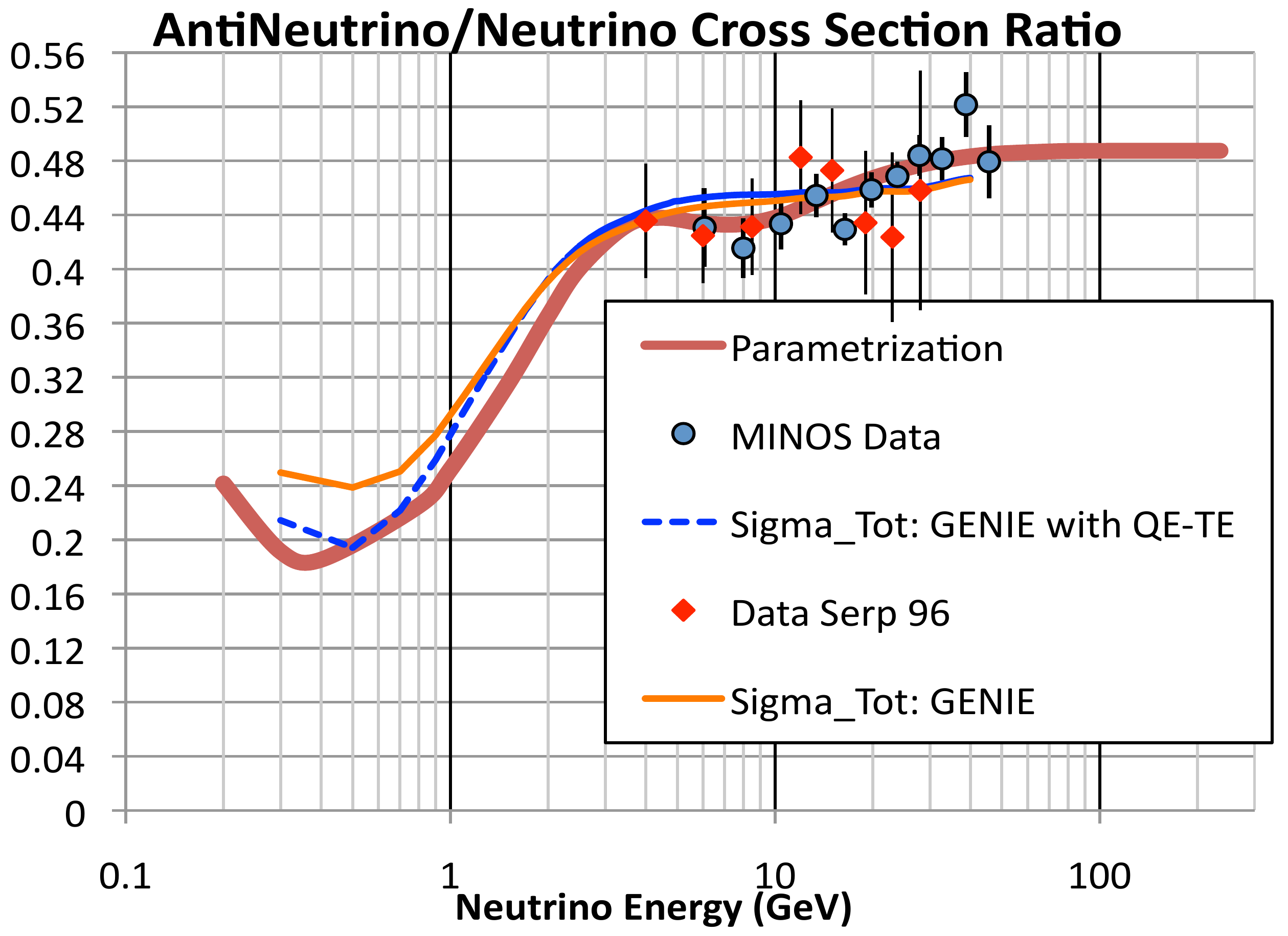}
\caption{Same as Fig. \ref{cross} but (a) for  the antineutrino charged current cross section, (b) for the ratio of antineutrino and neutrino total cross sections (color online).
 } 
\label{cross2}
\end{figure}
\subsection{Neutrino and antineutrino total cross sections}
The MINOS collaboration uses the criteria that the fraction of ``low-$\nu$'' events that are used for the determination of the relative neutrino flux in an energy bin should be  less than  60\% of the total number of charged current events.  In order to test for this fraction, we need to use a parameterization
to estimate the energy dependence of the  neutrino and antineutrino charged current total cross sections.

Fig. \ref{cross} and \ref{cross2}  show the  $\nu_\mu$ and $\nub_\mu$ total charged current cross sections  measured  on isoscalar nuclear targets by the    MINOS\cite{MINOS2} (iron),  NOMAD\cite{NOMAD}(carbon), and  Serpukov\cite{Serp96} (Serp96, aluminum)  experiments.  The  total  cross sections
per nucleon (divided by neutrino energy) are shown  in units of $10^{-38}\cm^2/$GeV (with statistical, systematic and normalization errors combined in quadrature).  The ratio
of the   $\nub_\mu$ and $\nu_\mu$  total charged current cross sections is shown in the bottom panel of Fig. \ref{cross2}. 
The cross sections reported by the  MINOS collaboration were measured using a neutrino
flux extracted from ``low-$\nu$'' samples with  $\nu$ less than 1, 3, and 5 GeV. 

Also shown in  Fig. \ref{cross} are low energy cross sections measured by at BNL\cite {BNL82} (BNL82).  Since the BNL82
cross sections were measured on a deuterium target we apply a correction
to account for nuclear effects. The  BNL82 points shown in the figure were increased 
by the difference  of the predictions of the TE  model for the QE cross section
(which is expected to describe the cross section on a heavy nuclear target)  and the  
 "Independent Nucleon (MA=1.014)"  model (which is expected to describe the QE cross sections on deuterium). 

The orange line shows the predictions of the GENIE Monte Carlo. The QE cross sections
in the GENIE MC are computed  using the independent nucleon model
with $M_A=0.99$ GeV.  The QE contribution to the cross section from GENIE is shown as a blue line. 
The QE contribution calculated with the TE model
is shown  as a green line.
The  curve labeled GENIE with QE-TE (shown as a dotted blue line)  
represents the GENIE cross section  increased 
by the difference  of the predictions of the TE  model for the QE cross section
(which is expected to describe the cross section on a heavy nuclear target)  and the  
 "Independent Nucleon (MA=0.99)"  model  (which is currently implemented in GENIE).

In our investigation of the  ``low-$\nu$'' technique, we use a parameterization to estimate
the  total  $\nu_\mu$, $\nub_\mu$ charged current cross sections. The parameterization, which is shown as the thick red line in  Fig. \ref{cross}, is given by
 
$$\frac{\sigma_\nu}{E_\nu} = [A + B~e^{-E_\nu/C1}+D~e^{-E_\nu^2/C2}](1-Ke^{-(E_\nu-0.1)/C3})$$ 
where for $\nu_\mu$ we use $A_\nu$=~0.675, $B_\nu$=~0.12, $C1_\nu$=~9~GeV, $D_\nu$~=0.4, $C2_\nu$=~3 GeV$^2$, $C3_\nu$=~0.22 GeV, and $K=~1.0$. 
For 
$\nub_\mu$ we use $A_{\nub}$=0.329, $B_{\nub}$=~-0.06 and $C1_{\nub}$=13~GeV $D_{\nub}$=~0.09,  $C2_{\nub}$=~30 GeV$^2$ , $C3_\nu$=~0.8 GeV, and $K=~0.8$. Here, $\frac{\sigma_\nu}{E_\nu}$ is 
total charged current  cross section per nucleon 
 in units of $10^{-38}\cm^2/$GeV.

The above form is constrained to yield the average world cross section measurements in the 30 to 50 GeV region of  
of  0.675 $10^{-38}\cm^2/$GeV, and 0.329 $10^{-38}\cm^2/$GeV  for  
$\nu_\mu$  and $\nub_\mu$, respectively.  

We only use this parameterization to estimate the fractional contribution of  
 ``low-$\nu$'' events to the total cross section  to determine the region where it is less than 60$\%$. 
 When improved total cross section measurements become available (e.g. from MINERvA), this parameterization can be updated to include the new data.
%
%
 %
  \begin{figure}
\includegraphics[width=3.7 in,height=3.0in]{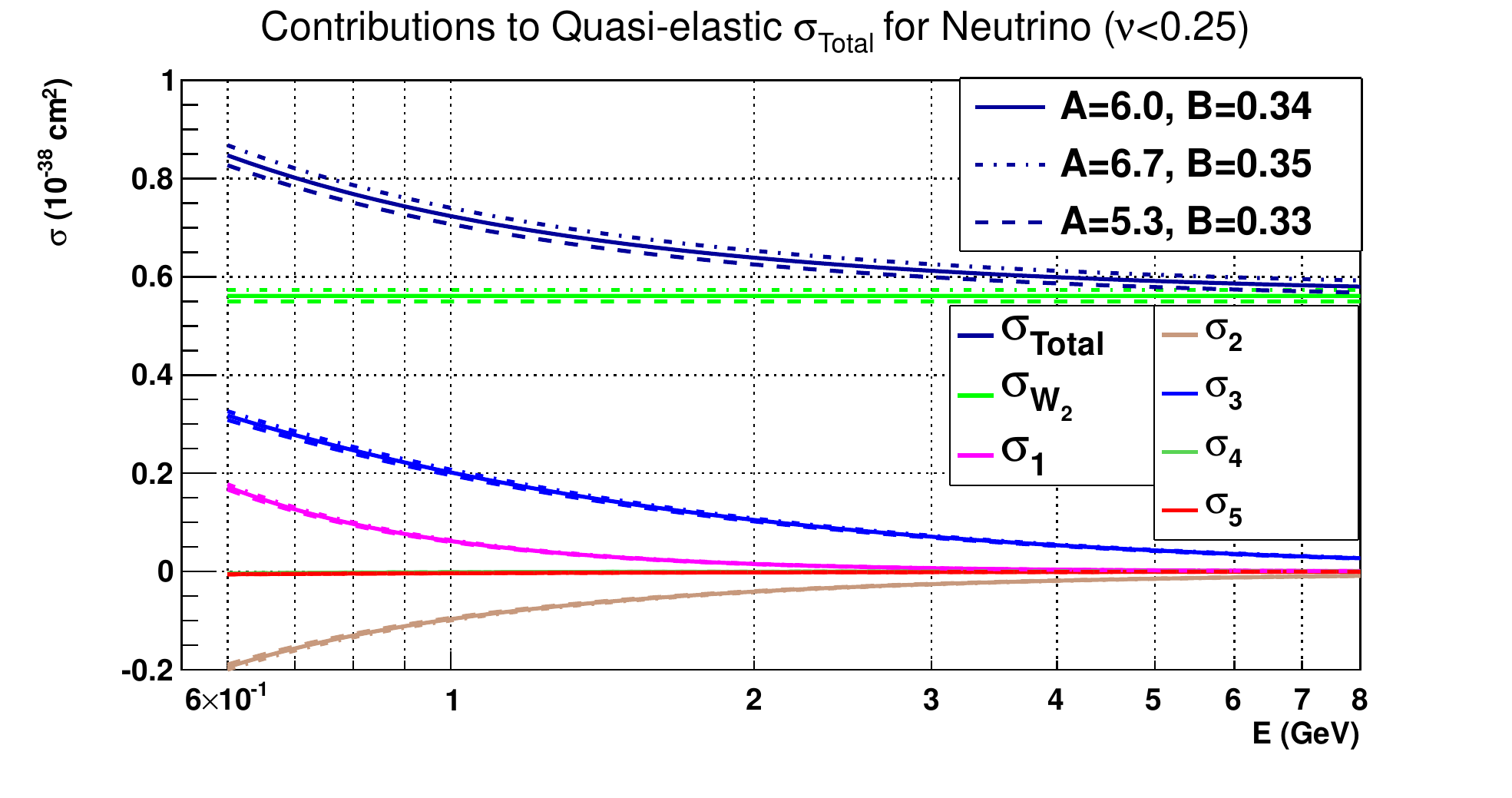}
     
     \vspace{-0.2in} 
     
\includegraphics[width=3.7 in,height=3.0in]{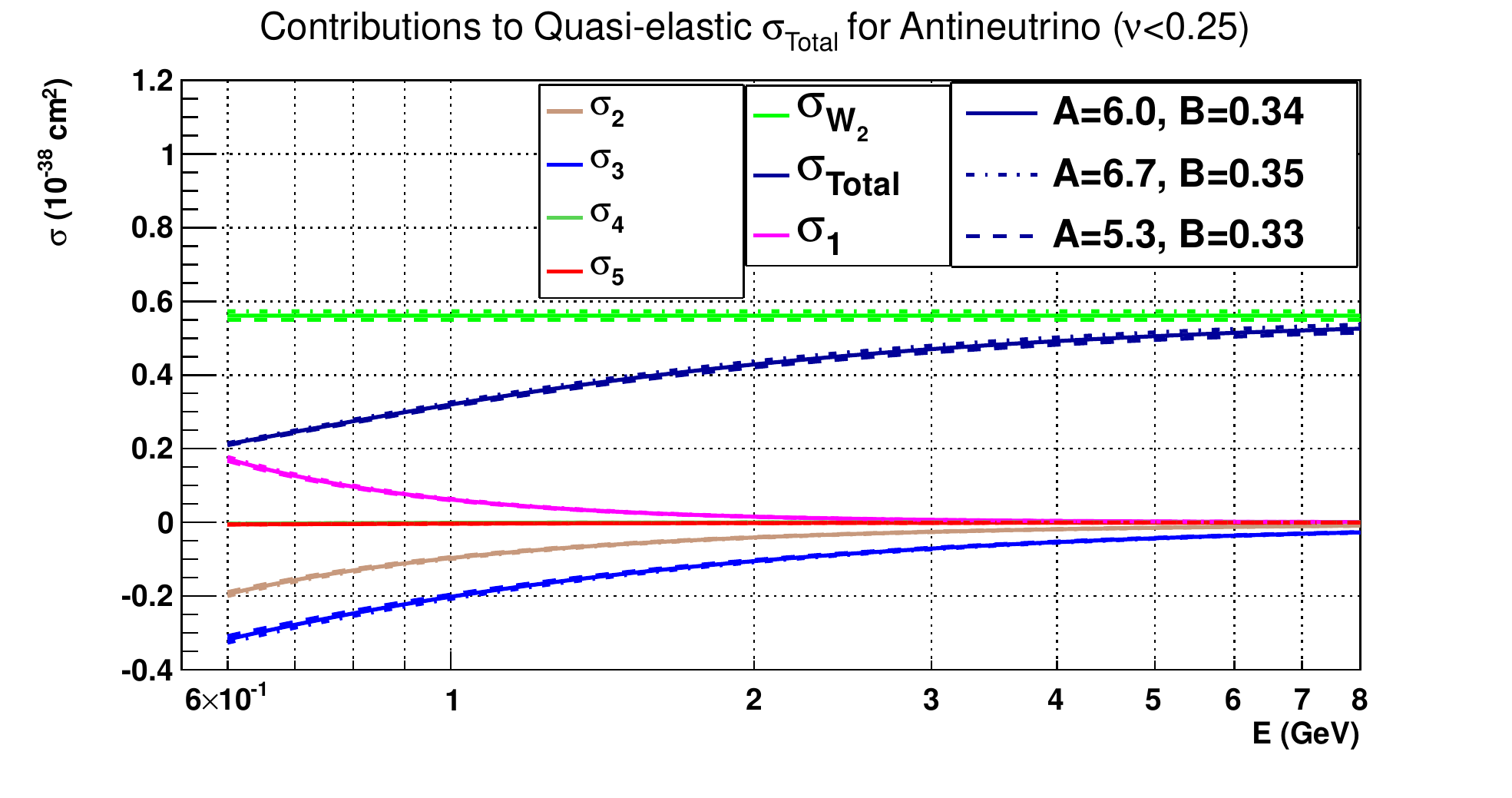}
\vspace{-0.3in}
\caption{Contribution of the various components
($\sigma_{W_2}$, $\sigma_{2}$,  $\sigma_{1}$, $\sigma_{3}$,  $\sigma_{4}$, $\sigma_{5}$)  to the $\nu< 0.25$ GeV partial charged current cross section. This sample  is dominated by QE $\nu_\mu N \rightarrow \mu^- P$ events.  Top panel: Neutrinos. Bottom panel: Antineutrinos (color online).
  }
\label{nu25comp}
\end{figure}

             \begin{figure}
\includegraphics[width=3.7in,height=3.0in]{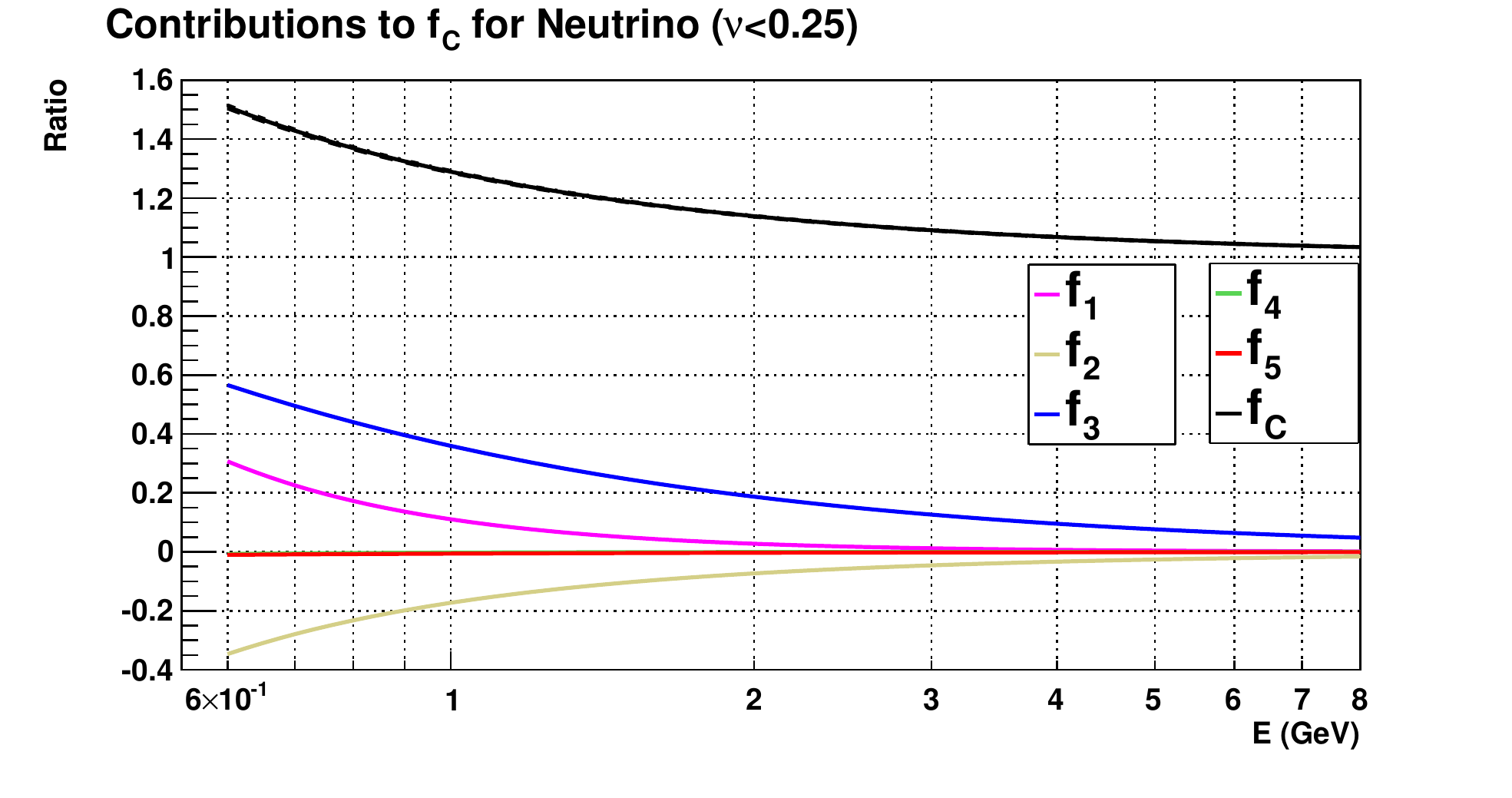}
     
     \vspace{-0.2in} 
     
\includegraphics[width=3.7in,height=3.0in]{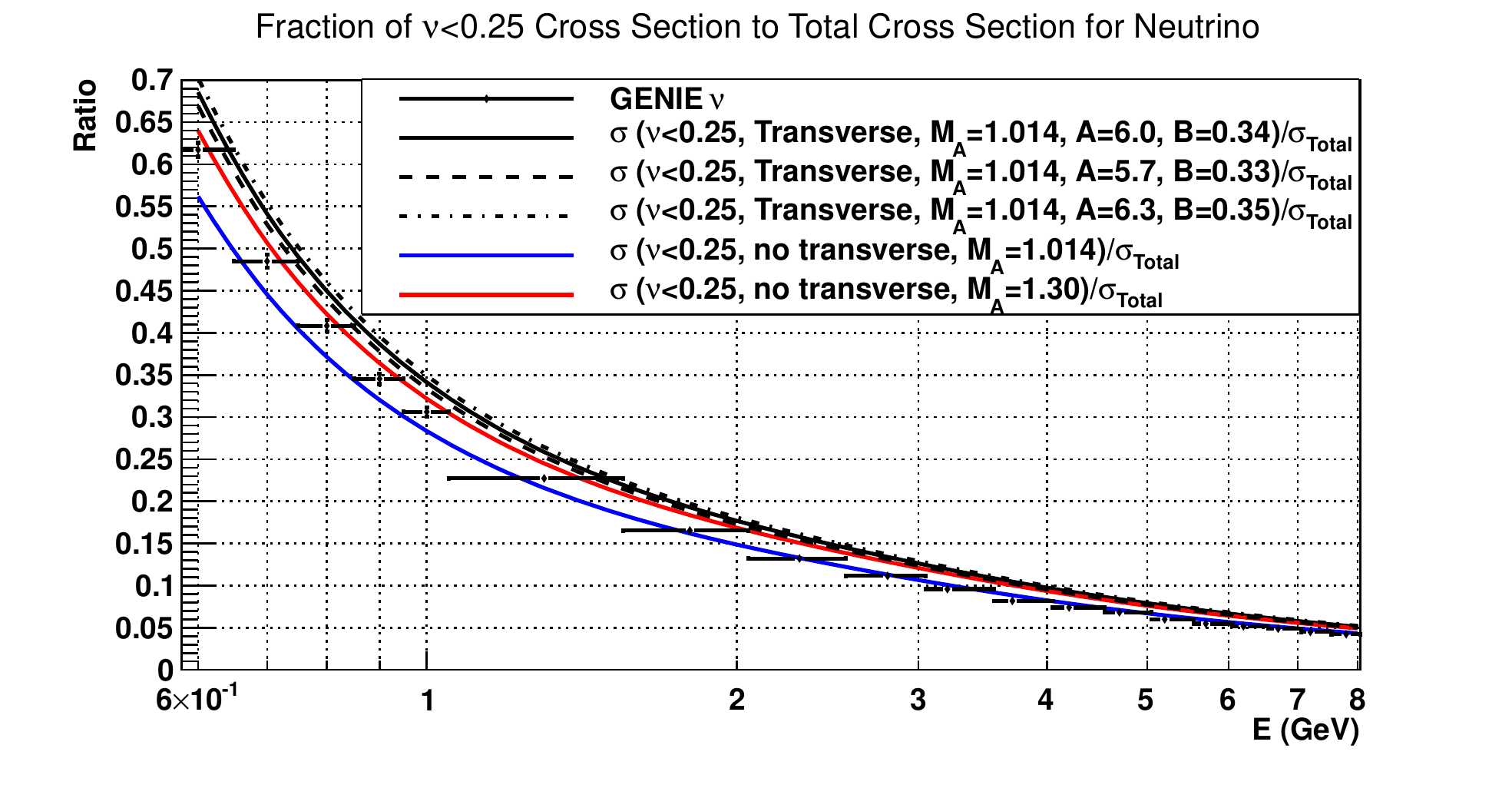} 
\vspace{-0.3in}          
\caption{The  $\nu<0.25$  GeV sample for $\nu_\mu$ scattering on carbon.
 Top panel: The total correction factor $f_C$  (black line),  the  contribution
of  the kinematic correction to ${\cal W}_2$ ($f_2$) (yellow line), the contributions  from ${\cal W}_1$ ($f_1$) (pink line), the contribution from ${\cal W}_3$ ($f_3$) (blue line),  and the very small contributions of ${\cal W}_4$ ($f_4$),  and ${\cal W}_5$ ($f_5$).   Bottom panel: The fractional contribution of  
 $\nu<0.25$ GeV events to the total $\nu_\mu$ charged current cross section. 
  (color online). 
  }
\label{neutrino25}
\end{figure}
%
        \begin{figure}
\includegraphics[width=3.7in,height=3.0in]{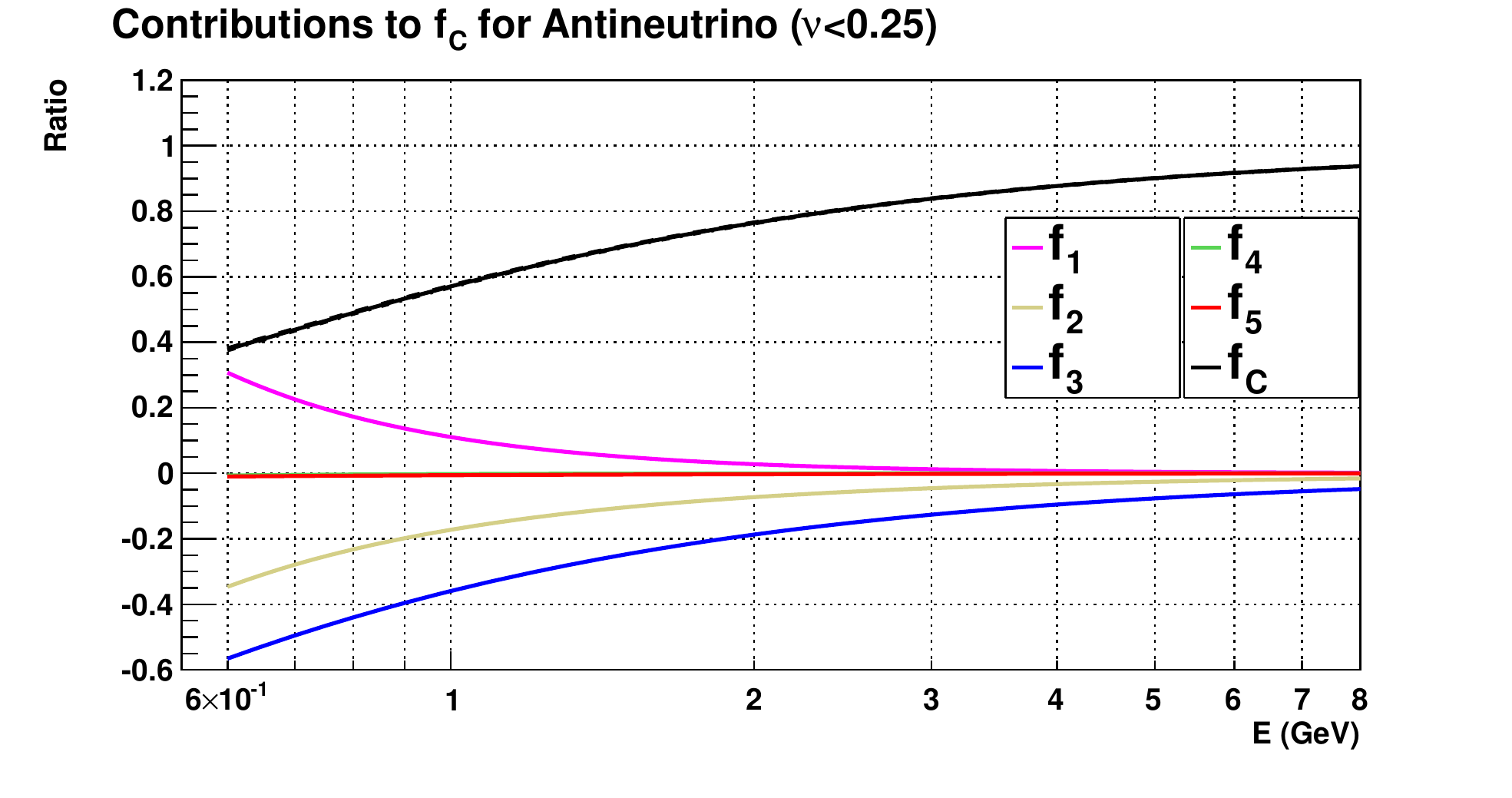}
     
     \vspace{-0.2in} 
     
\includegraphics[width=3.7in,height=3.0in]{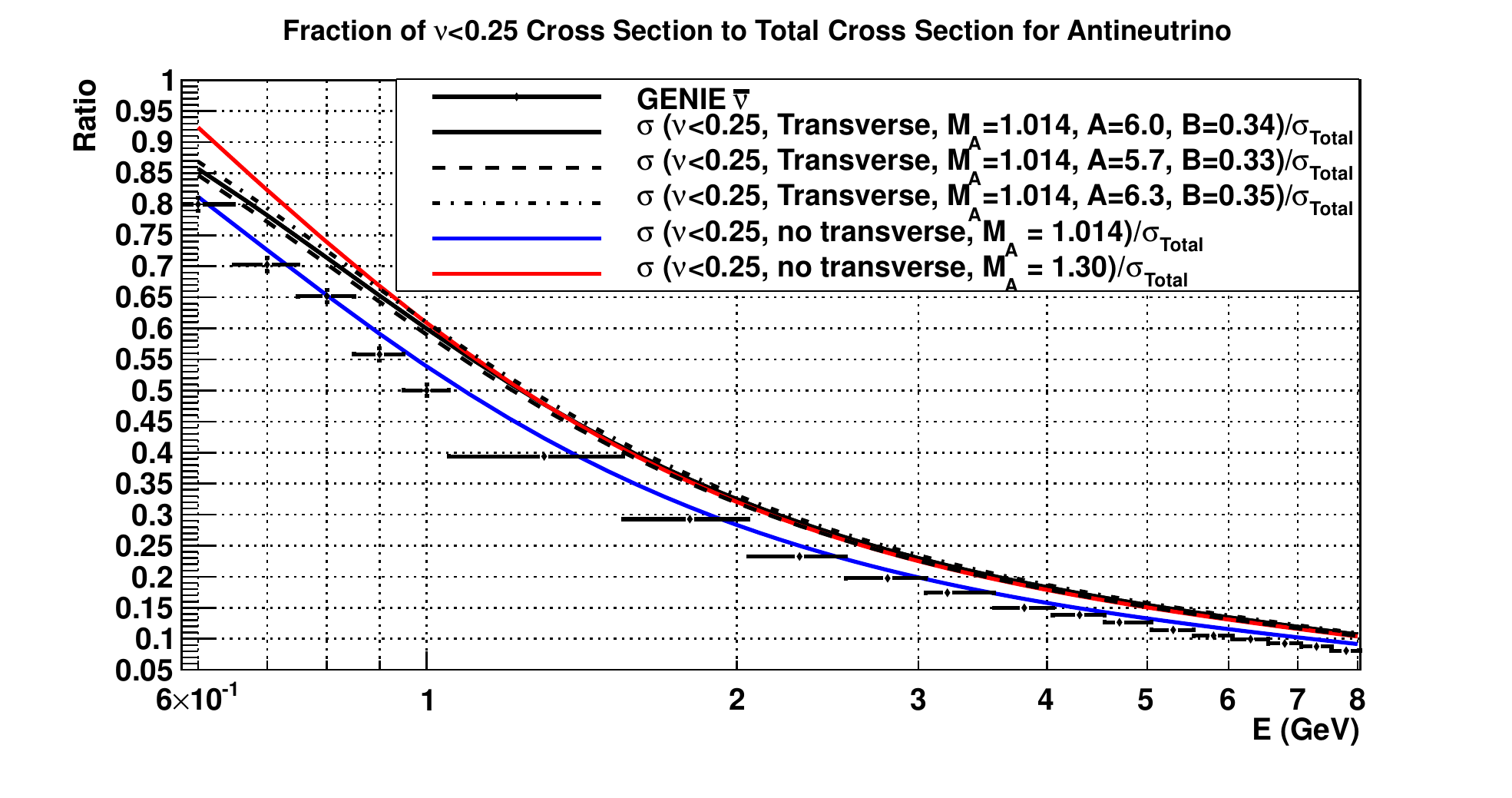}
\vspace{-0.3in} 
\caption{ Same as Fig. \ref{neutrino25} for the case of antineutrinos.}
\label{antineutrino25}
\end{figure}
 %
%
     \begin{figure}
     \includegraphics[width=3.7in,height=3.0in]{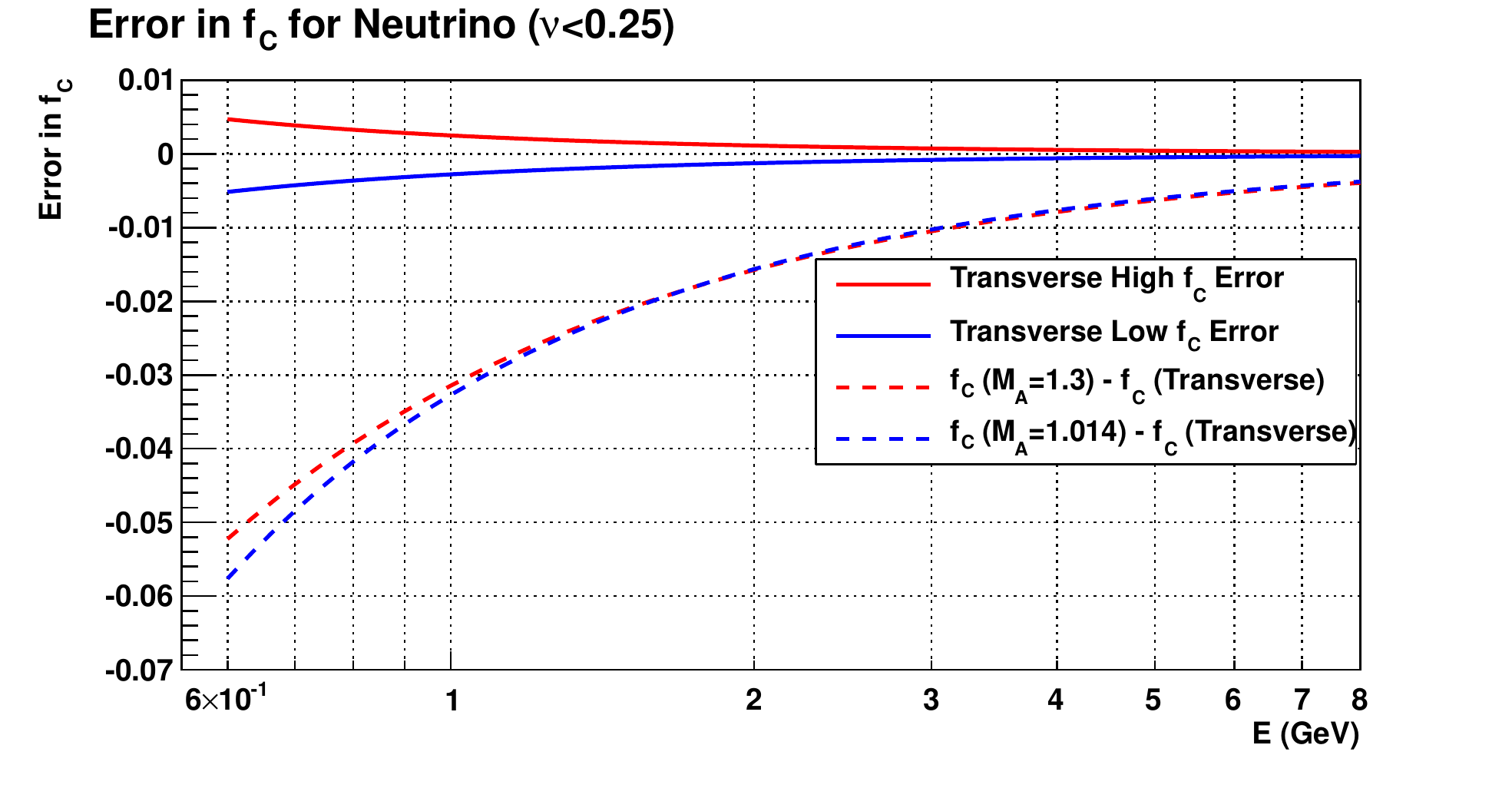}
     
     \vspace{-0.2in} 
     
     \includegraphics[width=3.7in,height=3.0in]{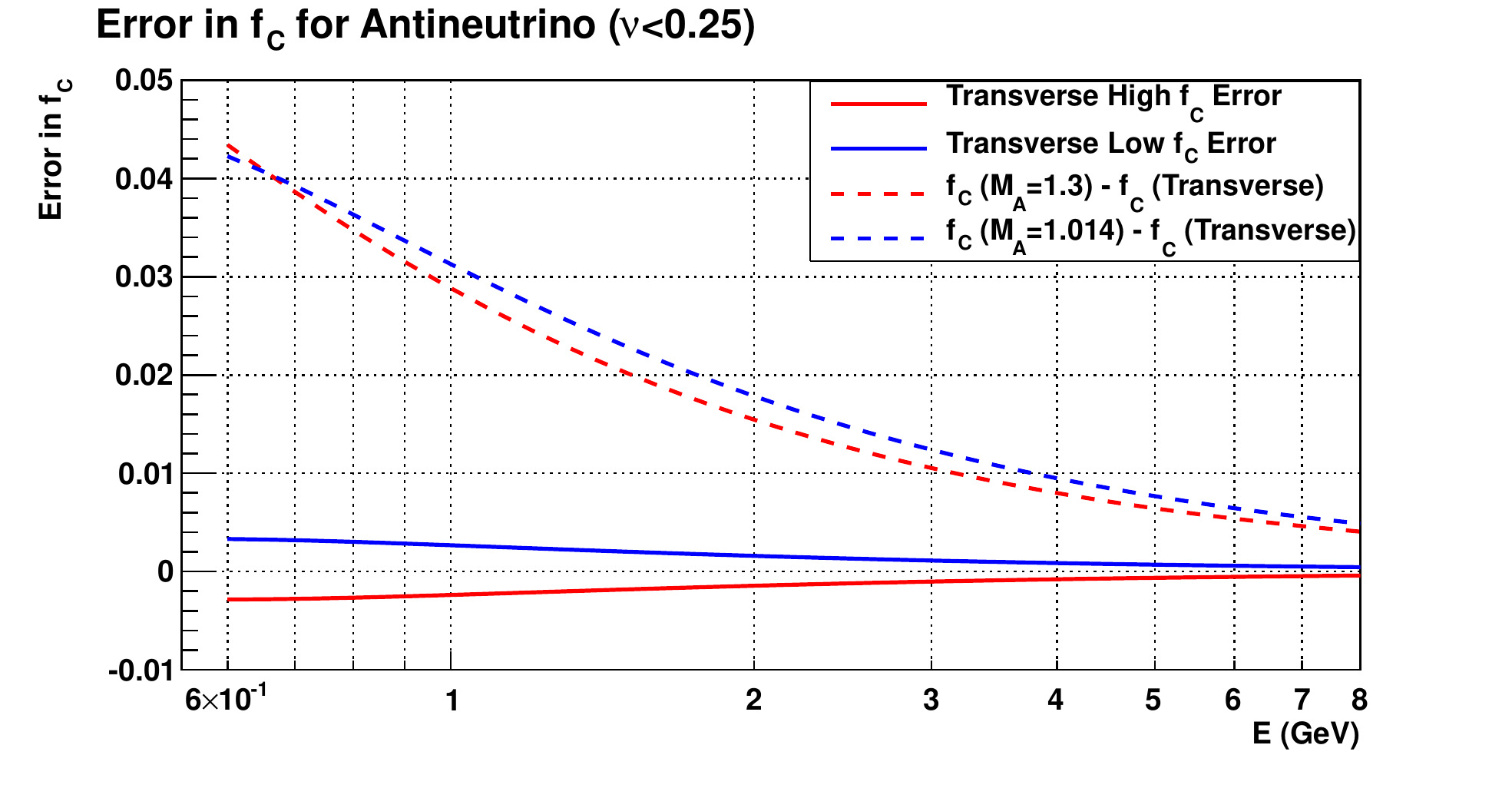}
     \vspace{-0.3in} 
\caption{ The error band in the correction factor $f_C$ for $\nu<0.25 $ GeV.  Top panel: Neutrinos. Bottom panel: Antineutrinos (color online).   }
\label{nudiff}
\end{figure}
%
%
         \begin{figure}
\includegraphics[width=3.7in,height=3.0in]{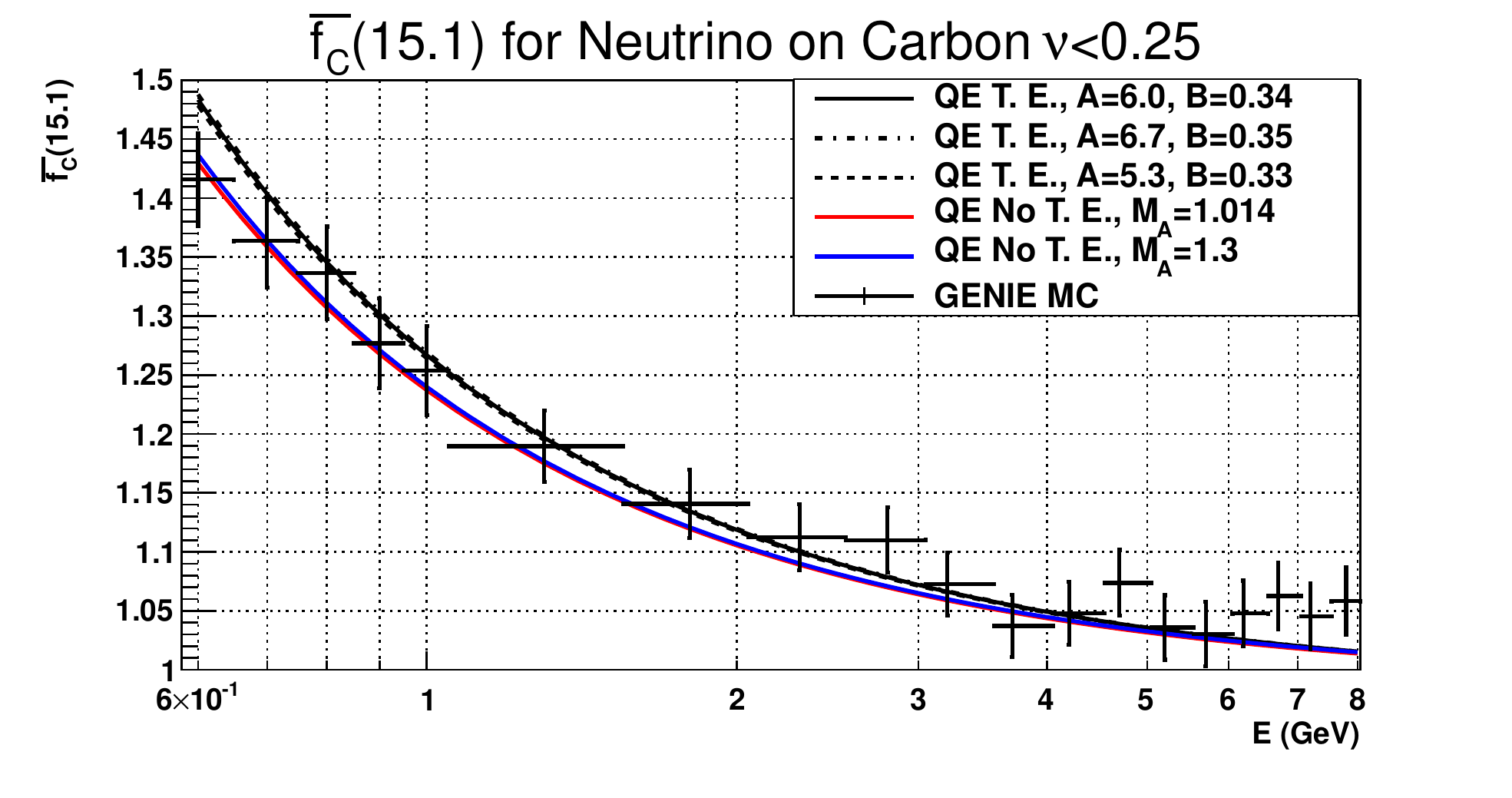}
     
     \vspace{-0.2in} 
     
\includegraphics[width=3.7in,height=3.0in]{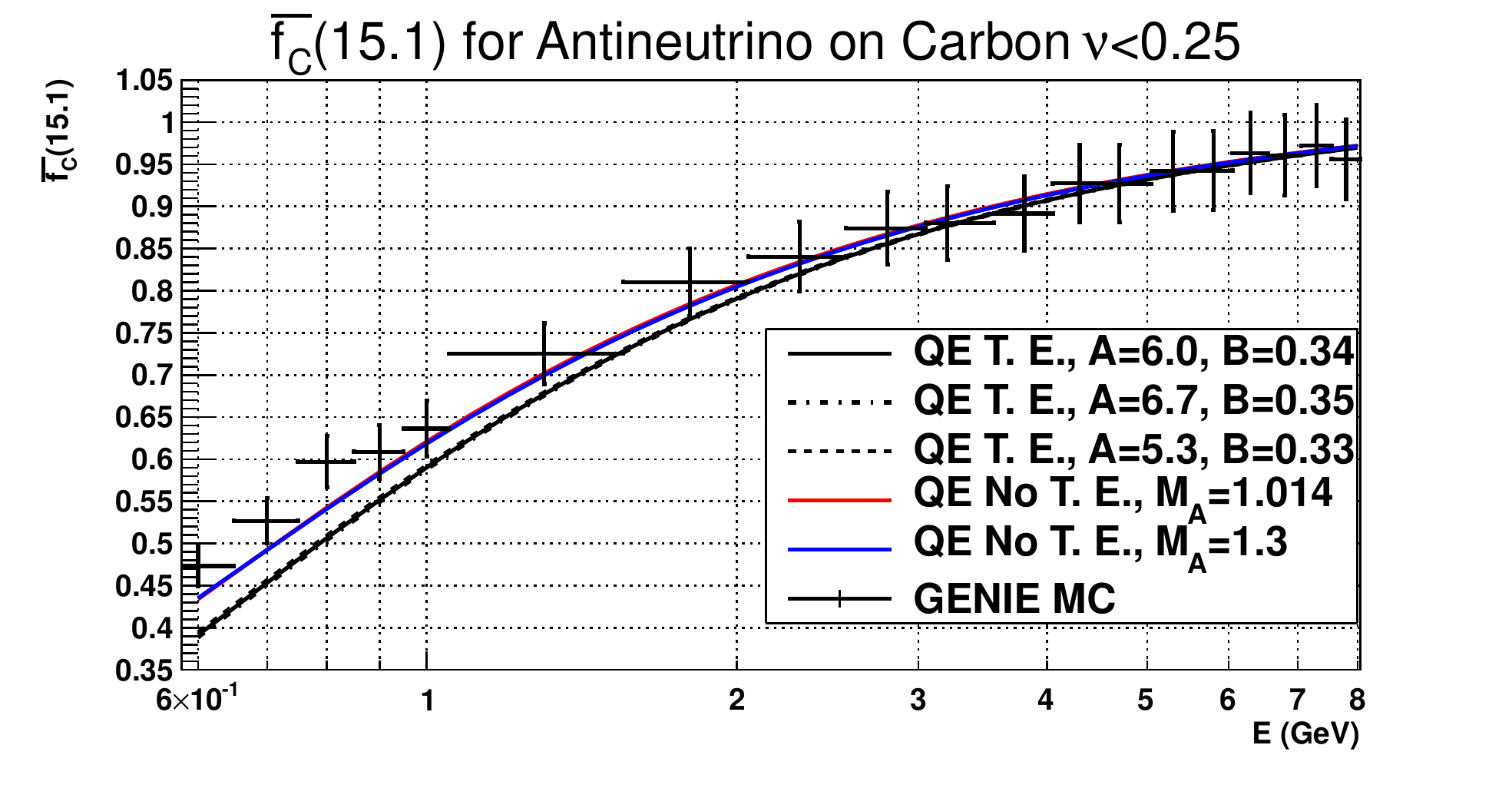}
\vspace{-0.3in} 
\caption{ Comparisons of our calculated values of  the normalized  
${\bar f_{C:\nu<0.25}} (15.1)(E)$ (=$\bar f_C (15.1) $ for $\nu<0.25 $ GeV) 
 to  values from the GENIE MC. 
The values calculated with the nominal TE model for QE scattering  ($M_A=1.014$ GeV) are shown in black. The values calculated assuming no transverse enhancement and $M_A=1.014$ GeV are shown in red. 
The  GENIE prediction (which has no transverse enhancement and  uses $M_A=0.99$ GeV)  is close to the red curve as expected. 
 (color online).
  }
\label{fcbar25}
\end{figure}
%
\subsection{Results with $\nu<0.25$ GeV}
Fig. \ref{nu25comp} shows the 
contribution of the various components
($\sigma_{W_2}$, $\sigma_{2}$,  $\sigma_{1}$, $\sigma_{3}$,  $\sigma_{4}$, $\sigma_{5}$)  to the  $\nu<0.25$ GeV partial cross section. This sample  is dominated by QE $\nu_\mu N \rightarrow \mu^- P$ events.  The partial cross section as a function of energy
for neutrinos is shown in the top panel and the partial cross section
for antineutrinos is shown in the bottom panel.  The partial cross section  (per nucleon) is calculated on a carbon target using the TE model.

The uncertainty in the relative values of the  $\nu<0.25$ GeV partial cross section 
as a function of energy determines the uncertainty in the determination of the relative fluxes.  Here
$f_C (E) $ is the ratio of the partial cross section to the value of the partial cross section
at $E=\infty$.

Fig. \ref{neutrino25}(a) (top) shows the correction factor
$f_C$ for the  $\nu<0.25$ GeV sample for neutrinos as a function  energy.  The error bands
in $f_C$ (originating from the uncertainty in the transverse enhancement)  are shown as the dotted lines,
and represent the lower limit on errors.  Also shown on the figure 
is the negative contribution from the kinematic correction $f_2$ (which is well known),  and the contributions
of $f_1$, $f_3$, $f_4$ and $f_5$.  Here the contribution of   $f_4$ and $f_5$ is
negligible. For the case of neutrino scattering, the  positive contributions of $f_1$ and $f_3$ partially cancel the negative contribution of $f_2$.  Fig. \ref{neutrino25}(b) (bottom) shows the  fractional contribution of the  
$\nu<0.25$  GeV sample to the  total  neutrino charged current cross section.  This fraction is  less than  60$\%$ for $\nu_\mu$ energies above 0.70 GeV. 

Fig. \ref{antineutrino25} is the same as Fig. \ref{neutrino25}  for the case of antineutrinos.
For the case of antineutrino scattering  $f_3$  changes sign, and both 
 $f_2$ and  $f_3$ are negative.    
 The  fractional contribution of the  
$\nu<0.25$  GeV sample to the total antineutrino charged current cross section is  less than  60$\%$ for $\nub_\mu$ energies above 1.0  GeV. 
%
\subsubsection{Uncertainty in the $f_C$ correction factors}
It has been traditional to use the value and  error in the effective  $M_A$ extracted
from neutrino scattering data as an estimate of various uncertainties. Typically, the  difference
between results with  
$M_A=1.014$  GeV and $M_A=1.3$ GeV are used an upper limit on  the error. 

We find that the values of the  $f_C$ correction factor 
 are insensitive to $M_A$.  This is because
 at small  $Q^2$, both ratios $f_1$, and $f_3$ are insensitive to $M_A$. Specifically,  both
 $$\frac{{\cal W}_1^{QE}}{{\cal W}_2^{QE}} = \frac{(1+\tau)|{\cal F}_A (Q^2)|^2+\tau |{\cal G}_M^V(Q^2)|^2}{|{\cal F}_A (Q^2)|^2+[{\cal F}_V(Q^2)]^2}$$
 $$\frac{{\cal W}_3^{QE}}{{\cal W}_2^{QE}} = \frac{|2 {\cal G}_M^V(Q^2) {\cal F}_A (Q^2)|}{|{\cal F}_A (Q^2)|^2+[{\cal F}_V(Q^2)]^2},$$
are insensitive to  $M_A$   because the  change in $F_A$  at small $Q^2$ is small. 
 Since  $f_C$ is insensitive to large variations in  $M_A$  one may naively surmise that the error in $f_C$ is small.

However, we find that the difference between the values $f_C$ calculated
with and without transverse enhancement is larger than the error estimate 
extracted from the uncertainty in $M_A$.    This is because  $\frac{{\cal W}_3^{QE}}{{\cal W}_2^{QE}}$ is sensitive
to   ${\cal G}_M^V(Q^2)$, which depends on the magnitude of the  
 transverse enhancement at small $Q^2$.

 Fig. \ref{nudiff} shows the errors in $f_C$ from the uncertainty in the TE parameters.  The 
 error originating
 from  uncertainties in the TE parameters is  also very small (less than 0.005).   
 
 We obtain a more conservative estimate of the systematic error in $f_C$
 originating from uncertainties in the   modeling
 the QE cross section by taking the difference
 between $f_C$ calculated with and without transverse enhancement.  At  the lowest energy of 
 0.7 GeV, this difference is  -0.05   for $\nu_\mu$.  Since at  0.7 GeV  $f_C^{\nu} \approx1.3$ this 
 corresponds to a maximum error in the determination of the $\nu_\mu$ flux  of 3.8\%. 
 
   For $\nub_\mu$ the difference
 between $f_C$ calculated with and without transverse enhancement at an energy of 
1.0 GeV is  +0.03.  Since at 1.0 GeV  $f_C^{\nub} \approx0.6$ 
this corresponds a maximum error in the determination of the $\nub_\mu$ flux  of 5\%. 

 \vspace{-0.1in} 
 %
\subsection{Comparison to GENIE and ${\bar f_{C:\nu<0.25}}$(15.1 GeV)}
 We have used a sample of events generated by the GENIE Monte Carlo.
Our studies are done at the generated
  level and therefore do not depend on the 
  detector parameters or energy resolutions of any specific experiment.
  
 We extract the energy dependence of the
 $\nu<0.25$ GeV cross section from the GENIE MC sample using
 the following expression: 
   
$$ \sigma_{\nu<0.25} ^{MC}(E) = \frac {N ^{MC}_{\nu<0.25} (E) }{N^{MC}_{QE}(E)} \times \sigma_{QE} ^{MC}(E) $$
where  the superscript $MC$ refers to events generated by the GENIE Monte Carlo.

  Here, $N ^{MC} (E)$ is the number of events generated by the Monte Carlo
  with neutrino energy E, and  $N ^{MC}_{\nu<0.25}(E)$ is the subset of these events
with $\nu<0.25$ GeV.
  
    As mentioned earlier, we propose that the neutrino cross sections at low energy
   be measured relative to the neutrino cross section at 15.1 GeV. 
   For any cross section
   model we can define the normalized quantity ${\bar f_{C:\nu<0.25}}$(15.1 GeV) as: 
     $${\bar f_{C:\nu<0.25}} (15.1)(E)=\frac {\sigma_ {\nu<0.25} (E)}{ \sigma_ {\nu<0.25} (E=15.1~ \GeV)}$$
    which is equivalent to
   $${\bar f_{C:\nu<0.25}} (15.1)(E)  = \frac{ f_{C:\nu<0.25} (E)}{f_{C:\nu<0.25}(E=15.1~\GeV)}$$
   We compare the values of 
    ${\bar f_{C:\nu<0.25}} (15.1)(E) $ predicted by the GENIE MC to our calculations.
  
  For completeness, we give the  values of  $f_{C:\nu<0.25}$(15.1)  that can be used to convert between ${\bar f_{C:\nu<0.25}}(E)$ and ${f_{C:\nu<0.25}} (E)$. 
  
  For the TE QE model we find 
    $f_{C:\nu<0.25}$(15.1)=1.018 (for $\nu$) and 0.966 (for $\nub$). For QE models without TE we find similar values of $f_{C:\nu<0.25}$(15.1)=1.016 ($M_A$=1.014) and $f_{C:\nu<0.25}$(15.1)=1.014 ($M_A$=1.03) for $\nu$. For  $\nub$  we find  $f_{C:\nu<0.25}$(15.1)=0.969 for models without TE.

  Comparisons of our calculated values of the normalized
    ${\bar f_{C:\nu<0.25}} (15.1)(E)$  to  values from the GENIE MC are shown in   Fig. \ref{fcbar25}.
     The top panel shows the comparison for neutrinos and the bottom panel shows the comparison
     for antineutrinos. 
        Our calculation for the TE model is  shown in black.  Our calculation assuming no transverse enhancement and $M_A=1.014$ GeV is shown in red.
        As mentioned earlier, the values for $M_A=1.014$ GeV (red line) and $M_A=1.3$ GeV  (blue line) are very close to each other.   The  GENIE prediction (which has no transverse enhancement and  uses $M_A=0.99$ GeV)  is close to the red curve as expected.  The GENIE predictions include a contribution from coherent pion
        production. As shown in Appendix II, for the $\nu<0.25$ GeV sample, the contribution from
coherent pion production is less than 0.1\% for neutrinos and less than  0.6\% for antineutrinos.
        
        \subsection{Conclusions of the studies with  $\nu<$0.25 GeV}     
  In conclusion, we find that the method works very well for $\nu<$ 0.25 GeV. 
  If one takes the average of all the models,  a conservative upper
 limit of the model uncertainty in the relative flux extracted from the $\nu<0.25$ GeV sample is   1.9\% for $\nu_\mu$ energies  above 0.7 GeV   and 2.5\% for  $\nub_\mu$ energies above 1.0 GeV. The GENIE Monte Carlo
 is in reasonable agreement with the models and therefore can be used to obtain a first order neutrino flux. 
 
     A study of the $Q^2$ distributions of QE events in MINERvA can
 be used to constrain the $Q^2$  dependence of  the  QE differential cross sections and thus reduce  the model  dependence in the determination of the  relative flux to a negligible level. A GENIE Monte Carlo which
 is tuned to agree with the new data can be used to extend the technique to lower energies. 
   %
        %
        %
        %
            \begin{figure}
 \includegraphics[width=3.7in,height=3.0in]{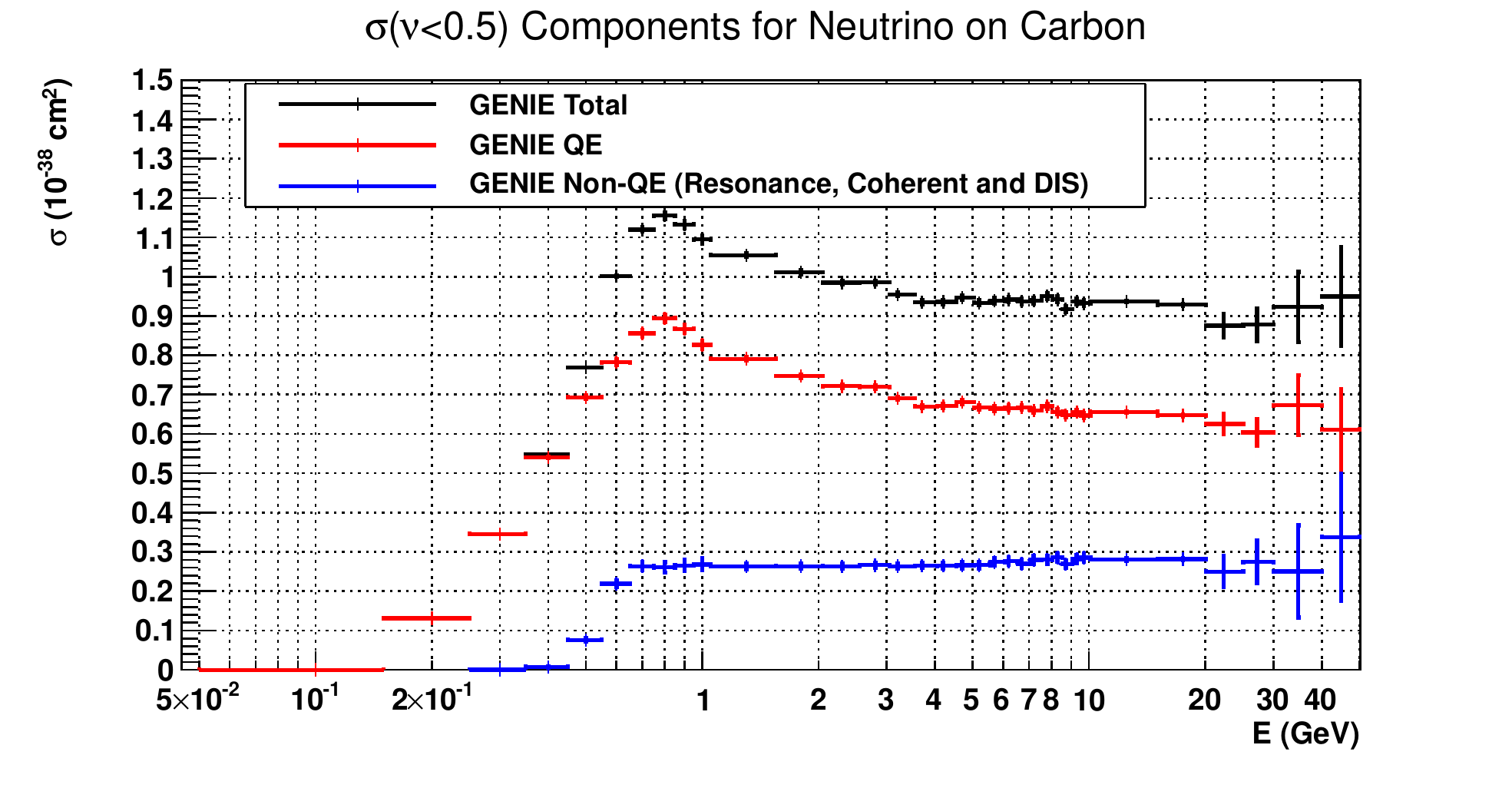}
     
     \vspace{-0.2in} 
     
\includegraphics[width=3.7in,height=3.0in]{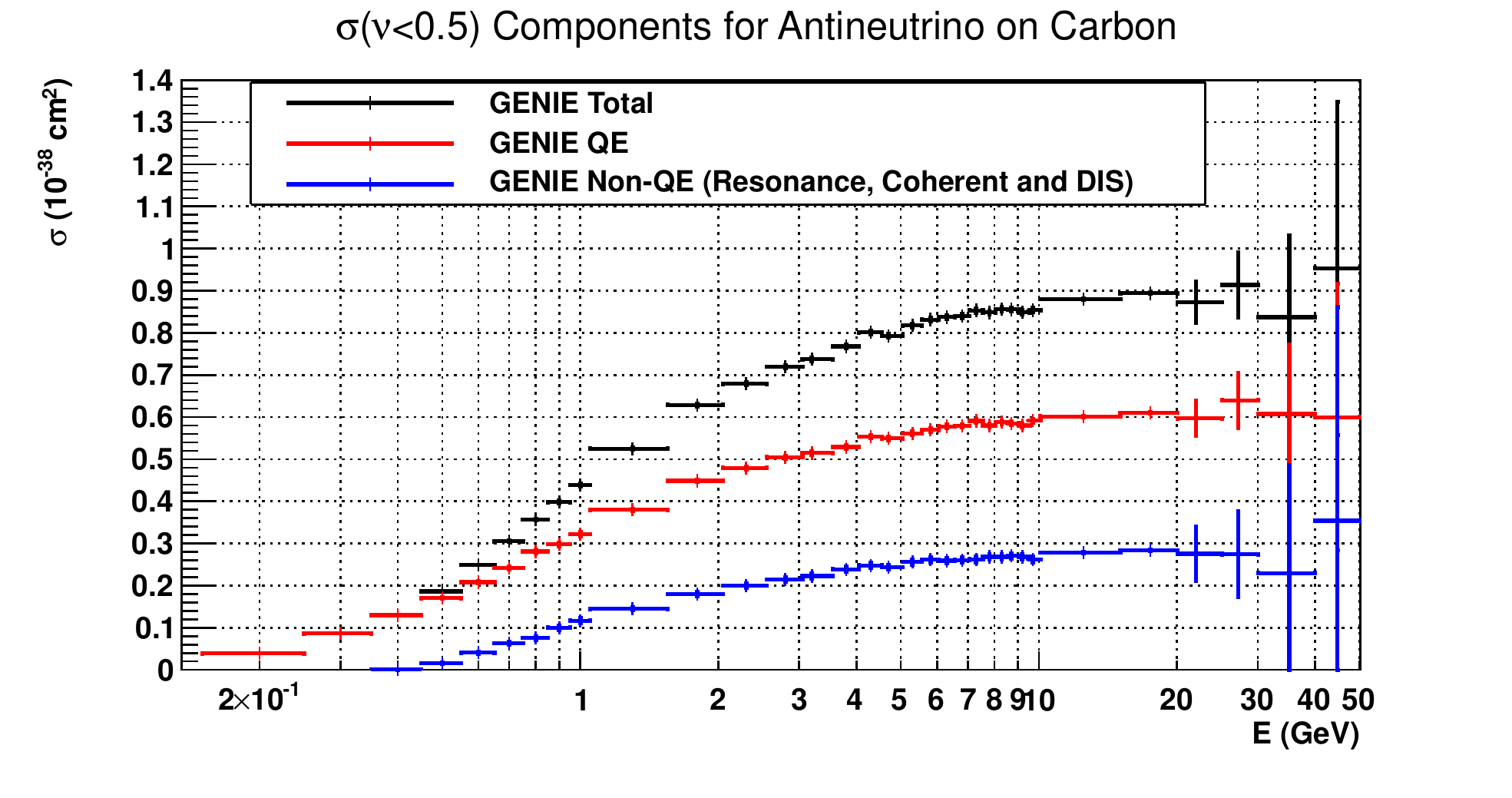}    
\vspace{-0.3in}        
\caption{The $\nu<0.5$  GeV partial charged current cross section as a function of energy from the GENIE Monte Carlo.
The QE contribution is shown in red, the contribution from pion production process ($\Delta$, inelastic and coherent pion production) is shown in blue,  and
 the total $\nu<0.5$ GeV GeV partial cross section is shown in black.   The $\nu<0.5$ GeV GeV partial cross section for $\nu_\mu$
 is shown in the top panel, and the $\nu<0.5$ GeV  partial cross section for $\nub_\mu$
 is shown in the bottom panel. 
 (color online).
   }
\label{QEvsDelta50}
\end{figure}
\vspace{-0.1in}   
 %
 %
\section{Using  ``low-$\nu$'' events with $\nu<0.5$ GeV }
The  $\nu<0.5$ GeV   $\nu_\mu$ and  $\nub_\mu$ samples  have close to twice the number of events
as the  $\nu<0.25$ GeV samples.  
These samples for scattering are also  dominated by QE events,   but include  a significant fraction  (about 1/3) of events in which a single pion is produced in the final state.
As seen in  Fig. \ref{nuvsq22},  the $\nu<0.5$ GeV samples are composed of 
QE events with $Q^2<0.9$ GeV$^2$, and $\Delta$(1232) events with 
$Q^2<0.3$ GeV$^2$.
%
%
%
         \begin{figure}
\includegraphics[width=3.7in,height=2.9in]{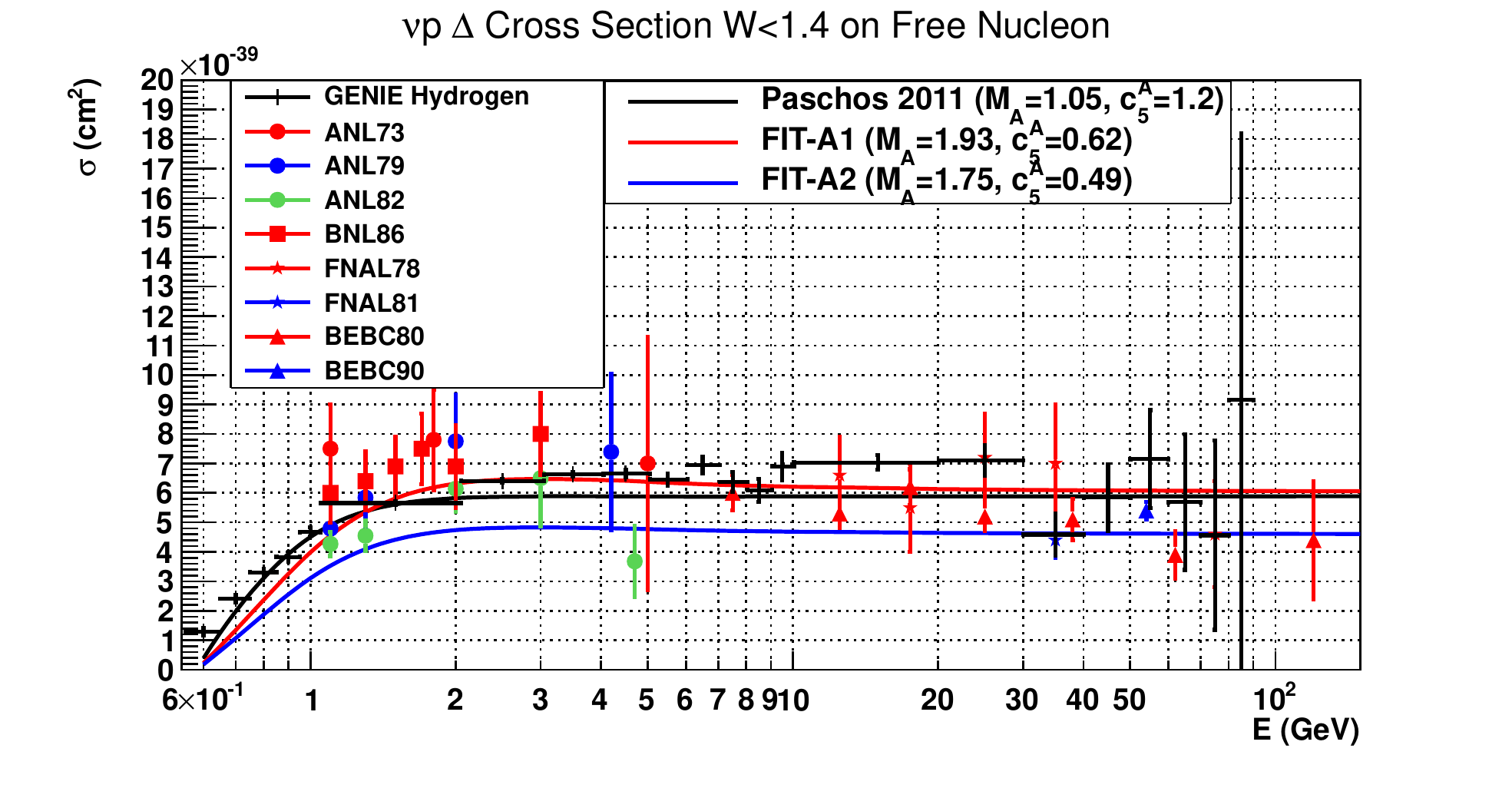}
     
     \vspace{-0.2in} 
     
\includegraphics[width=3.7in,height=2.9in]{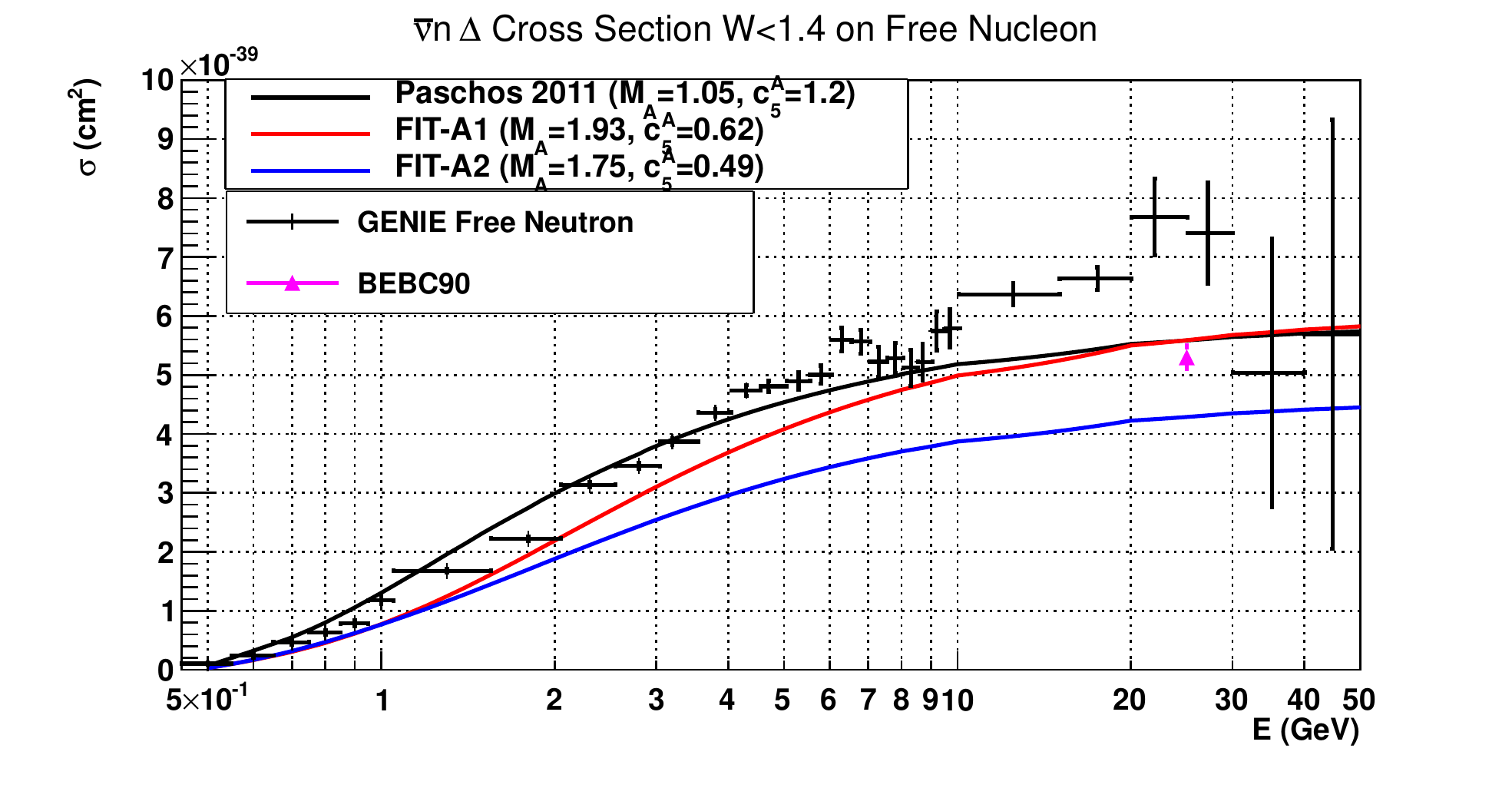}
\vspace{-0.3in} 
\caption{ $\nu_\mu P \to \mu^- \Delta^{++}$ (top panel)  and $\nub_\mu N \to \mu^+  \Delta^-$ (bottom panel)  
cross sections  (for $W<1.4$ GeV) measured
 on free nucleons (H and D), compared to predictions from the GENIE MC (black points with errors).  
(color online). 
 }
\label{deltanup}
\end{figure}
%
         \begin{figure}
\includegraphics[width=3.7in,height=2.9in] {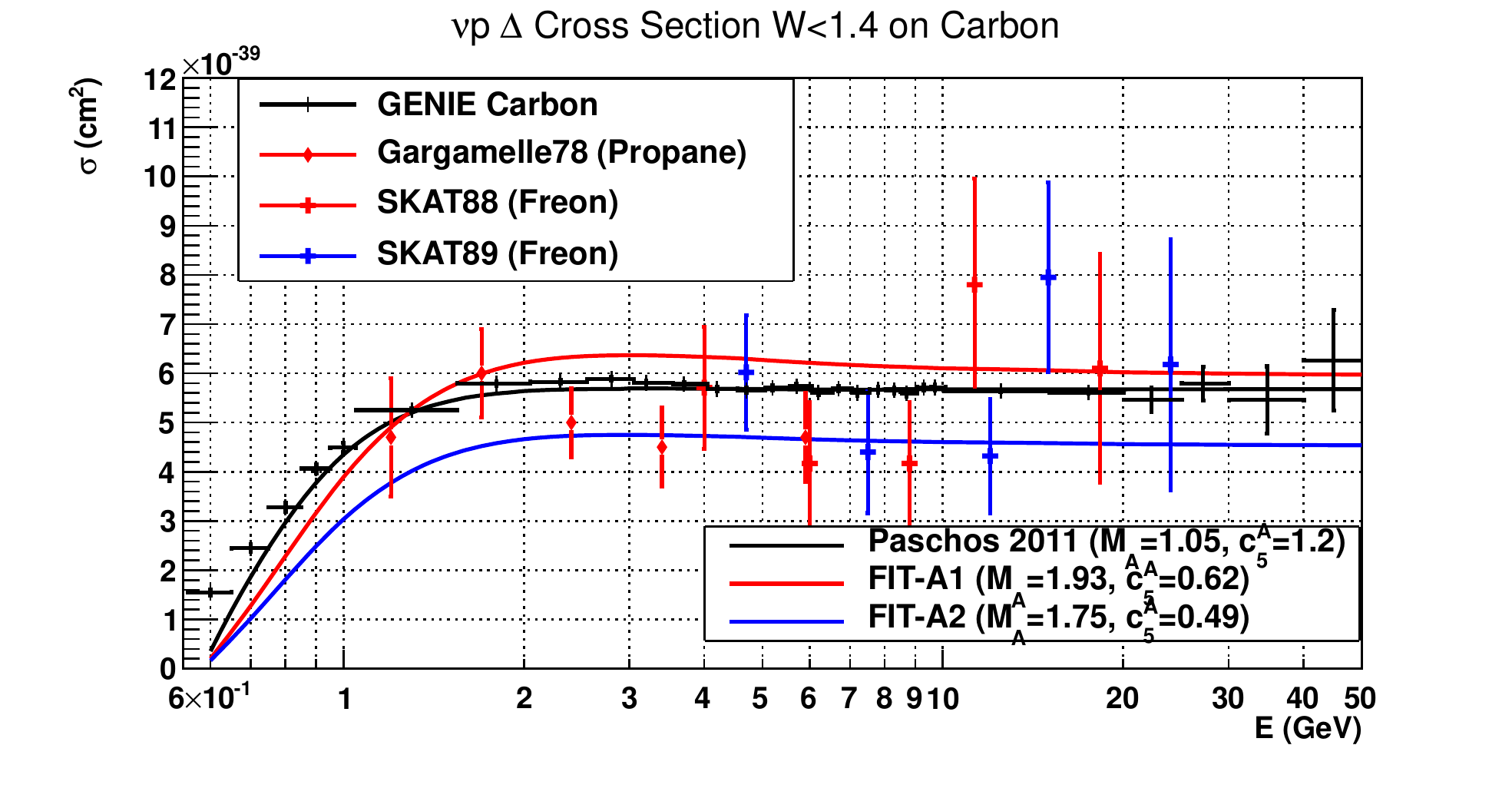}
     
     \vspace{-0.2in} 
     
\includegraphics[width=3.7in,height=2.9in] {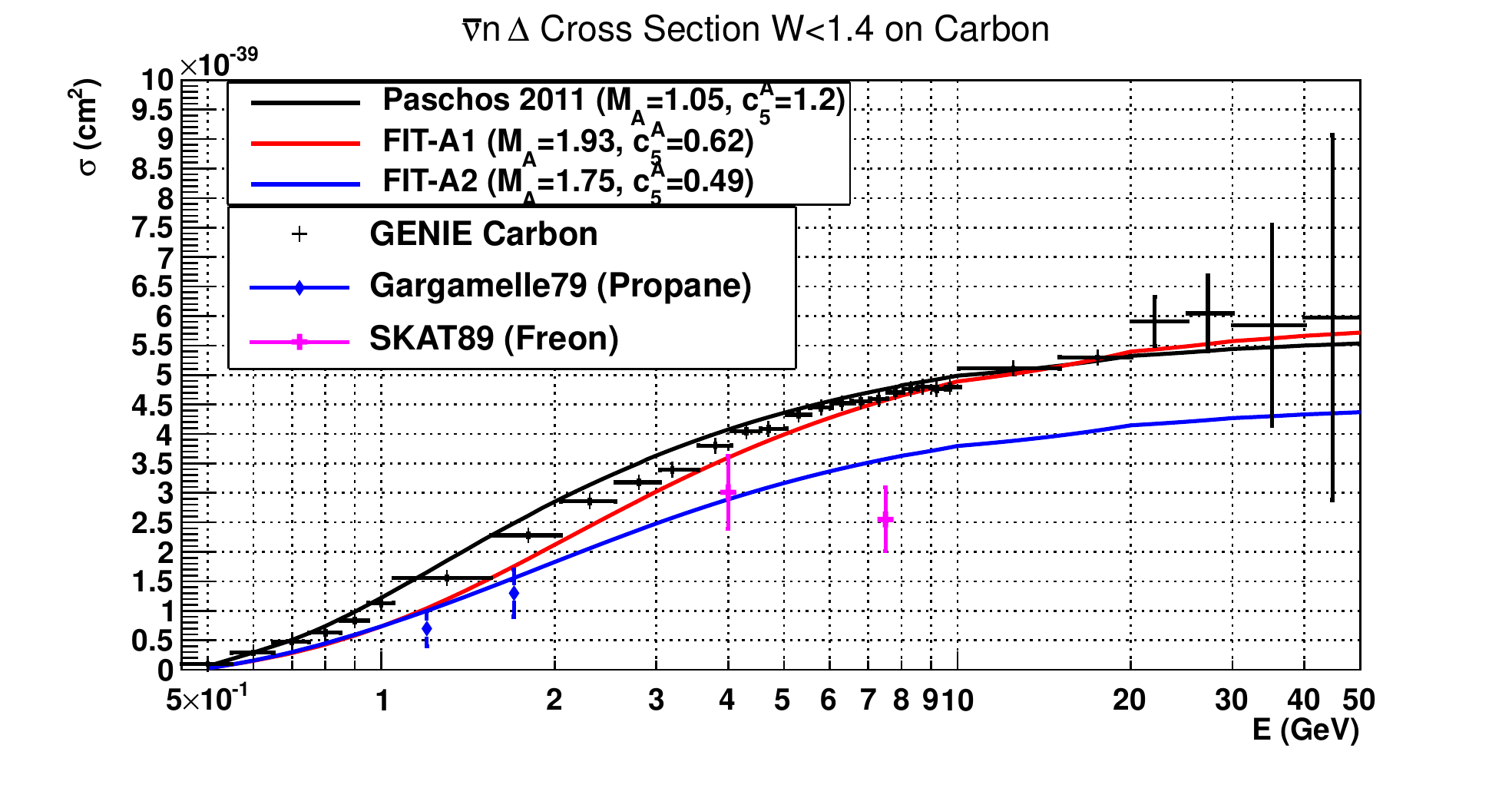}
\vspace{-0.3in} 
\caption{  Same as Fig. \ref{deltanup} for the case of nuclear targets.
%
(color online).
}
\label{deltanup-pauli}
\end{figure}
%
%
%

Fig.  \ref{QEvsDelta50} shows 
the $\nu<0.5$ GeV partial charged current  cross sections as a function of energy. The partial cross sections
 extracted from the  GENIE Monte Carlo are shown as black points
with MC statistical errors.   The $\nu<0.5$ GeV partial cross section for $\nu_\mu$ scattering 
 is shown on the top panel, and the $\nu<0.5$ GeV partial cross section for $\nub_\mu$ scattering
 is shown on the bottom panel. 
 The QE contribution to the $\nu<0.5$  GeV partial  cross section   is shown in red,  and  the contribution from pion production processes ($\Delta$, inelastic and coherent pion production) is shown in blue.      
 
As seen in Fig.  \ref{QEvsDelta50}, the  pion production contribution to the   $\nu<0.5$ GeV   partial
cross section is relatively constant with energy, while the QE contribution has some  energy
dependence.   Therefore,  the  energy dependence of the
sum of the two contributions to the  $\nu<0.5$  GeV partial  cross section requires modeling of the relative magnitude of  QE and pion production processes (specifically at low $Q^2$).
    
As shown in  Fig. \ref {deltanup} and  \ref {deltanup-pauli},  the consistency 
among the experimental measurements of pion production cross sections in the
region of the $\Delta (1232$) resonance  is about  20\%
(depending on the neutrino energy and the nuclear target).   We use this  variation to get an
estimate of the model uncertainty in the determination of the neutrino flux from the  $\nu<0.5$ GeV samples. This uncertainty can be greatly reduced when more precise measurements of the QE and pion production cross sections become available (e.g. from MINERvA).    
%
\section{Pion production with  $W<1.4$ GeV}
In this section we describe the uncertainties
in the modeling of pion production
cross sections for $W<1.4$ GeV.
The antineutrino structure functions are related to the
neutrino structure functions by the following relationship.
\begin{eqnarray}
{\cal F}_{i} ^{\bar{\nu} n} & =&  {\cal F}_{i} ^{\nu p}  \\
{\cal F}_{i} ^{\bar{\nu} p} & =&  {\cal F}_{i} ^{\nu n}  \nonumber
\label{iso}
\end{eqnarray}

\subsection{$\nu_\mu P \to \mu^- \Delta^{++}$ and $\nub_\mu N \to \mu^+  \Delta^-$ (FIT-A)}

We define the cross section for $\nu_\mu P \to \mu^- \Delta^{++}$  as the integrated
cross section  for  $W<1.4$ GeV for the following single final state:
$$\nu_\mu P \to \mu^-  P \pi ^{+}$$
We define the cross section for $\nub_\mu N \to \mu^+  \Delta^-$  as the integrated
cross section  for $W<1.4$ GeV for the following single final state:
$$\nub_\mu N \to \mu^+  N \pi^-$$

Therefore, our definition includes the sum of the contributions of the resonant cross section
and the non-resonant continuum.

The structure functions (form factors)  for the reactions $\nu_\mu P \to \mu^- \Delta^{++}$ and $\nub_\mu N \to \mu^+  \Delta^-$  defined above are the same (except that for antineutrinos the structure function $W_3$ changes sign). 
It has been experimentally determined\cite{BEBC90} that $\nu_\mu P$ cross section for $W<1.4$ GeV is  dominated by the resonant   $\Delta^{++}$  production 
process.  Similarly,  the  $W<1.4$ GeV  cross section for $\nub_\mu N$ is  dominated by the 
resonant $\Delta^-$ production process. 

As discussed in the Appendix, we parametrize the $\Delta^{++}$ and  $\Delta^-$   production cross sections in terms of form factors as given by Paschos and Lalakulich\cite{paschos}, with the form factors of Paschos and  Schalla\cite{paschos}.   In order to obtain predictions for the   $W<1.4$ GeV region, we divide all theoretical $\Delta$ production cross sections by a factor of 1.2 (because 20\% of the resonant cross section is above  $W=1.4$ GeV).  We vary two of the parameters in the model, specifically  $M_A^{\Delta}$ and $C_5^A$ to obtain a band that span the experimental data. We extract $M_A^{\Delta}$ from the measured
$Q^2$ distributions and use $C_5^A$ to set the overall normalization.
%
%
         \begin{figure}
\includegraphics[width=3.7in,height=2.9in]{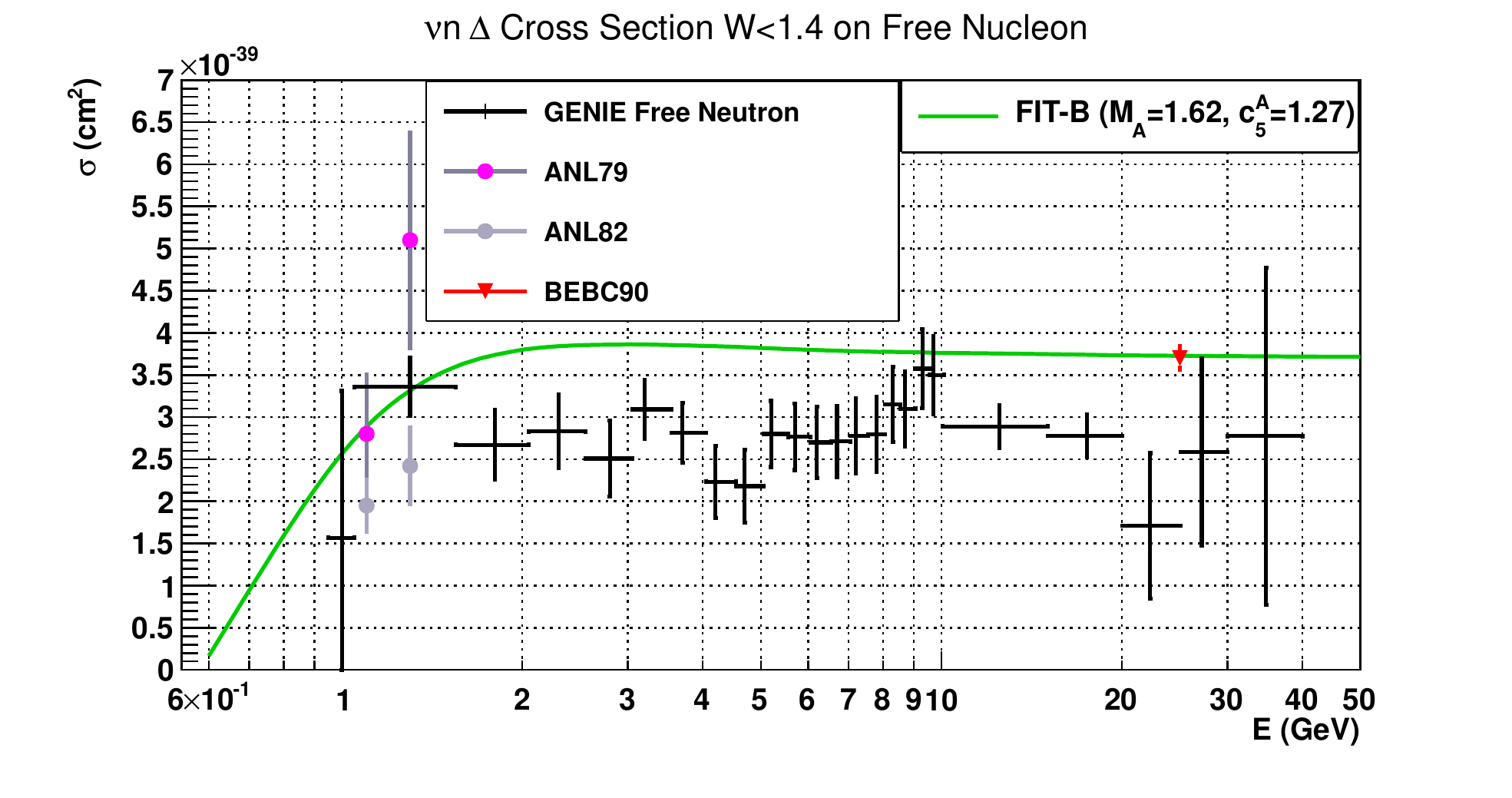}
     
     \vspace{-0.2in} 
     
\includegraphics[width=3.7in,height=2.9in]{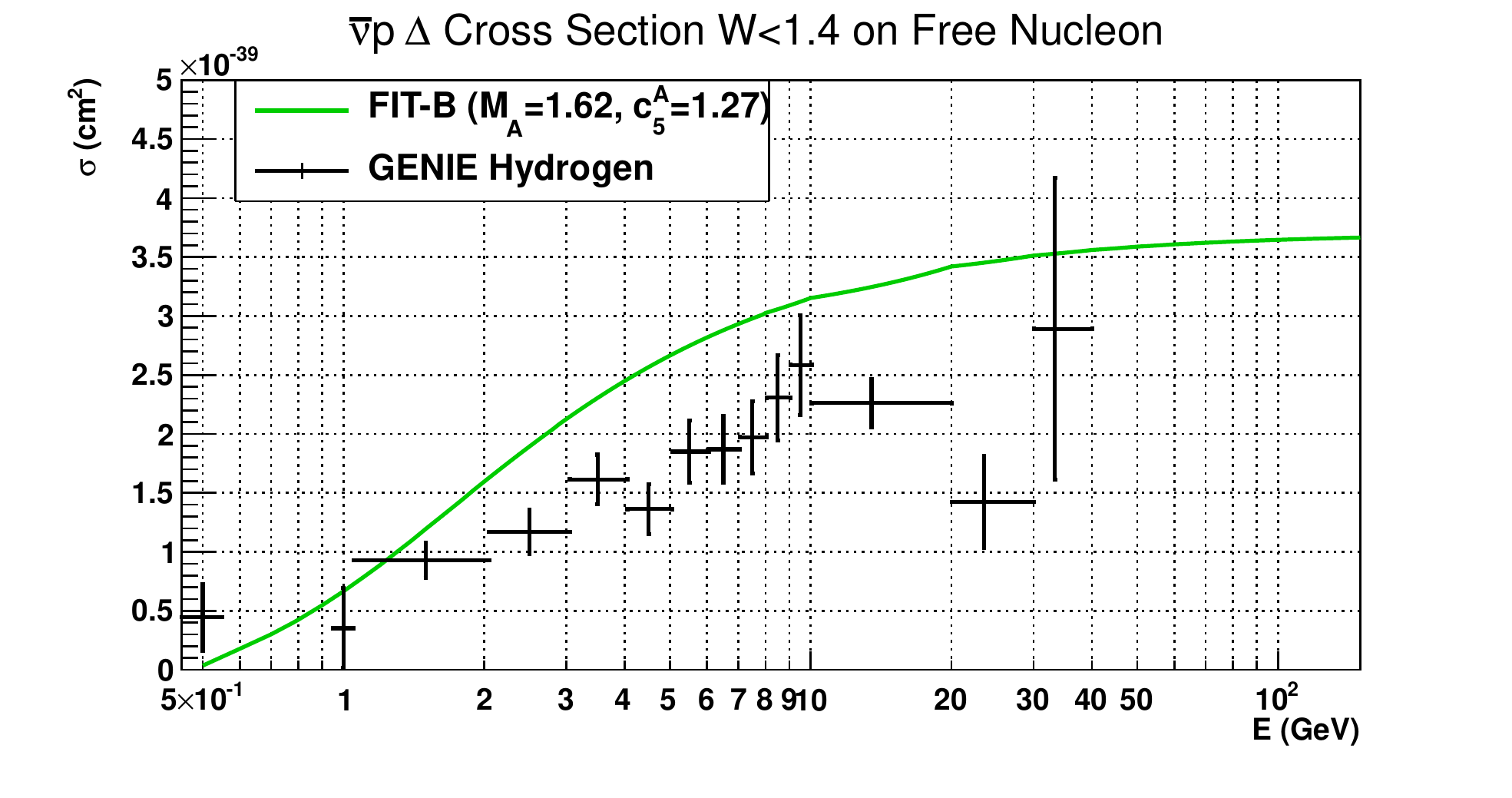}
\vspace{-0.3in} 
\caption{ $\nu_\mu N \to \mu^- \Delta^{+}$ (top panel)  and $\nub_\mu P \to \mu^+  \Delta^0$ (bottom panel)     cross sections $(W<1.4~GeV)$ measured  on free nucleons (H or D). The predictions from the GENIE MC  are shown as black points with errors.   
 (color online). 
 }
\label{deltanun}
\end{figure}
%
         \begin{figure}
\includegraphics[width=3.7in,height=2.9in]{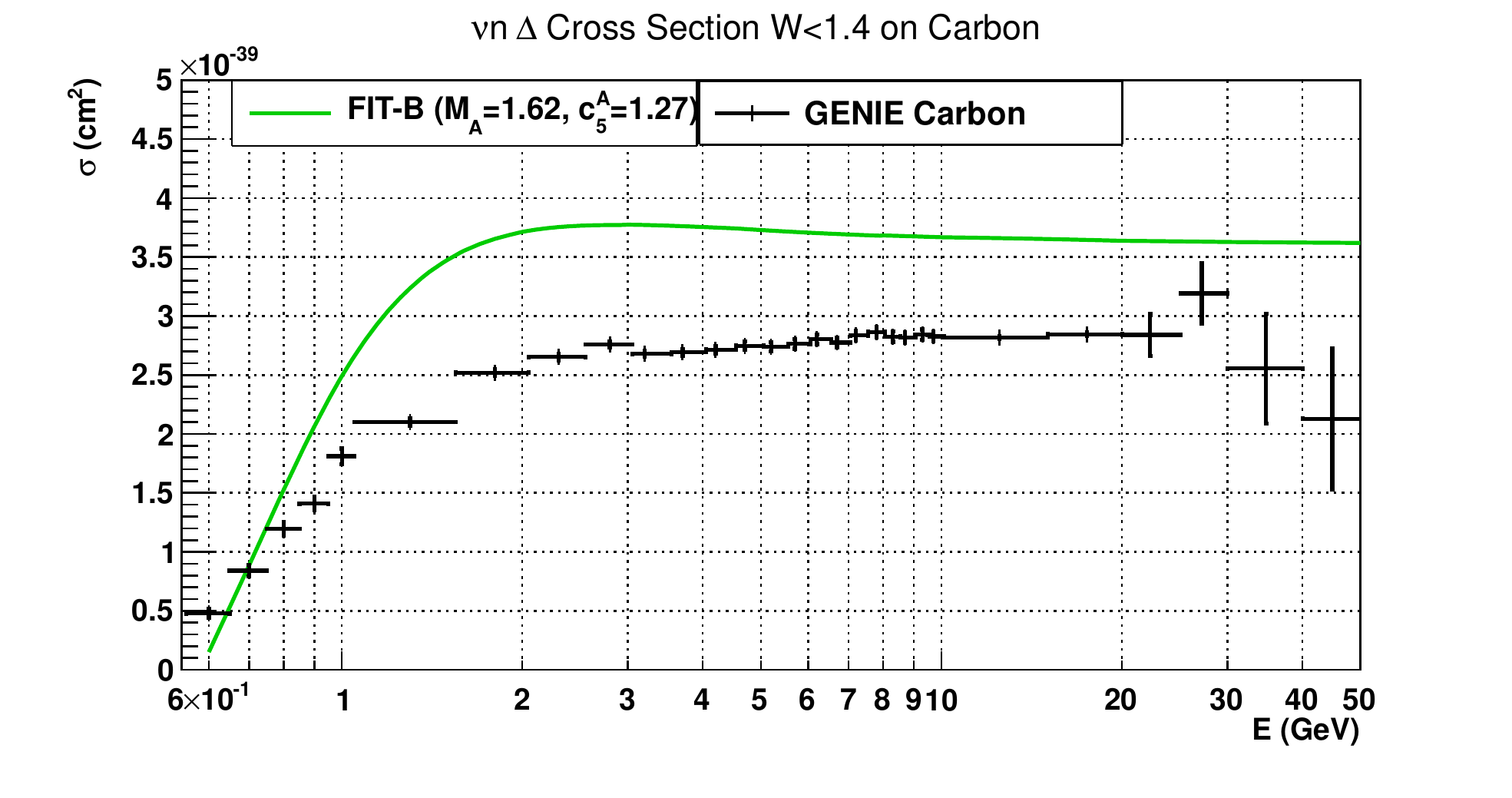}
     
     \vspace{-0.2in} 
     
\includegraphics[width=3.7in,height=2.9in]{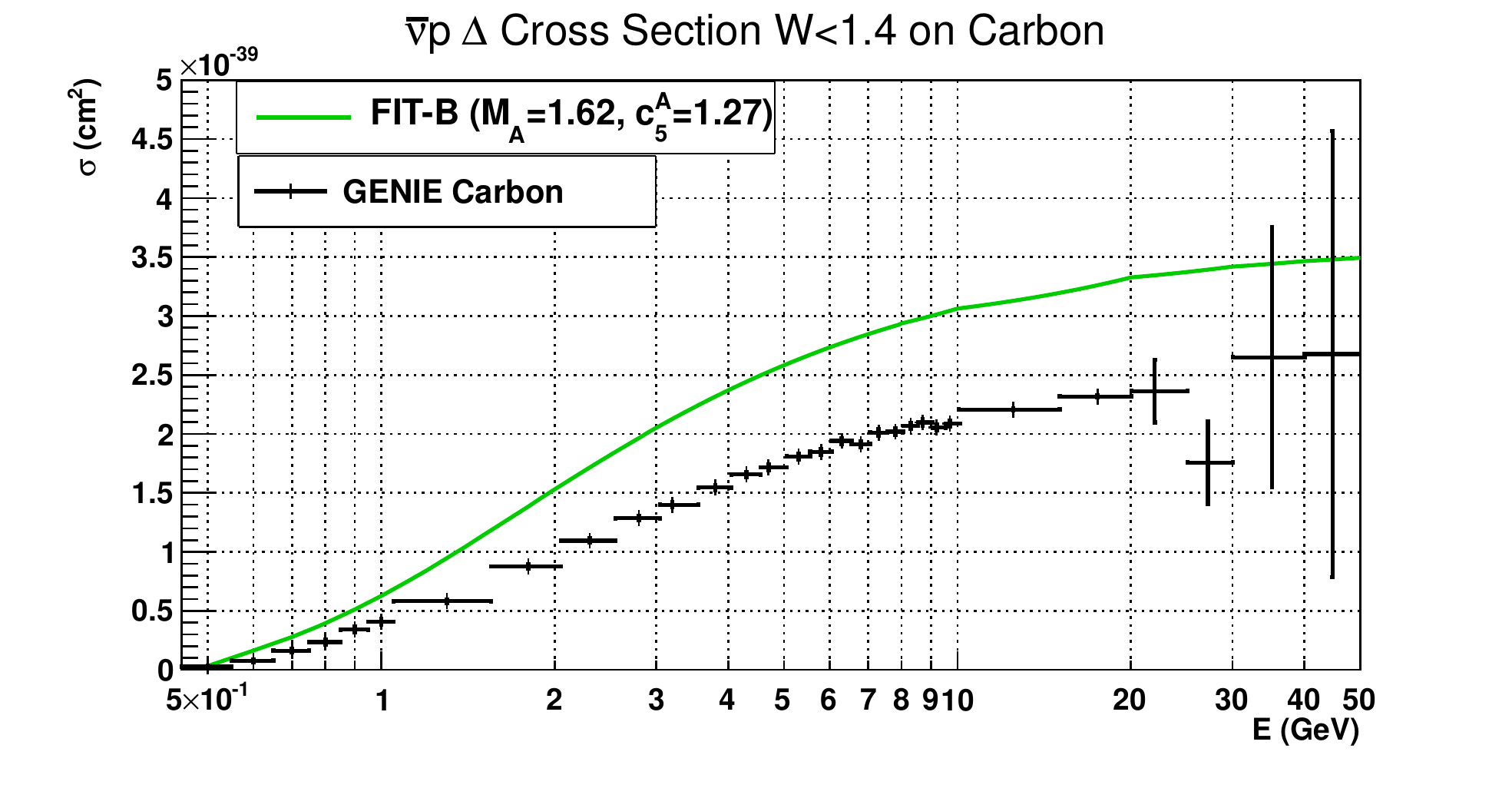}
\vspace{-0.3in} 
\caption{  Same as Fig. \ref{deltanun} for nuclear targets.
(color online).   
}
\label{deltanun-pauli}
\end{figure}

The top panel in Fig. \ref {deltanup} shows a summary of cross section measurements for  $\nu_\mu P \to \mu^- \Delta^{++}$  on free nucleons (hydrogen or deuterium targets.)  Shown are bubble chamber measurements at low energy from Argonne (ANL73\cite{ANL73}, ANL79\cite{ANL79}, ANL82\cite{ANL82}) and measurement at low energy from Brookhaven (BNL86\cite{BNL86}). Also shown are measurements at higher energies from the  Fermilab bubble chamber (FNAL78\cite{FNAL78},  FNAL81\cite{FNAL81}) and high energy data from  CERN (BEBC80\cite{BEBC80}, BEBC80\cite{BEBC90}).   
 The bottom panel in Fig. \ref {deltanup}   shows  the BEBC90\cite{BEBC90}
cross section measurements for  $\nub_\mu N \to \mu^+  \Delta^-$  on free nucleons (deuterium target). The predictions from the GENIE MC  on free nucleons  (shown as black points with MC statistical errors)
are near the upper bound of our three parameterizations.  

The black curve labeled Paschos-2011 ($M_A^{\Delta}$=1.05, $C_5^A$ = 1.2) uses the original
values of  $M_A^{\Delta}$ and  $C_5^A$ from the paper\cite{paschos} by Paschos and Lalakulich.
These values were obtained from fits to cross sections and $Q^2$ distributions
measured at low energies at Brookhaven and Argonne.
The red curve
labeled  FIT-A1 ($M_A^{\Delta}$=1.93, $C_5^A$ = 0.62)  is derived from
a fit to the cross sections and $Q^2$ distribution of the higher energy
BEBC90\cite{BEBC90} data for $\nub_\mu N \to \mu^+  \Delta^-$.  The blue curve labeled
FIT-A2 ($M_A^{\Delta}$=1.75, $C_5^A$ = 0.49) is derived from
a fit to the cross sections and $Q^2$ distribution of the higher energy
BEBC90\cite{BEBC90} data for $\nu_\mu P \to \mu^- \Delta^{++}$.

The top panel in  Fig. \ref {deltanup-pauli} shows a summary of cross section measurements for  $\nu_\mu P \to \mu^- \Delta^{++}$ data on nuclear targets.   Shown are the measurements of  Gargamelle78\cite{GGM78} (Propane),  SKAT88\cite{SKAT88} (Freon), and SKAT89\cite{SKAT89} (Freon).
The bottom panel  shows measurements  of
 $\nub_\mu N \to \mu^+  \Delta^-$  cross sections on nuclear  targets from
 Gargamelle78\cite{GGM79} (Propane) and   SKAT89\cite{SKAT89} (Freon).

Aside from Pauli suppression and final state interaction, 
 the structure functions (form factors)  for the processes in
Fig. \ref {deltanup} and \ref {deltanup-pauli} are the same.
The black (Paschos-2011),  red (FIT-A1) and blue (FIT-A2)  curves shown in Fig. \ref {deltanup} and  \ref {deltanup-pauli} use the free nucleon form factors (but include the Pauli suppression for the case of nuclear targets).  The calculations 
do not include the effect of final state interaction for the nuclear targets.   
The three  curves (Paschos-2011, FIT-A1 and FIT-A2)  conservatively
span all the available  $\Delta^{++}$ and $\Delta^-$ production  cross sections on hydrogen, deuterium  and nuclear targets, as shown in  Fig. \ref {deltanup} and  \ref {deltanup-pauli}.  The cross sections for
the production of $\Delta^{++}$ and $\Delta^-$ on nuclear targets 
predicted by GENIE are near the upper bound of our three parameterizations. 
Additional details are given in the Appendix.


\subsection{$\nu_\mu N \to \mu^- \Delta^{+}$ and $\nub_\mu P \to \mu^+  \Delta^0$ (FIT-B)}

We define the cross section for $\nu_\mu N \to \mu^- \Delta^{+}$   as the sum of the integrated
cross sections  for  $W<1.4$ GeV for the following two  final states:
$$\nu_\mu P \to \mu^-  N \pi ^{+}$$
$$\nu_\mu P \to \mu^-  P \pi ^{0}$$
We define the cross section for $\nub_\mu P \to \mu^+  \Delta^0$   as the sum of the  integrated
cross sections  for $W<1.4$ GeV for the following two final states:
$$\nub_\mu N \to \mu^+  P \pi^-$$
$$\nub_\mu N \to \mu^+  N \pi^0$$

Therefore, our definition includes the sum of the contributions of the resonant cross section
and non-resonant continuum.

The structure functions (form factors)  for the reactions   $\nu_\mu N \to \mu^- \Delta^{+}$ and $\nub_\mu P \to \mu^+  \Delta^0$  defined above  
are the same  (except that for antineutrinos the structure function $W_3$ changes sign). 
Because of Clebsch-Gordan coefficients\cite{paschos}
the resonant cross section for  $\Delta^{+}$ production  in $\nu_\mu N$ collisions is a third of the resonant cross section for  $\Delta^{++}$ production  in $\nu_\mu P$ collisions.  Similarly, the resonant cross section for  $\Delta^{0}$ production  in $\nub_\mu P$ collisions is a third of the cross section
for resonant production of   $\Delta^{-}$  in $\nub_\mu N$ collisions.

However, unlike the case for  $\nu_\mu P$ ($\Delta^{++}$) and   $\nub_\mu N$ ($\Delta^{-}$), where
the cross sections are  dominated by the resonant process, there is
a significant  contribution from the non-resonant continuum to the $W<1.4$ GeV cross section
in  $\nu_\mu N$   and  $\nub_\mu P$ collisions.  
         \begin{figure}
\includegraphics[width=3.7in,height=2.9in]{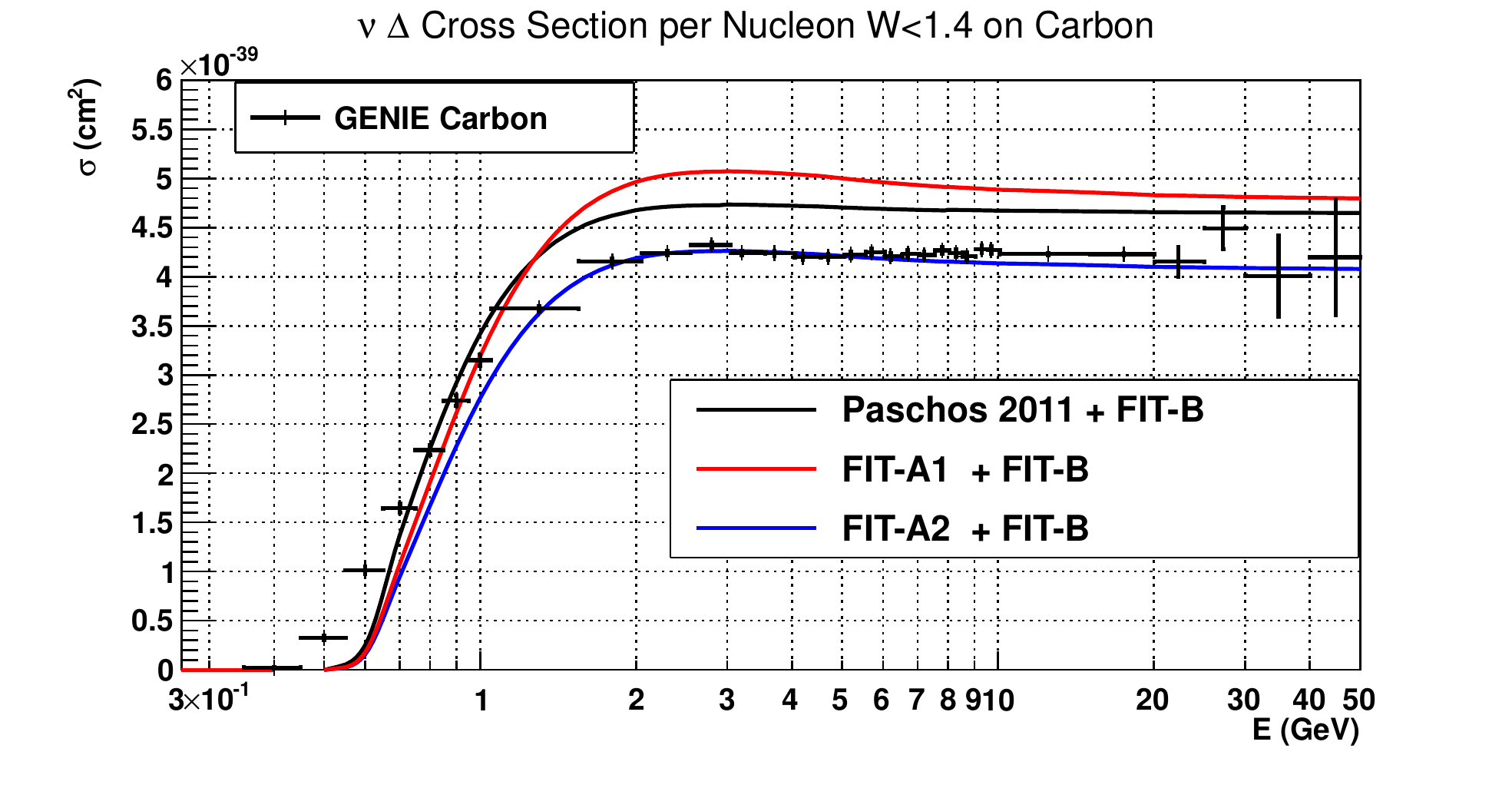}
     
     \vspace{-0.2in} 
     
\includegraphics[width=3.7in,height=2.9in]{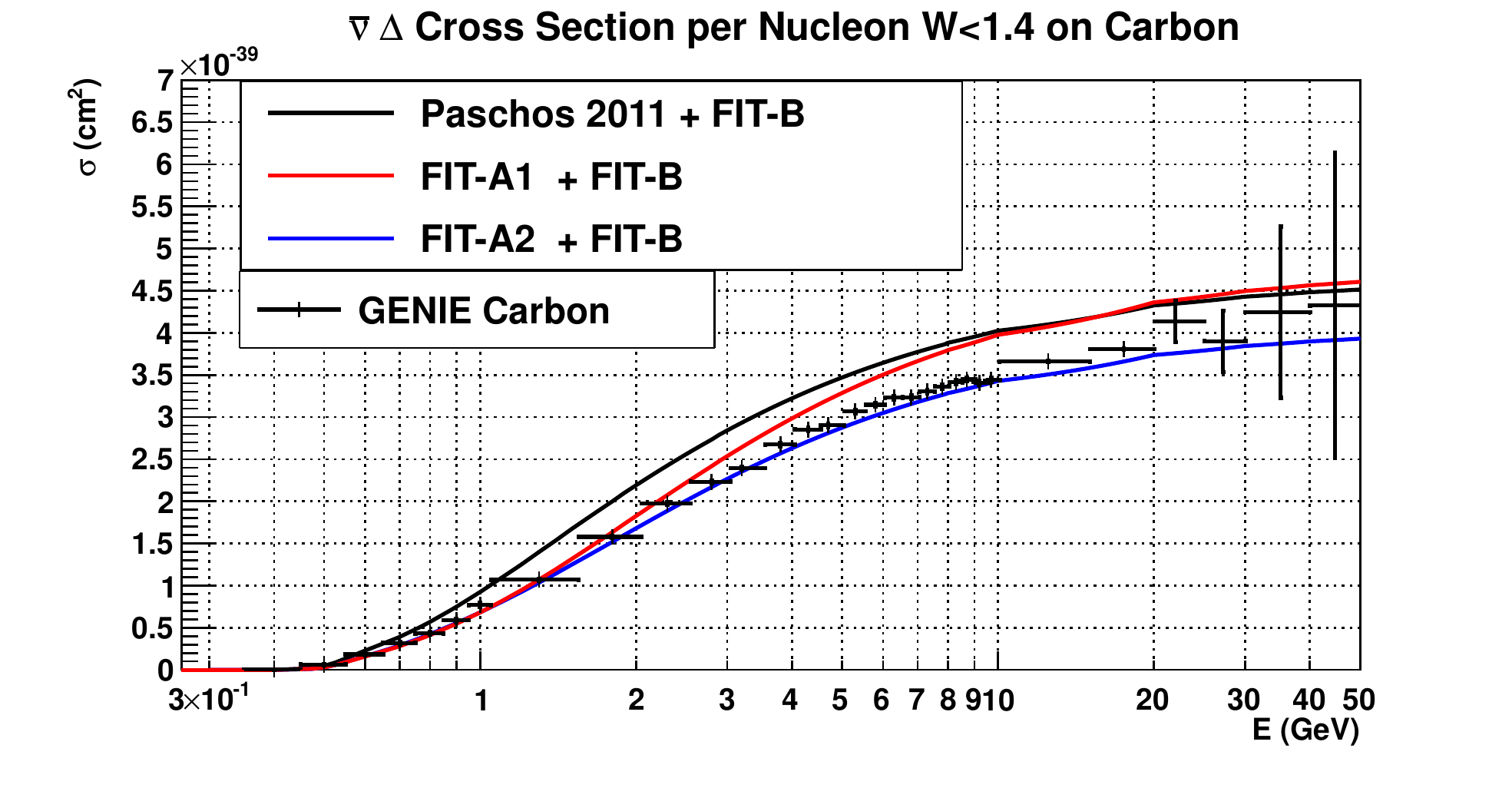}
\vspace{-0.3in} 
\caption{  The total cross sections on carbon (per nucleon) predicted by GENIE for $W<1.4$ GeV (black points with MC statistical errors) for $\nu_\mu C \to (\mu^- \Delta^{++}$ or  $\Delta^{+}$) are shown on the top panel, and for $\nub_\mu C \to \mu^+  (\Delta^0$ or $\Delta^{-}$) are shown on the bottom panel.
    (color online).}
\label{deltan-carbon}
\end{figure}
 
 The top panel of Fig. \ref{deltanun} shows the  $\nu_\mu N \to \mu^- \Delta^{+}$  cross sections ($W<1.4~$GeV) measured  on free nucleons (deuterium).  Shown
 are measurements from  ANL79\cite{ANL79}, ANL82\cite{ANL82},  and BEBC90\cite{BEBC90}. 
 The predictions from the GENIE MC  are shown as black points with MC statistical  errors.   
 In order to describe the data (which has a large  non-resonant contribution)
   we changed the parameters in the
 Paschos and Lalakulich\cite{paschos} resonance model to fit the
 observed $Q^2$ distribution and total $W<1.4$ GeV 
 cross sections. The green curve labeled FIT-B ($M_A^{\Delta}$=1.62, $C_5^A$ = 1.27)  is derived from
a fit to the $W<1.4$ GeV cross sections and $Q^2$ distribution of the
BEBC90\cite{BEBC90} data for $\nu_\mu N \to \mu^- \Delta^{+}$.
 This curve provides a parameterization which describe the experimental data for
 the production of $\Delta^{+}$ (with neutrinos)  and $\Delta^0$ (for antineutrinos)  on  free nucleons.   The GENIE MC cross sections  for
 the production of $\Delta^{+}$ on free nucleons are lower than the fit.  

The structure functions (form factors)  for the reactions  $\nu_\mu N \to \mu^- \Delta^{+}$ and $\nub_\mu P \to \mu^+  \Delta^0$  ($W<1.4~$GeV) are are same.   The bottom panel of Fig. \ref{deltanun} shows a comparison
of  the predictions of FIT-B ($M_A^{\Delta}$=1.62, $C_5^A$ = 1.27) (green curve) for the
 $\nub_\mu P \to \mu^+  \Delta^0$ cross sections on free nucleons compared to the predictions from the GENIE MC  which are shown as black points with MC statistical  errors.   The GENIE MC cross sections  for
 the production of $\Delta^{0}$ on free nucleons are lower than the fit.  
  
          \begin{figure}
\includegraphics[width=3.7in,height=2.9in]{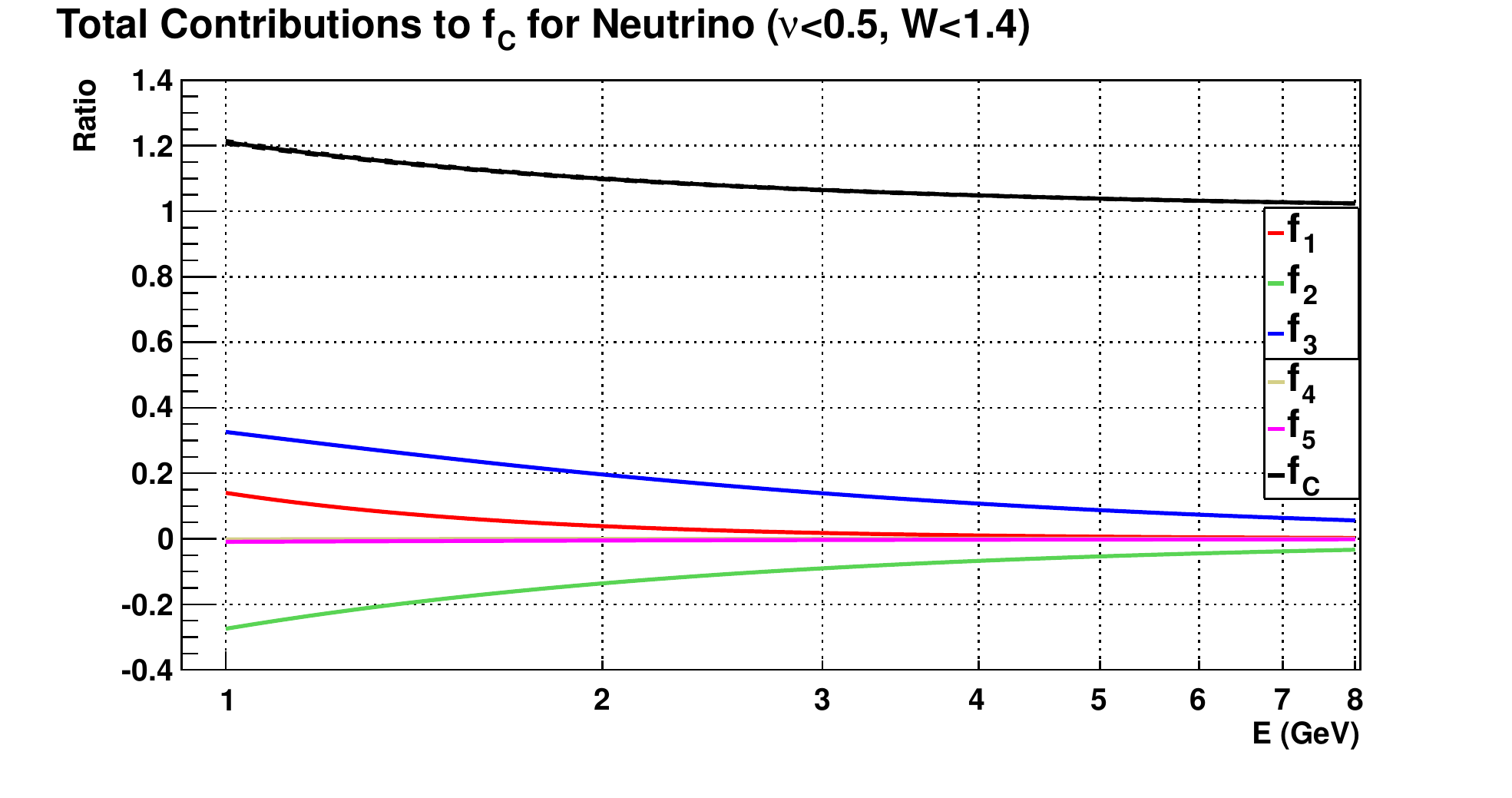}
     
     \vspace{-0.2in} 
     
\includegraphics[width=3.7in,height=2.9in]{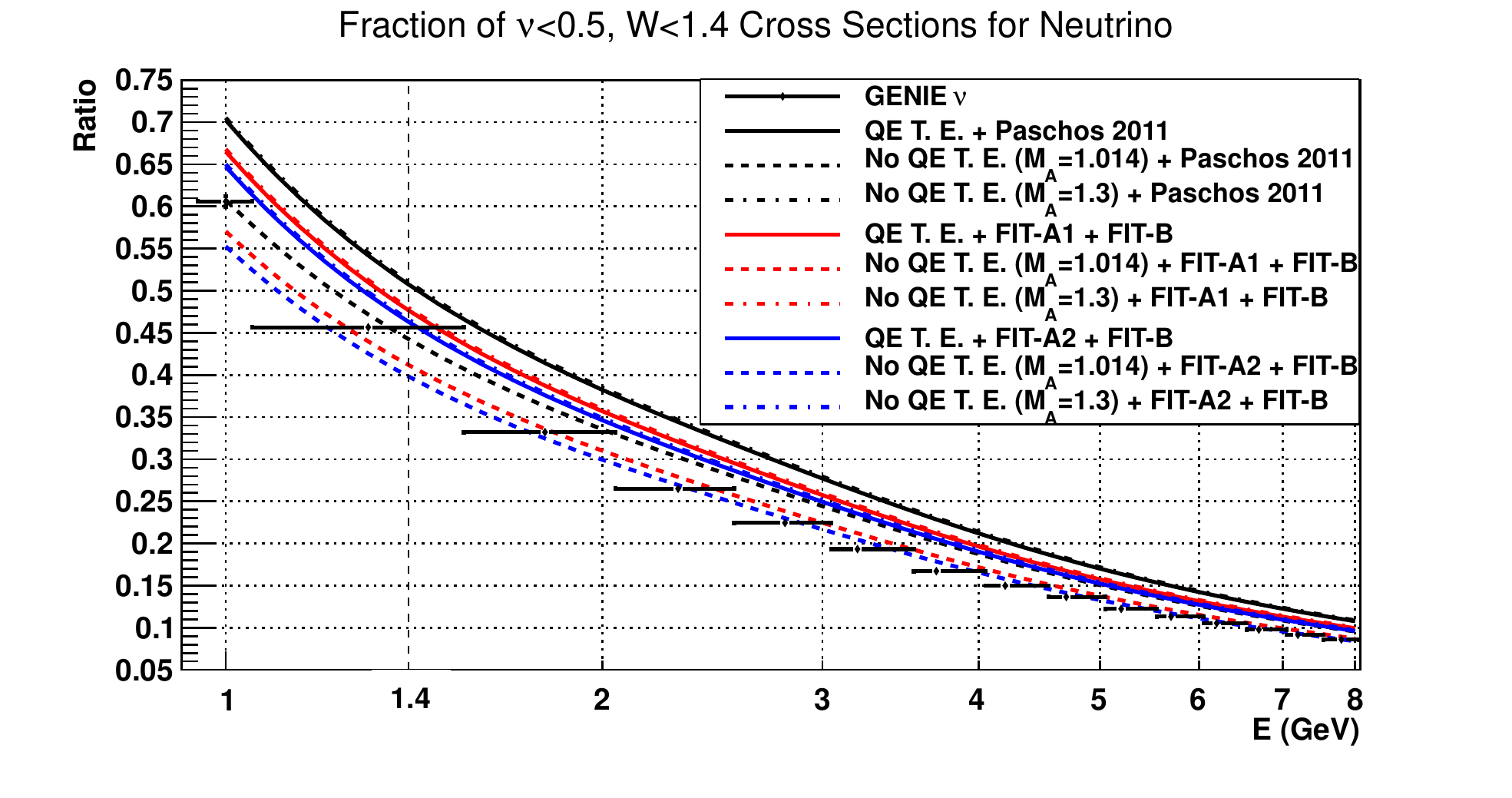} 
\vspace{-0.3in}           
\caption{The  $\nu<0.5$ GeV sample for $\nu_\mu$. This sample  includes  both   QE  $\nu_\mu N \rightarrow \mu^- P$ events ($\approx$66\%)   and $\Delta$ production events ($\approx$ 33\%).   Top panel: The total corrections factor $f_C$ (with error bands) and  the  contributions
of  the kinematic correction to ${\cal W}_2$ ($f_2$), and the contributions  from ${\cal W}_1$ ($f_1$), ${\cal W}_3$ ($f_3$), ${\cal W}_4$ ($f_4$),  and ${\cal W}_5$ ($f_5$).   Bottom panel: The fractional contribution of  
$\nu<0.5$ GeV events to the total cross section. 
 (color online). 
  }
\label{neutrino50}
\end{figure}
%
        \begin{figure}
\includegraphics[width=3.7in,height=2.9in]{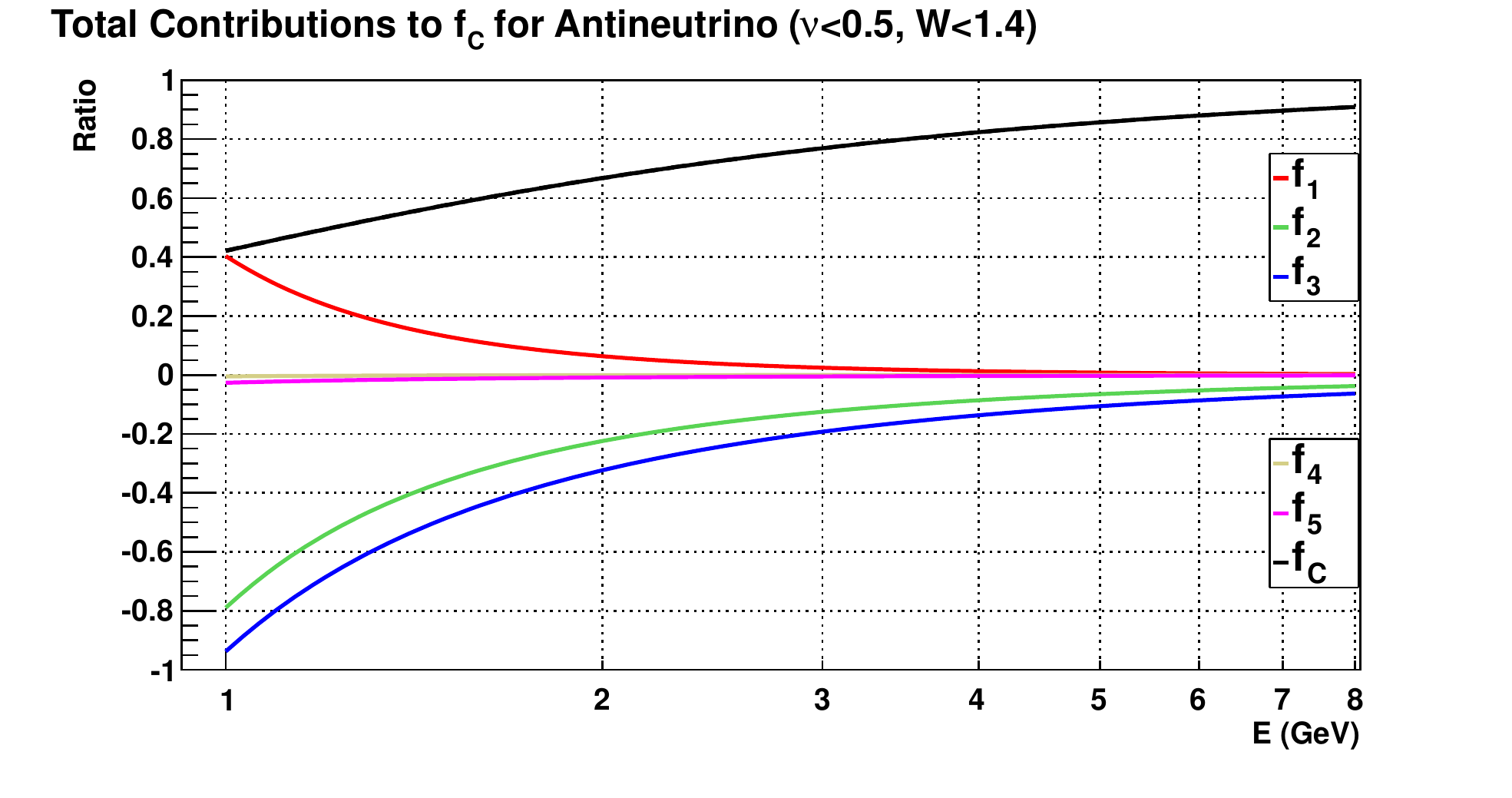}
     
     \vspace{-0.2in} 
     
\includegraphics[width=3.7in,height=2.9in]{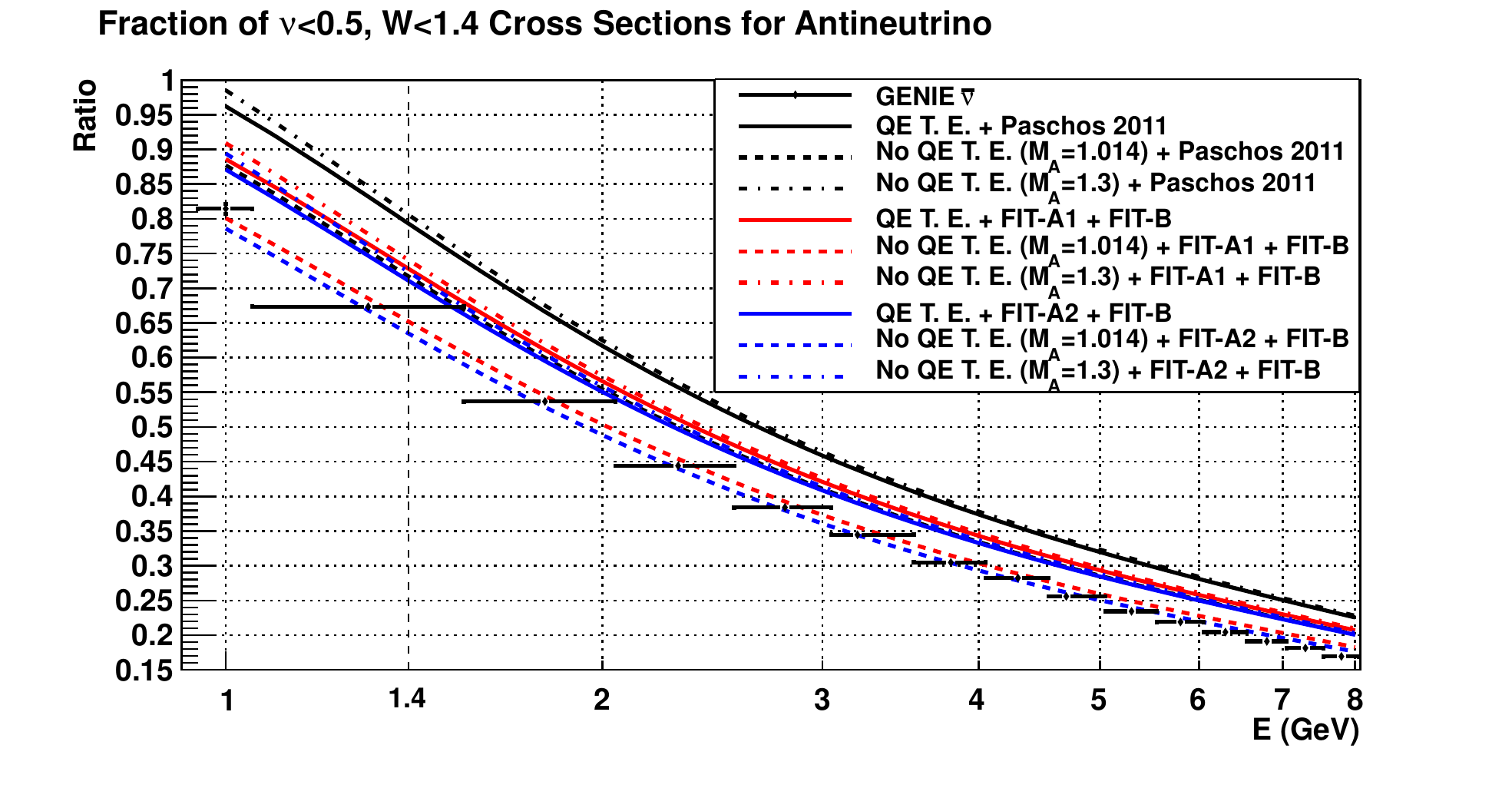}
\vspace{-0.3in} 
\caption{ Same as Fig. \ref{neutrino50} for the case of antineutrinos.
(color online).
}
\label{antineutrino50}
\end{figure}

Fig. \ref{deltanun} shows the prediction of  FIT-B ($M_A^{\Delta}$=1.62, $C_5^A$ = 1.27) (green curve) for the  $\nu_\mu N \to \mu^- \Delta^{+}$ (top panel)  and $\nub_\mu P \to \mu^+  \Delta^0$ (bottom panel)   $W<1.4$ GeV cross sections on nuclear targets  compared to  predictions from the GENIE MC (black points with MC statistical errors).
The cross sections  on nuclear targets are expected to be somewhat lower than the cross sections
on free nucleons (which are shown in Fig. \ref {deltanun}). 
Here, FIT-B  includes the effect of
Pauli suppression (but not final state interaction). The GENIE MC cross sections  for
 the production of $\Delta^{+}$ and $\Delta^{0}$ on nuclear targets  are lower than the fit. 
 Additional details are given in the Appendix.

\subsection{Comparisons of $W<1.4$ GeV cross sections on carbon}

A more relevant comparison is to determine how well the GENIE Monte Carlo describes
the sum of the proton and neutron cross sections on carbon, since 
it is the  total number of   $\nu<0.5$ GeV events on carbon that are used in the determination
 of the neutrino flux. 
 
 Fig.  \ref {deltan-carbon}  shows  the predictions from the GENIE MC
 for total $\Delta$ production cross section for $W<1.4$ GeV  on carbon (per nucleon).  The neutrino cross sections for $\nu_\mu C \to \mu^- (\Delta^{++}$ or $\Delta^{+}$) are shown in the  top panel,   and the antineutrino cross sections $\nub_\mu C \to (\mu^+ \Delta^0$ or $\Delta^{-}$) are shown in the bottom panel.  The cross sections which are predicted by GENIE are compared to our three parameterizations. 
(Paschos-2011, FIT-A1 and FIT-A2 for $\Delta^{++}$ and $\Delta^{-}$, and FIT-B for 
$\Delta^{+}$ and $\Delta^{0}$).
The GENIE cross section predictions for the  total $\Delta$ production cross sections
    on carbon (which use the Rein and Seghal model\cite{rs} for resonance production) fall near the lower bound of our three parameterizations of the experimental data.

         \begin{figure}
\includegraphics[width=3.7in,height=2.9in]{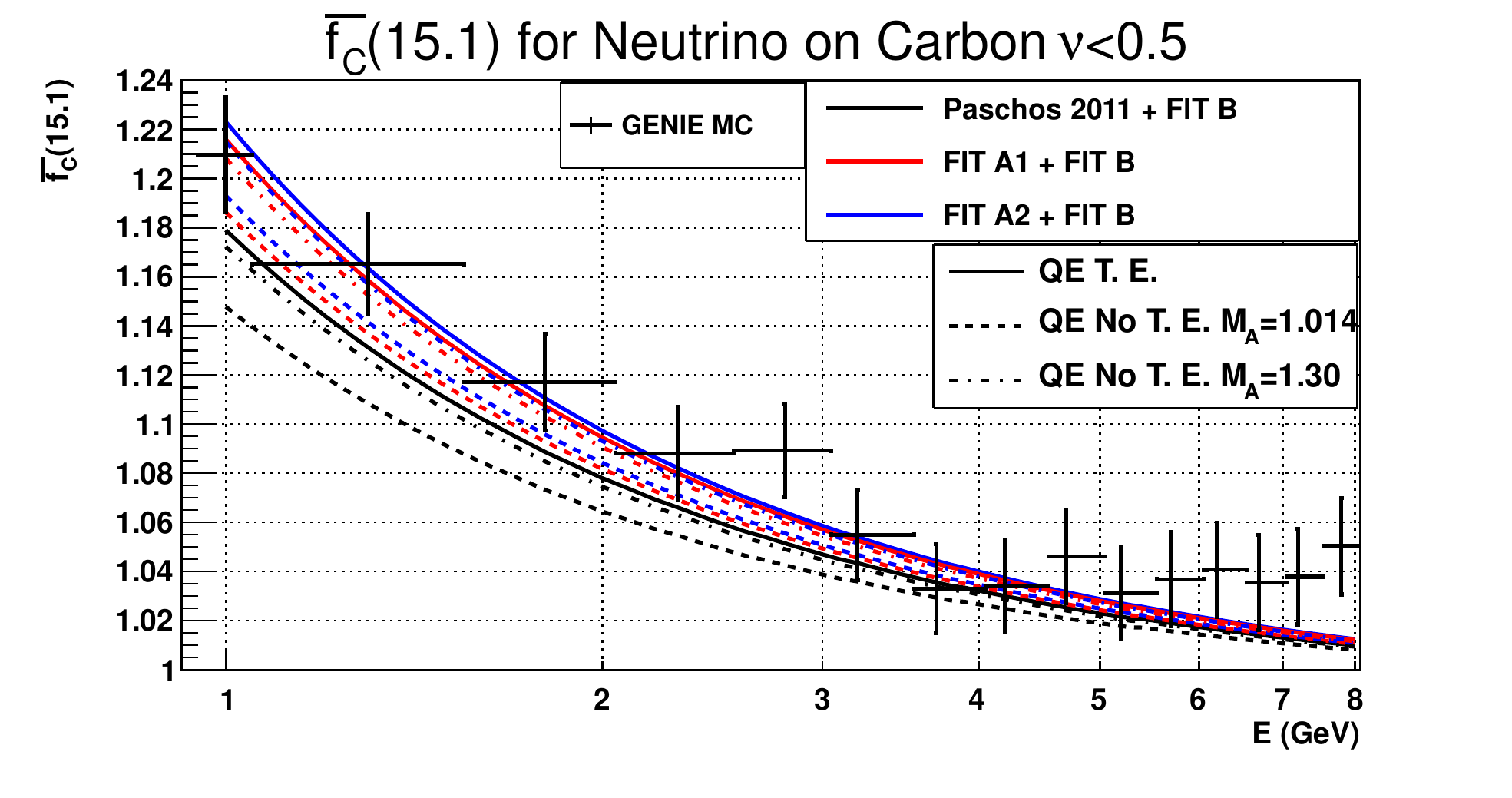}
     
     \vspace{-0.2in} 
     
\includegraphics[width=3.7in,height=2.9in]{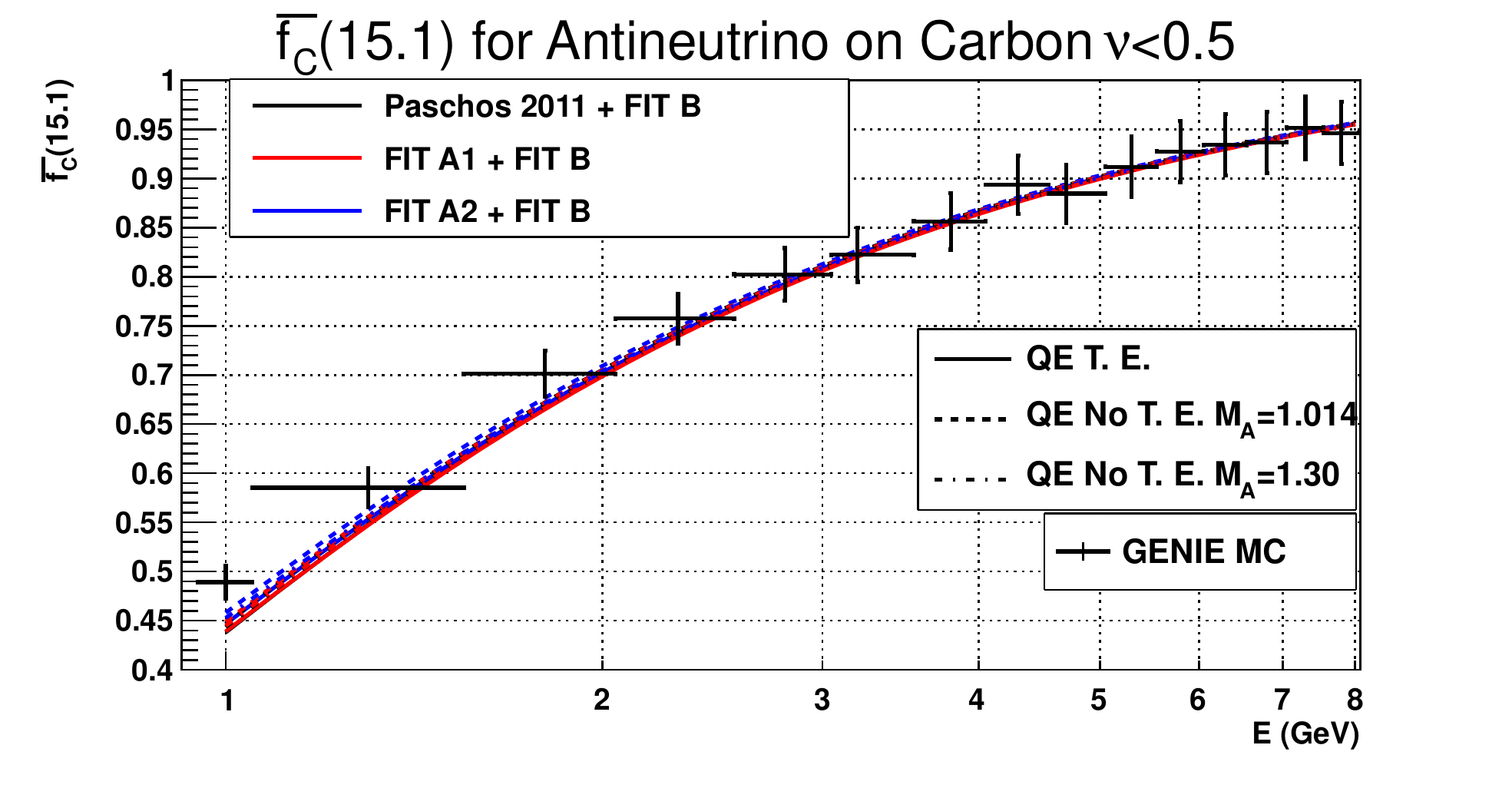}
\vspace{-0.3in} 
\caption{ Comparisons of our calculated values of  the normalized ${\bar f_{C:\nu<0.5}} (15.1)(E)$ (=$\bar f_C (15.1) $ for $\nu<0.5 $ GeV) 
to  values from the GENIE MC.  
Our nominal model (shown as the solid black line) uses the TE model for QE scattering and the Paschos 2011 model for $\Delta$ production. Neutrinos are shown on the top panel and antineutrinos are shown on the bottom panel (color online).     }
\label{fcbar50}
\end{figure}

     \begin{figure}
     \includegraphics[width=3.7in,height=2.9in]{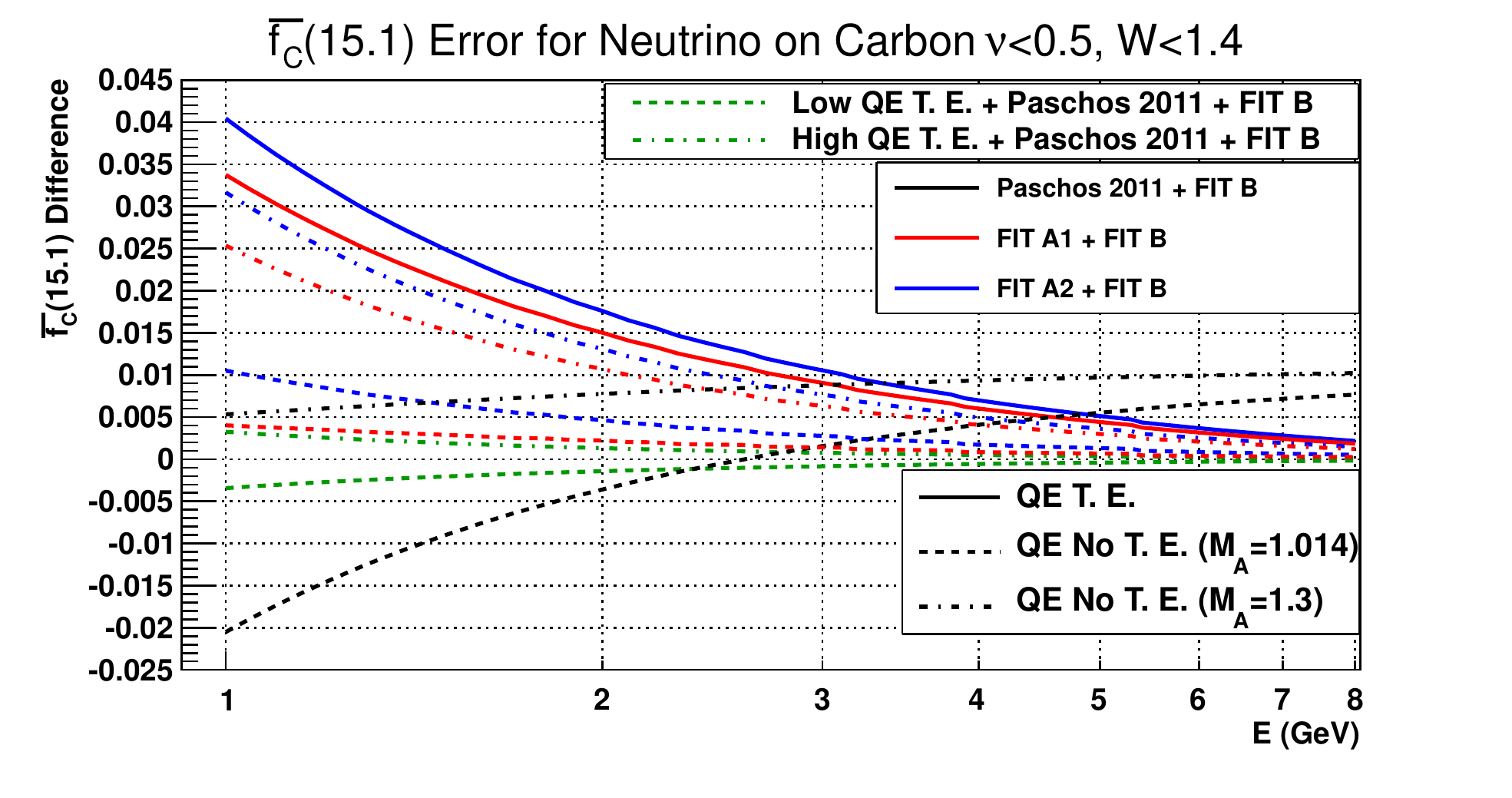}
     
     \vspace{-0.2in} 
     
     \includegraphics[width=3.7in,height=2.9in]{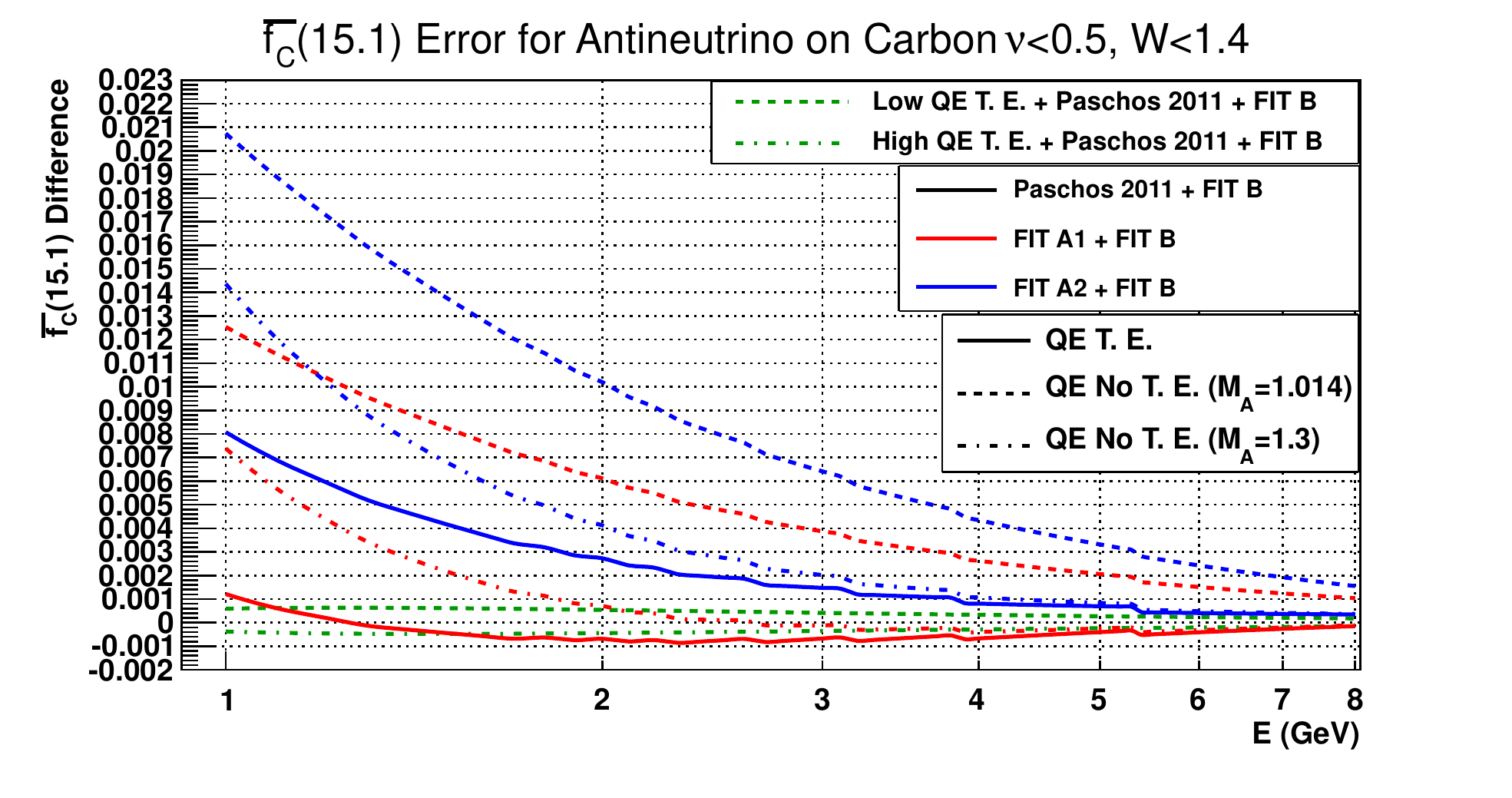}
     \vspace{-0.3in} 
\caption{ The error band in the normalized correction factor  ${\bar f_{C:\nu<0.5}} (15.1)(E)$ (=$\bar f_C (15.1) $ for $\nu<0.5 $ GeV).   Our nominal model is  QE with transverse enhancement and the Paschos 2011 model for $\Delta$ production.  Shown are the differences between our nominal model
and other model assumptions for neutrinos (top panel) and for antineutrinos (bottom panel).  
(color online). 
 }
\label{nudiff50}
\end{figure}
     %
     
     As described below,   the uncertainties in the measurements
     of the  $\Delta$ production cross sections do not place a serious limitation on the
      flux extractions using the low $\nu$ method.

\subsection{Determination of neutrino and antineutrino flux using $\nu<0.5$ GeV samples on carbon}
The $\nu<0.5$ GeV  sample  includes both  QE  $\nu_\mu N \rightarrow \mu^- P$ events ($\approx$ 66\%)   and $\Delta$ production events ($\approx$ 33\%).  

The top panel of Fig. \ref {neutrino50} shows the  total correction factor $f_{C}(E) $ for the  $\nu<0.5$ GeV  sample (defined as $f_{C:\nu<0.5}(E)$)  for $\bf {neutrino}$  running.
Also shown are the various contributions to  $f_{C:\nu<0.5}(E)$  including
the kinematic correction to ${\cal W}_2$ ($f_2$), and the contributions  from ${\cal W}_1$ ($f_1$), ${\cal W}_3$ ($f_3$), ${\cal W}_4$ ($f_4$),  and ${\cal W}_5$ ($f_5$).  The bottom panel shows the fractional contribution of   $\nu<0.5$ GeV events to the charged current neutrino total cross section.  Using our nominal model (TE model for QE scattering and the Paschos 2011 model for $\Delta$ production)  we find that the fraction of   $\nu<0.5$ GeV events  is less than  60\% for $\nu_\mu$ energies above  1.2 GeV. 

The top panel of Fig. \ref {antineutrino50} shows the  total correction factor  $f_{C:\nu<0.5}(E)$  for $\bf {antineutrino}$ running. Also shown are the various contributions to  $f_{C:\nu<0.5}$ including
the kinematic correction to ${\cal W}_2$ ($f_2$),
 and the contributions  from ${\cal W}_1$ ($f_1$), ${\cal W}_3$ ($f_3$), ${\cal W}_4$ ($f_4$),  and ${\cal W}_5$ ($f_5$).  The bottom panel shows the fractional contribution of   $\nu<0.5$ GeV events to the charged current antineutrino total cross section.  Using our nominal model (TE model for QE scattering and the Paschos 2011 model for $\Delta$ production) we find that the fraction of  $\nu<0.5$ GeV events  is less than  60\% for $\nub_\mu$ energies above  2  GeV.

   As for the  $\nu<0.25$ sample,  we propose that the neutrino  and antineutrino cross sections at low energy
   be measured relative to the cross sections at 15.1 GeV. 
   Therefore, we  define normalized quantity ${\bar f_{C:\nu<0.5}} (15.1)(E)$  for the $\nu<0.5$ sample  as: 
     $${\bar f_{C:\nu<0.5}} (15.1)(E)=\sigma_ {\nu<0.5} (E)/ \sigma_ {\nu<0.5} (E=15.1~ \GeV)$$
    which is equivalent to
   $${\bar f_{C:\nu<0.5}} (15.1)(E)  =  f_C (E) /f_C(E=15.1~\GeV)$$

Here,    $f_{C:\nu<0.5}$(E=15.1~GeV)=1.0113 (for  $\nu$)  and 0.9507 (for $\nub$). 
 These values can be used to convert between ${\bar f_{C:\nu<0.5}}(E)$ and ${f_{C:\nu<0.5}} (E)$.

Fig. \ref {fcbar50} shows  comparisons of our calculated values of the normalized ${\bar f_{C:\nu<0.5}} (15.1)(E)$ (shown as the solid black line) to  values extracted from the GENIE MC.  The GENIE predictions include a contribution from coherent pion   production. As shown in Appendix II, for the $\nu<0.50$ GeV sample, the contribution from
coherent pion production is less than 0.7\% for neutrinos and less than  3\% for antineutrino

Our  values are calculated from our nominal model which uses the  TE model for QE scattering and the Paschos 2011 model for $\Delta$ production. Neutrinos are shown on the top panel and antineutrinos are shown on the bottom panel (color online). 

%
%
 %
%
Figure \ref {nudiff50} shows the error band in the correction factor ${\bar f_{C:\nu<0.5}} (15.1)(E)$ for
neutrinos (top panel) and  antineutrinos (bottom panel).   
The error band is defined as the differences between our nominal model and other model assumptions.  For neutrinos with energies greater than 1.2 GeV, the error in 
$\bar f_C (15.1) $ is less than 0.03, which corresponds to a 2.6\% upper limit on  the model uncertainty in the neutrino flux extracted from the $\nu<0.5$ GeV sample.  For antineutrinos with energies greater than 2 GeV  the error in $\bar f_C (15.1) $ is less than 0.01 (which corresponds to a 1.4\% upper limit on the model  uncertainty in the antineutrino flux extracted from the $\nu<0.5$ GeV sample).

 In order to go to lower neutrino
and antineutrino energies we need to use the  $\nu<0.25$ GeV sample.  The model uncertainty  in the relative flux extracted  from the $\nu<0.25$ GeV sample is 1.9\% for $\nu_\mu$ energies  above 0.7 GeV   and 2.5\% for  $\nub_\mu$ energies above 1.0 GeV.  With improved determination of $QE$ and $\Delta$ production
cross sections (e.g. in MINERvA), the model uncertainties can be further reduced, and the method
may be extended to lower energies. 

         \begin{figure}
\includegraphics[width=3.7in,height=2.9in]{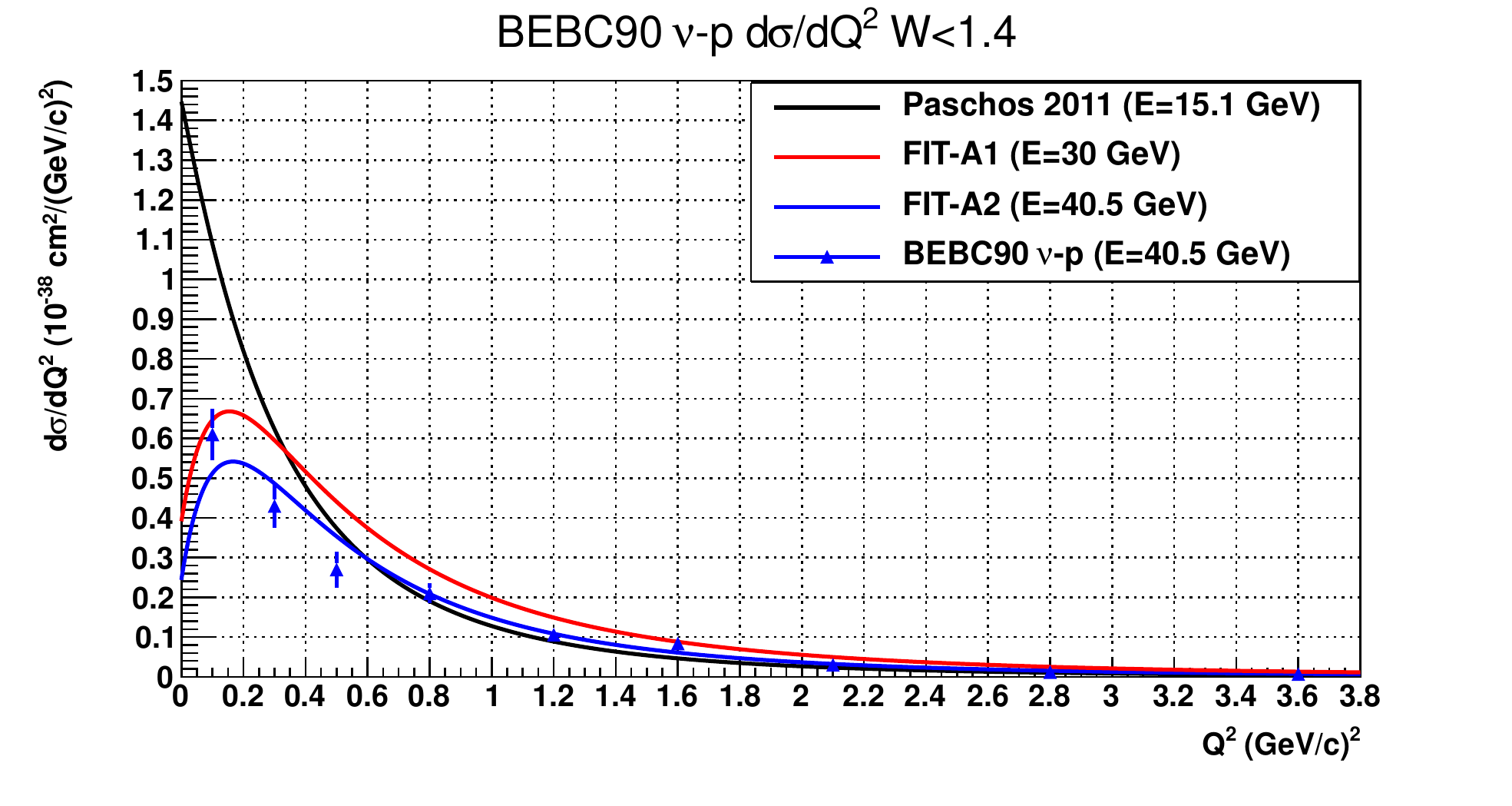}
     
     \vspace{-0.2in} 
     
\includegraphics[width=3.7in,height=2.9in]{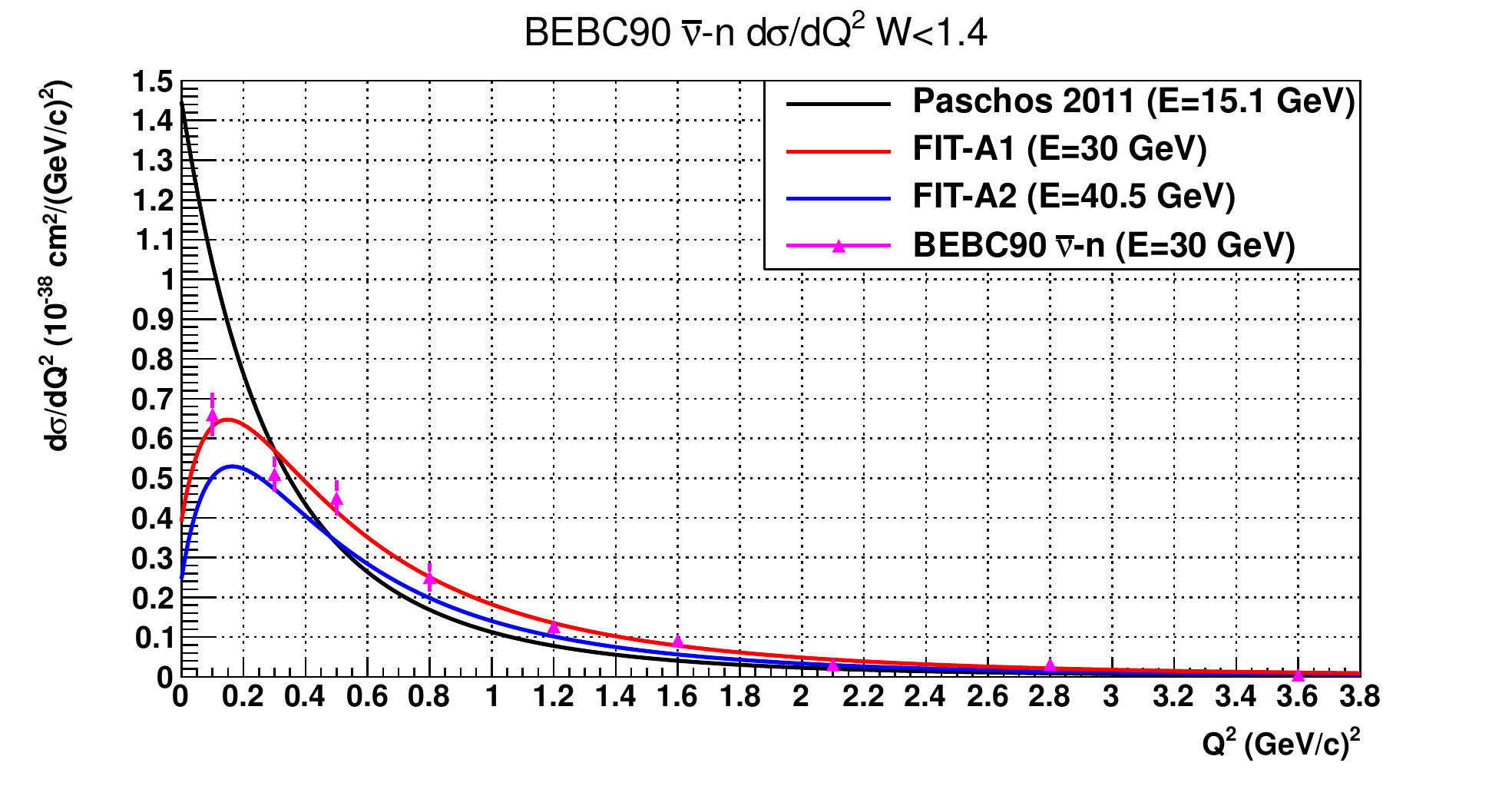}
\vspace{-0.3in} 
\caption{  $d\sigma/dQ^2$  cross sections  (for $W<1.4$ GeV) measured
 on deuterium at high energies by Allasia et al. (BEBC90\cite{BEBC90}). The cross sections for for $\nu_\mu P \to \mu^- \Delta^{++}$ are shown on the top panel and the cross sections for $\nub_\mu N \to \mu^+  \Delta^-$  are shown on the bottom panel.  
(color online). 
}
\label{dsdq2allasia}
\end{figure}

         \begin{figure}
\includegraphics[width=3.7in,height=2.9in]{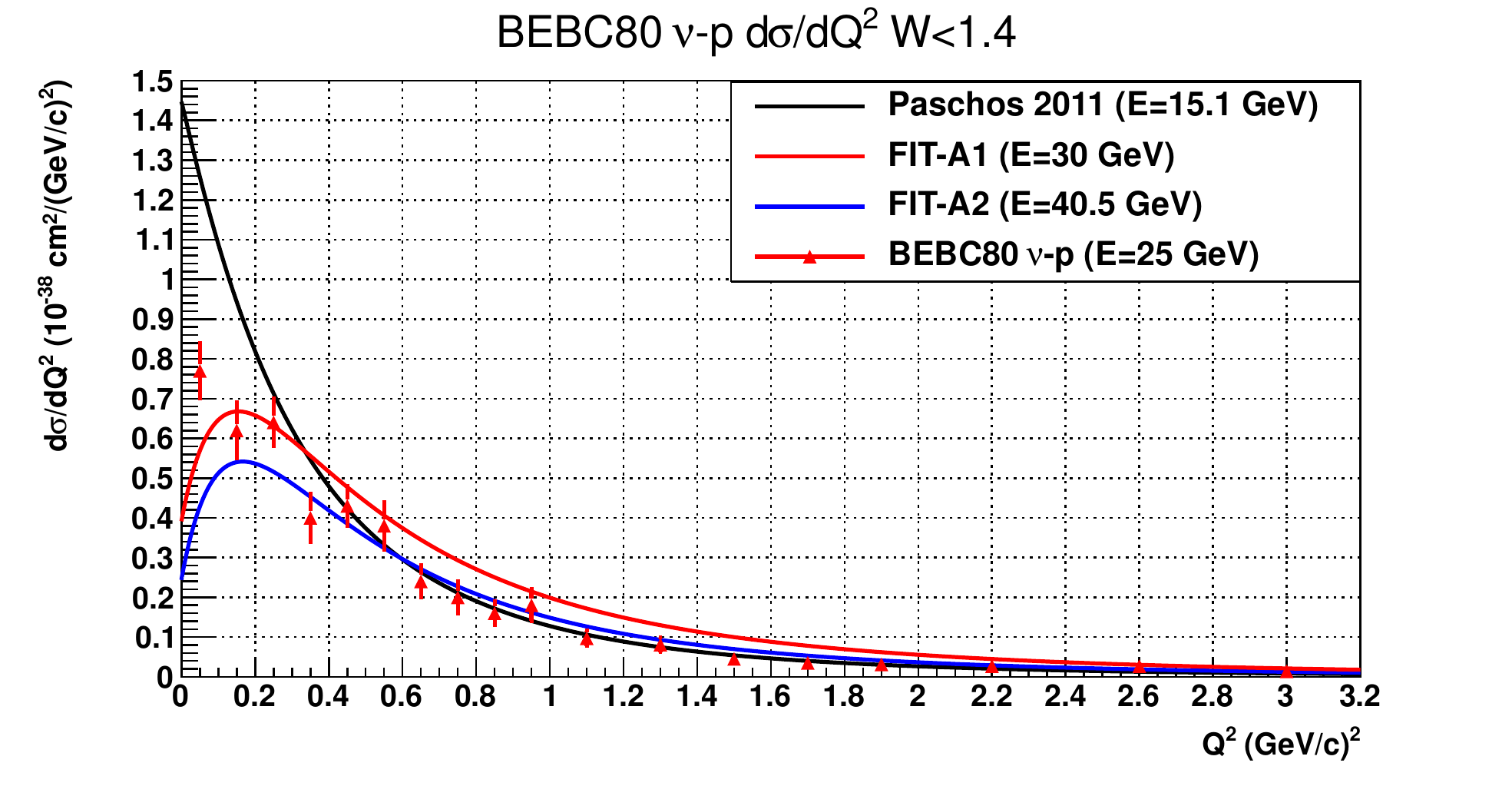}
     
     \vspace{-0.2in} 
     
\includegraphics[width=3.7in,height=2.9in]{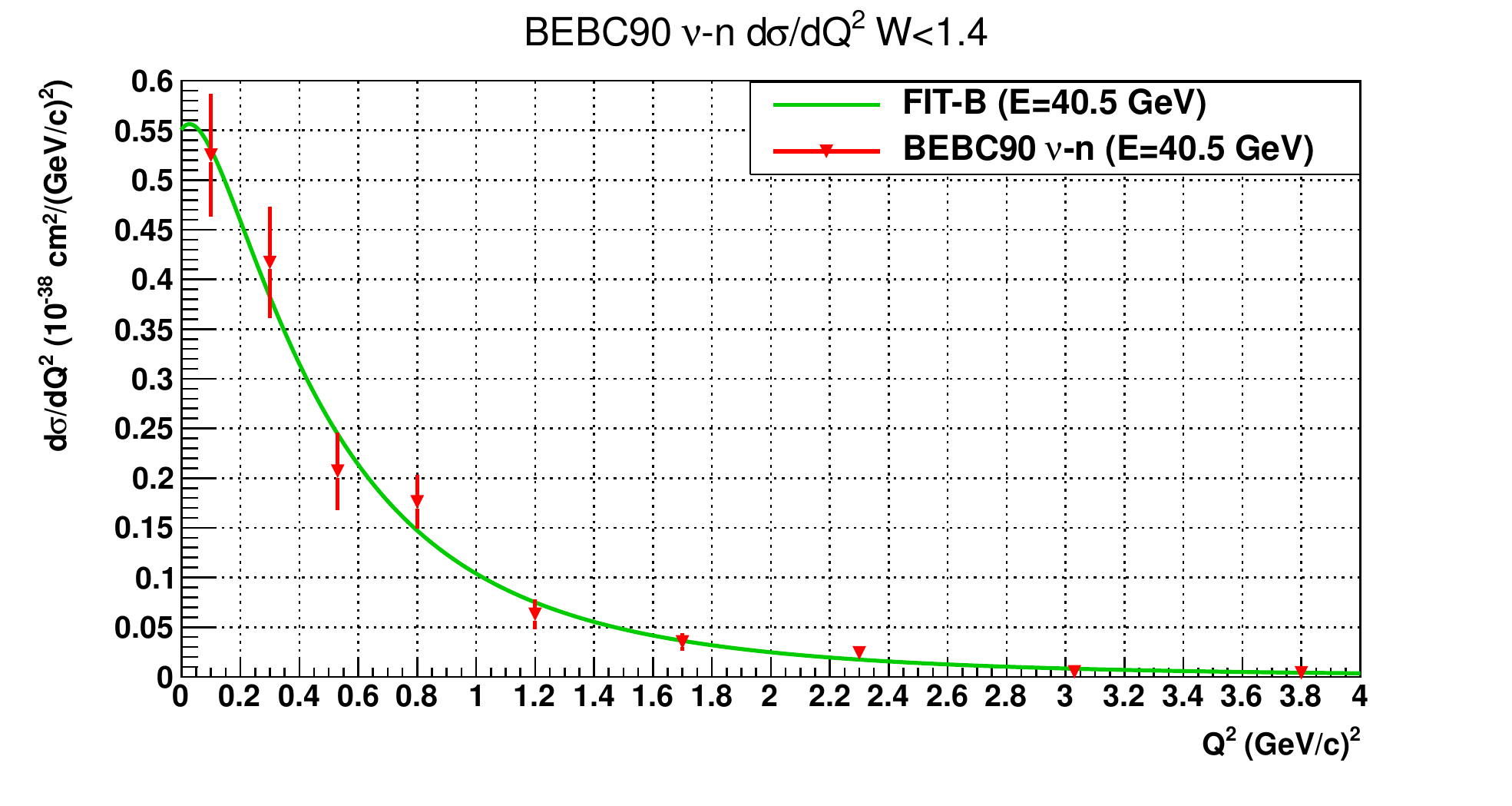}
\vspace{-0.3in} 
\caption{  Top panel:  $d\sigma/dQ^2$ for   $\nu_\mu P \to \mu^- \Delta^{++}$    
cross sections  (for $W<1.4$ GeV) measured on hydrogen at high energies by Allen et. al. (BEBC80\cite{BEBC80}). 
Bottom panel:  $d\sigma/dQ^2$ for  $\nu_\mu N \to \mu^-  \Delta^+$   
cross sections  (for $W<1.4$ GeV) measured by BEBC90 
 on free nucleons on deuterium. 
 (color online).
}
\label{dsda2allen}
\end{figure}

\subsection{Resolution, acceptance and radiative corrections}

The  ${\nu<0.25}$ GeV  events are primarily QE events with $Q^2<2M\times 0.25 \approx 0.45$ GeV$^2$. We can  select  either all events with  ${\nu<0.25}$ GeV or only QE events with  $Q^2<0.5$ GeV$^2$

 The ratio of the  number of reconstructed events with 
${\nu<0.25}$ GeV (or $Q^2<0.5$ GeV$^2$) in data and MC as a function
of energy  is proportional to the ratio of the true flux to the simulated flux in
the MC.  This ratio provides a measure of the relative neutrino flux
as a function of energy. A complete Monte Carlo should include the small contributions from
coherent pion production (shown in Appendix II), strange particle production such as QE production of
hyperons\cite{E180},  
 and radiative corrections\cite{radcor,rujula}. The  effects of experimental resolution and acceptance should also be simulated.

At present the
GENIE Monte Carlo includes coherent pion production, but
does not include the QE production of hyperons, nor radiative effects. 

If the GENIE Monte Carlo is used, then one may wish to weight the
rate of QE events (as a function of $Q^2$)  by the ratio  of events expected in the TE model to the number
 of events predicted by the  model
 which is implemented in GENIE  (i.e. the  "Independent Nucleon" model with $M_A=0.99$ GeV).
 In addition,   QE production of hyperons and  radiative effects need to be added.
 
 \section{Conclusions}
 
 We find that the model uncertainties in using the ``low-$\nu$'' event samples with $\nu<0.25$  and  $\nu<0.5$ 
 GeV are well under control (less than 3\%).  Therefore, the ``low-$\nu$'' technique can be used at low energies (0.7 GeV for neutrinos and 1 GeV for antineutrinos). 
 Once data from MINErVA on QE  scattering and resonance production becomes available,
 the model uncertainties can be made even smaller, and the technique may be extended to 
 even lower energies. 
 
 Since the model uncertainties are under control, the dominant systematic error originates
 from how well the detector response is understood, Specifically,  the mis- reconstruction
 of high $\nu$ events as  ``low-$\nu$'' events  must be modeled reliably. 
 This is because at  high energies (as shown in  Fig.~\ref{nuvsq24}) mis-reconstruction 
  of the hadron energy of high $\nu$ events can 
 increase the number of  ``low-$\nu$'' events, while at low energies there
 are fewer high $\nu$ events that can be mis-reconstructed at low $\nu$. 
 
 The dominant uncertainty in the method
 comes from the calibration and  resolution smearing in the measurement
 of the  hadronic energy.  This was the dominant error when this method
 was used in MINOS because of the poor resolution of
 the MINOS target calorimeter at low hadronic energy.  
 
 As mentioned in the introduction,
 the  standard method for the determination of the neutrino flux requires the
 modeling of pion production as well as the  complicated magnetic focusing elements.
 The determination of the flux for the 
  Fermilab NUMI beam with the standard method is limited at present
  by the uncertainties in pion production cross sections.
  The resulting error in the flux  is about 5\% at low energies (1-2 GeV) and
 10\%-15\% at the higher energies (10-20 GeV). Therefore, having 
 the  "low $\nu$" method which yields the relative neutrino flux
 as a function of energy very useful. 
 In principle, the uncertainties in the standard method can be improved
 with better measurements of pion and kaon production cross sections.
  Plans for such future measurements at the
  CERN Laboratory are currently under discussion.
 
A second  method, which requires the measurement of the muon rate downstream
 of the decay pipe,  can not determine the energy
 dependence of the flux.  It mostly  constrains the overall level of the flux.
 At present, the uncertainties in the overall calibration of the muon chambers
 yields an uncertainty in the flux of  about 10\%. . 
 
A third method  uses inverse muon
 decay  $\nu_\mu+  e \to \mu^- + \nu_e$ events in the detector. 
The threshold for this reaction is
 about 12 GeV. Therefore, this  method can only be used at higher energies.
  Inverse muon decay
 was used by NOMAD to constrain their neutrino flux at high energies.  
 In addition to being statistically limited, the final state
 energy of inverse muon decay events is not fully 
measured since there is a neutrino in the final state. This places a limitation
on the determination of  the energy dependence
of the neutrino  fluxes. This method cannot be used for the determination of the
 flux for antineutrinos. 
 
A fourth  method uses the neutral current 
reaction $\nu_\mu+  e \to \nu_\mu+  e$.  In  addition to being statistically limited, the final state energy
in $\nu_\mu+  e \to \nu_\mu+ e$ events  is not fully 
measured since there is a neutrino in the final state. This places a limitation
on the determination of  the energy dependence
of the fluxes. In this method 
only  the sum of the fluxes for
neutrinos and antineutrinos is measured  because
calorimetric detectors such as MINERvA cannot determine
the charge of final state electrons.

         \begin{figure}
\includegraphics[width=3.7in,height=2.9in]{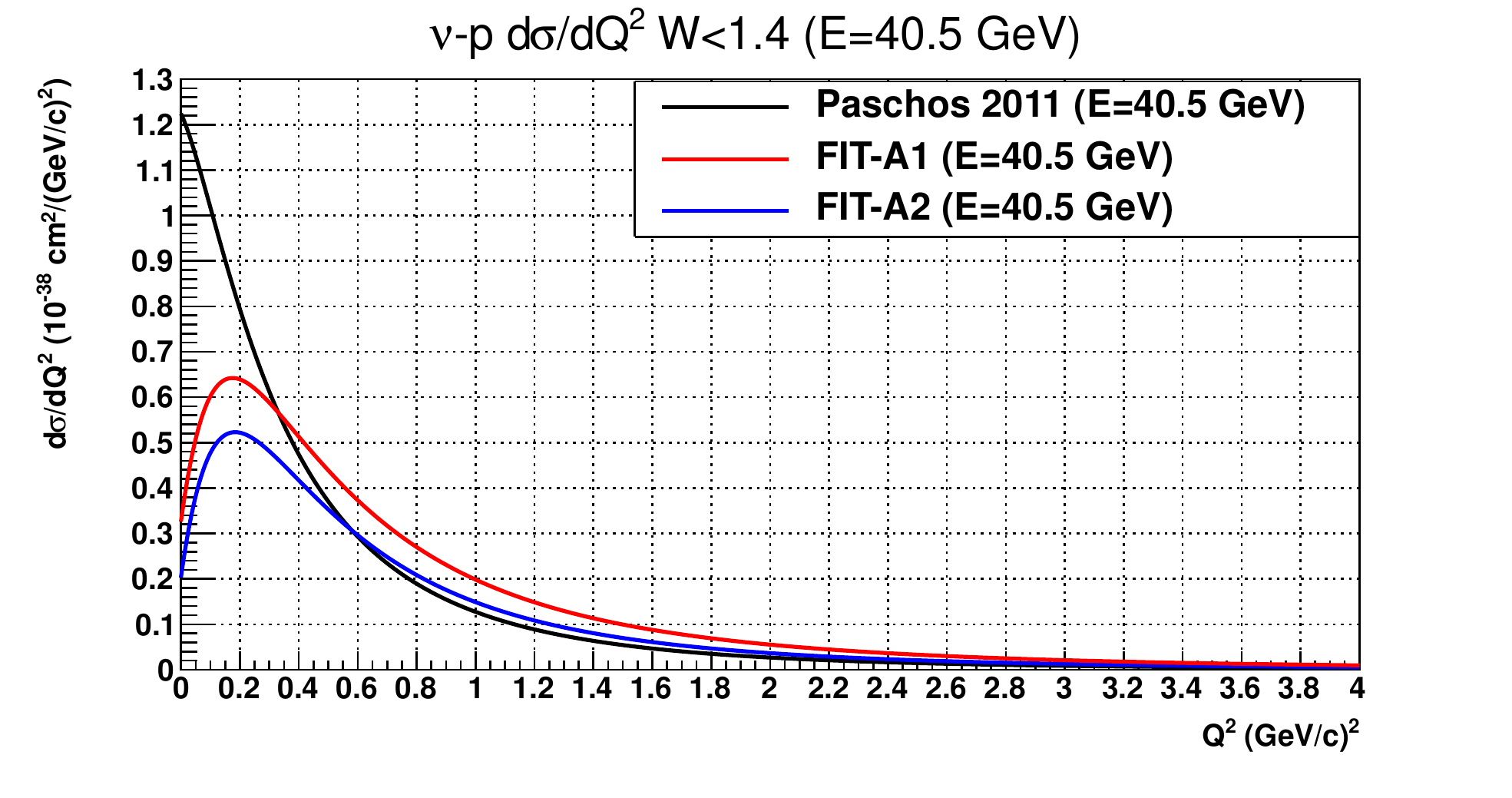}
     
     \vspace{-0.2in} 
     
\includegraphics[width=3.7in,height=2.9in]{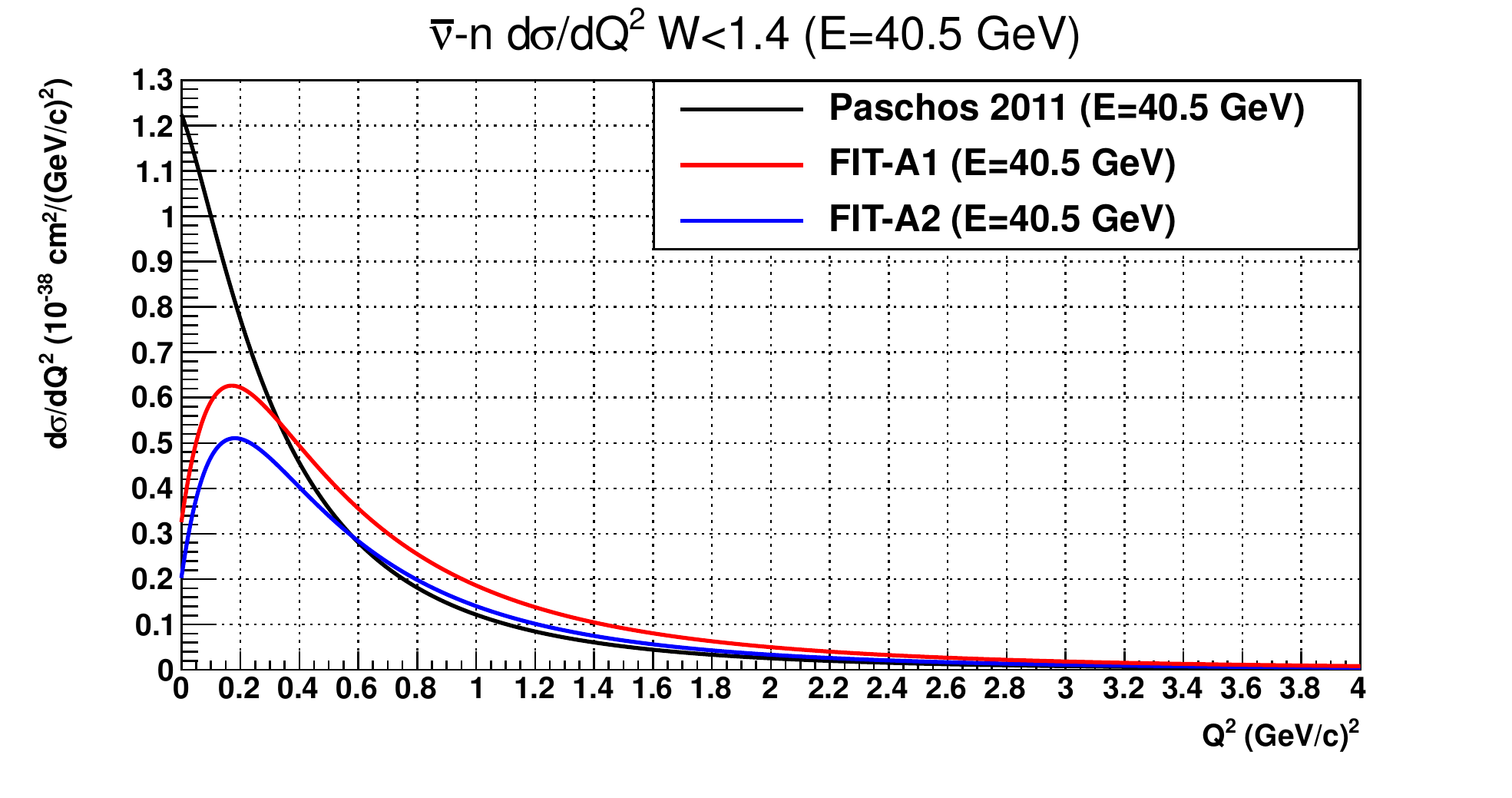}
\vspace{-0.3in} 
\caption{  The three  $\nu_\mu P/ \nub_\mu N$  $d\sigma/dQ^2$  cross sections  models  (for $W<1.4$ GeV) with Pauli suppression for nuclear
targets at an energy of 40.5 GeV.  The cross sections for for $\nu_\mu P \to \mu^- \Delta^{++}$ are shown on the top panel and the cross sections for $\nub_\mu N \to \mu^+  \Delta^-$  are shown on the bottom panel.  
(color online).
 }
\label{dsdq2Amodel}
\end{figure}

         \begin{figure}
\includegraphics[width=3.7in,height=2.9in]{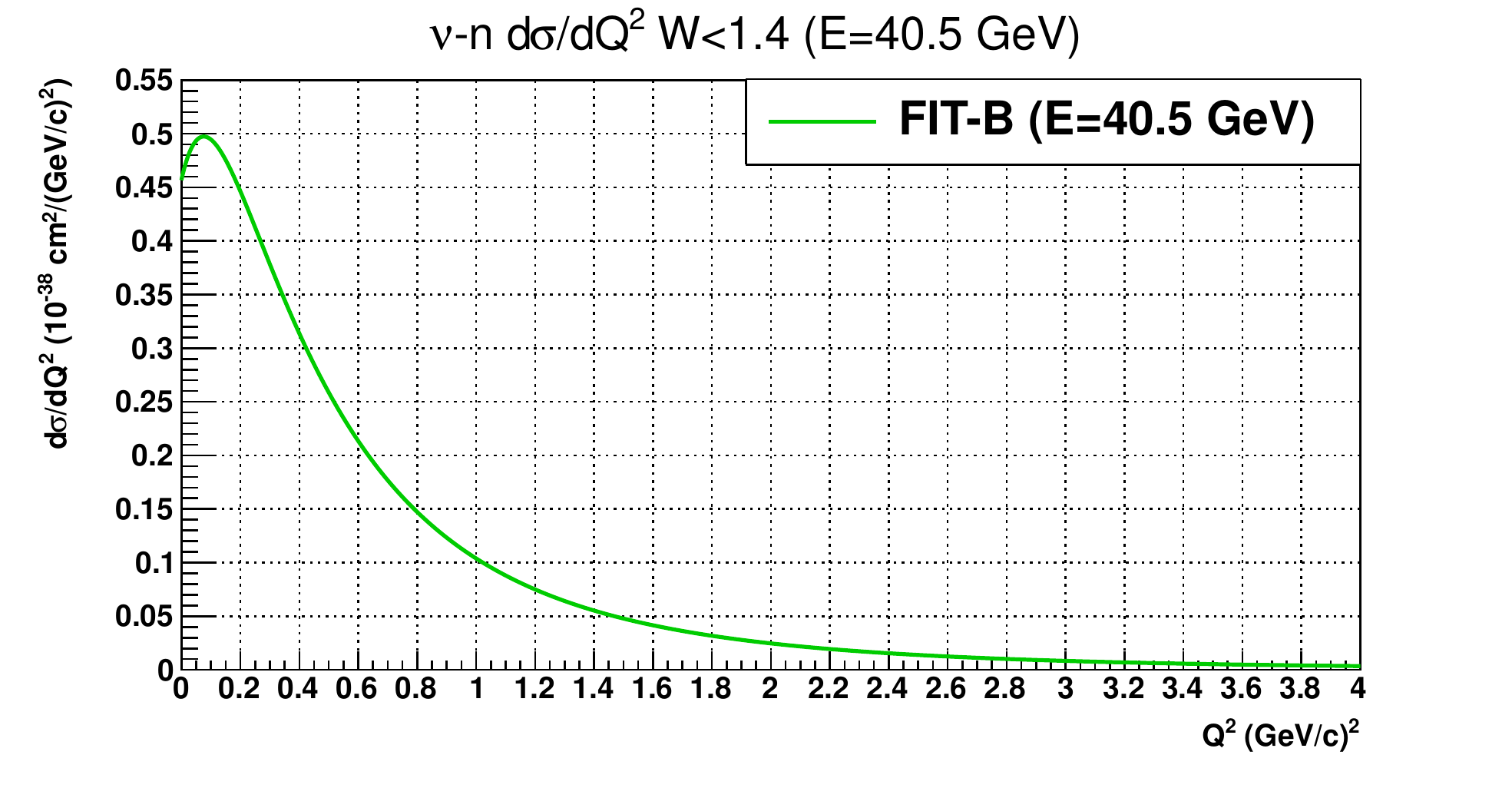}
     
     \vspace{-0.2in} 
     
\includegraphics[width=3.7in,height=2.9in]{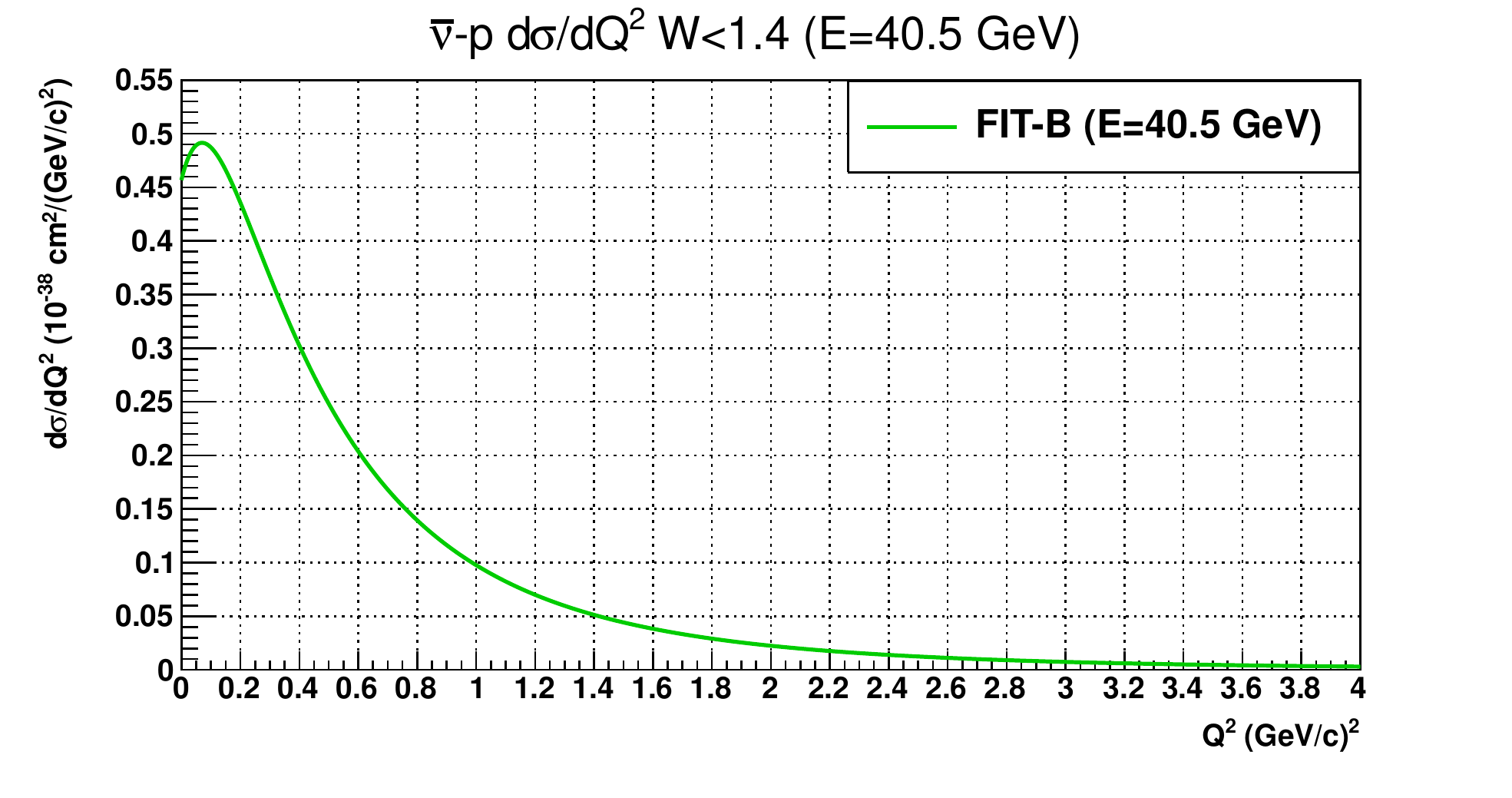}
\vspace{-0.3in} 
\caption{  Our $d\sigma/dQ^2$  cross sections  model  (for $W<1.4$ GeV) with Pauli suppression for nuclear
targets at an energy of 40.5 GeV.  The cross sections for for $\nu_\mu N \to \mu^+ \Delta^{-}$ are shown on the top panel and the cross sections for $\nub_\mu P \to \mu^+  \Delta^0$  are shown on the bottom panel.
The green curve labeled 
FIT-B ($M_A$=1.62, $C_5^A$ = 1.27) represents a fit to the BEBC90 $\nu_\mu N$  free nucleon data  (color online).
}
\label{dsdq2Bmodel}
\end{figure}

  \section{Appendix I:  $\Delta$ production cross sections }  
\subsection{$\Delta$ production form factors}

For the vector contribution we use the formulae for the structure functions
${\cal W}_1$, ${\cal W}_2$, ${\cal W}_3$, ${\cal W}_4$, and ${\cal W}_5$ on free nucleons from Lalakulich and Paschos\cite{paschos}. We neglect the effect of Fermi motion.  The form factors that we use
are taken from Paschos and Schalla\cite{paschos}. Specifically, the vector form factors are 

\begin{eqnarray}
 C_3^V (Q^2) = \frac{2.13 / D_V}{1+ \frac{Q^2}{4M_V^2}} \; & , & \; C_4^V (Q^2) = \frac{-1.51 / D_V}{1+ \frac{Q^2}{4M_V^2}} \\
 C_5^V (Q^2) = \frac{0.48 / D_V}{1+\frac{Q^2}{0.776 M_V^2}} \; & \textmd{ and } & \; D_V =\left( 1 + \frac{Q^2}{M_V^2} \right)^2
\end{eqnarray}
with $M_V$ =  0.84 GeV, which have been  extracted from electroproduction data.

For the vector-axial interference $W_3 (Q^2,\nu)$  Paschos and Schalla use the form factor $C_5^A (Q^2)$
where 
  $$C_5^A(Q^2)= \frac{C_5^A}{(1+Q^2/M_A^2)^2}\frac{1}{1+2Q^2/M_A^2},$$
  $$C_4^A= -\frac{1}{4} C_5^A(Q^2)$$  
  Here,  we define $C_5^A=C_5^A (0)$ 
  
Paschos and Schalla use low energy  $\pi^+ p \rightarrow \Delta^{++}$ where the non-resonant background is smallest.  With $M_A$=1.05 GeV they extract value of  $C_5^A (0) = 1.08$ from the data. Since this value is close to $1.20$ predicted by the Goldberger-Treiman relation, they chose to use  $C_5^A$=1.2.

Paschos and Schalla  mention that several recent articles also  calculate $C_5^A(0)$ by fitting experimental data~\cite{Hernandez:2007qq,Lalakulich:2010ss,Leitner:2008ue,Graczyk:2009qm,Hernandez:2010bx,AlvarezRuso:1998hi,SajjadAthar:2009rc} with values varying from 0.87 up to 1.20. Models with a resonant background~\cite{Hernandez:2007qq,Lalakulich:2010ss} prefer the power value, while the other articles~\cite{Leitner:2008ue,Graczyk:2009qm,Hernandez:2010bx,AlvarezRuso:1998hi,SajjadAthar:2009rc} prefer values closer to 1.20. The reasons for the differences is the treatment of the non-resonant background,  the form of the axial form factor that is used, and the exact kinematics at small $Q^2$.

For  $\Delta^{++}$ and $\Delta^{-}$  we define the   Pachos-2011 parameterization using above
form factors with     $C_5^A$=1.2  (extracted through PCAC),  $M_A$=1.05 GeV, and the vector form factors described above.   As mentioned earlier, FIT-A1 and FIT-A2 use the same form but with  different values of 
  $C_5^A$ and   $M_A$.

 For   $\Delta^{+}$ and $\Delta^{0}$ production our Fit-B uses the same form factors multiplied by a factor of $1/\sqrt(3)$ (as expected  from Clebsch-Gordan coefficients\cite{paschos}).    However, in order to account for the large non-resonance background, we use different values
   $C_5^A$ and   $M_A$.
       \begin{figure}
\includegraphics[width=3.7in,height=2.9in]{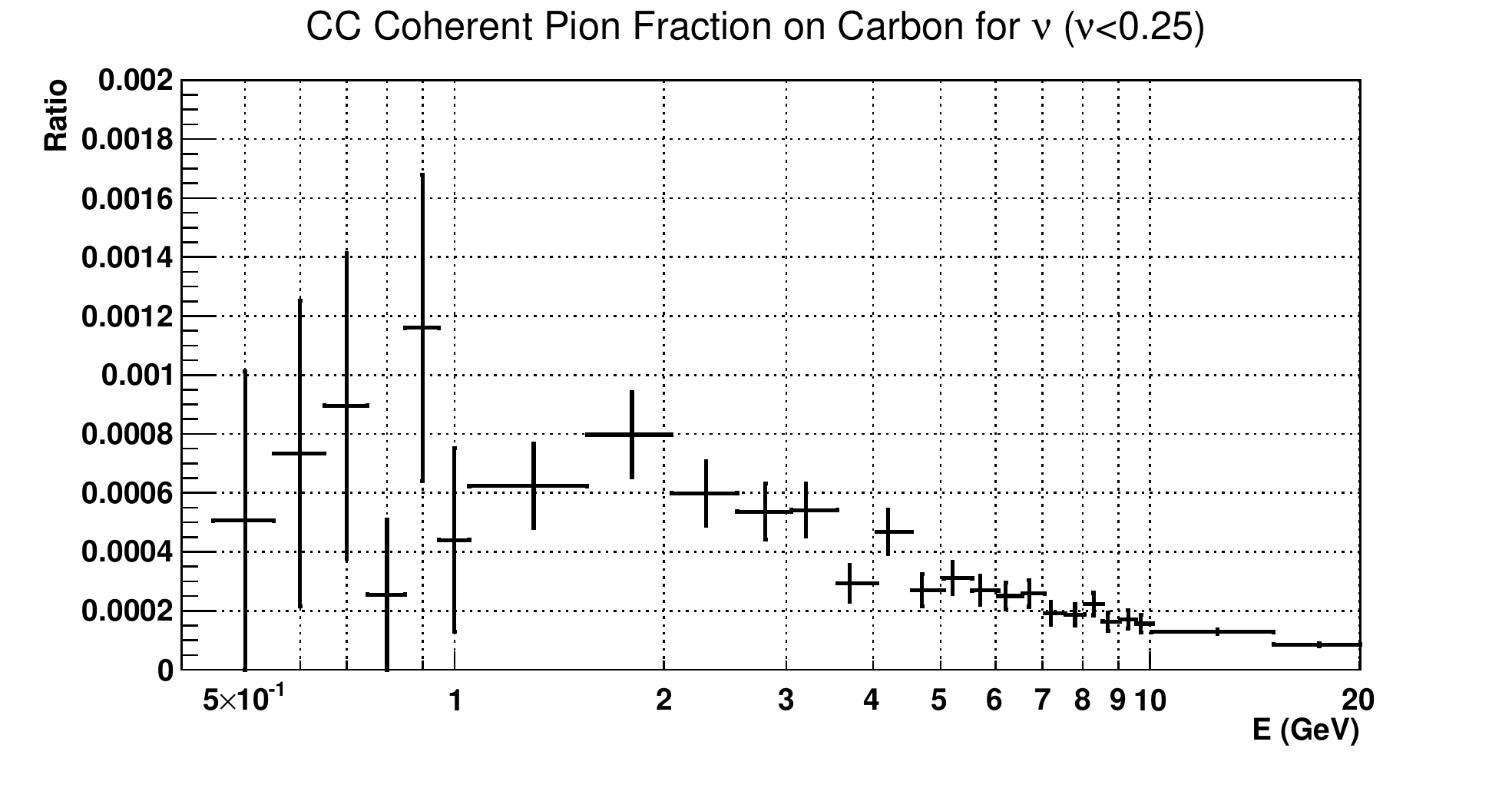}
     
     \vspace{-0.2in} 
     
\includegraphics[width=3.7in,height=2.9in]{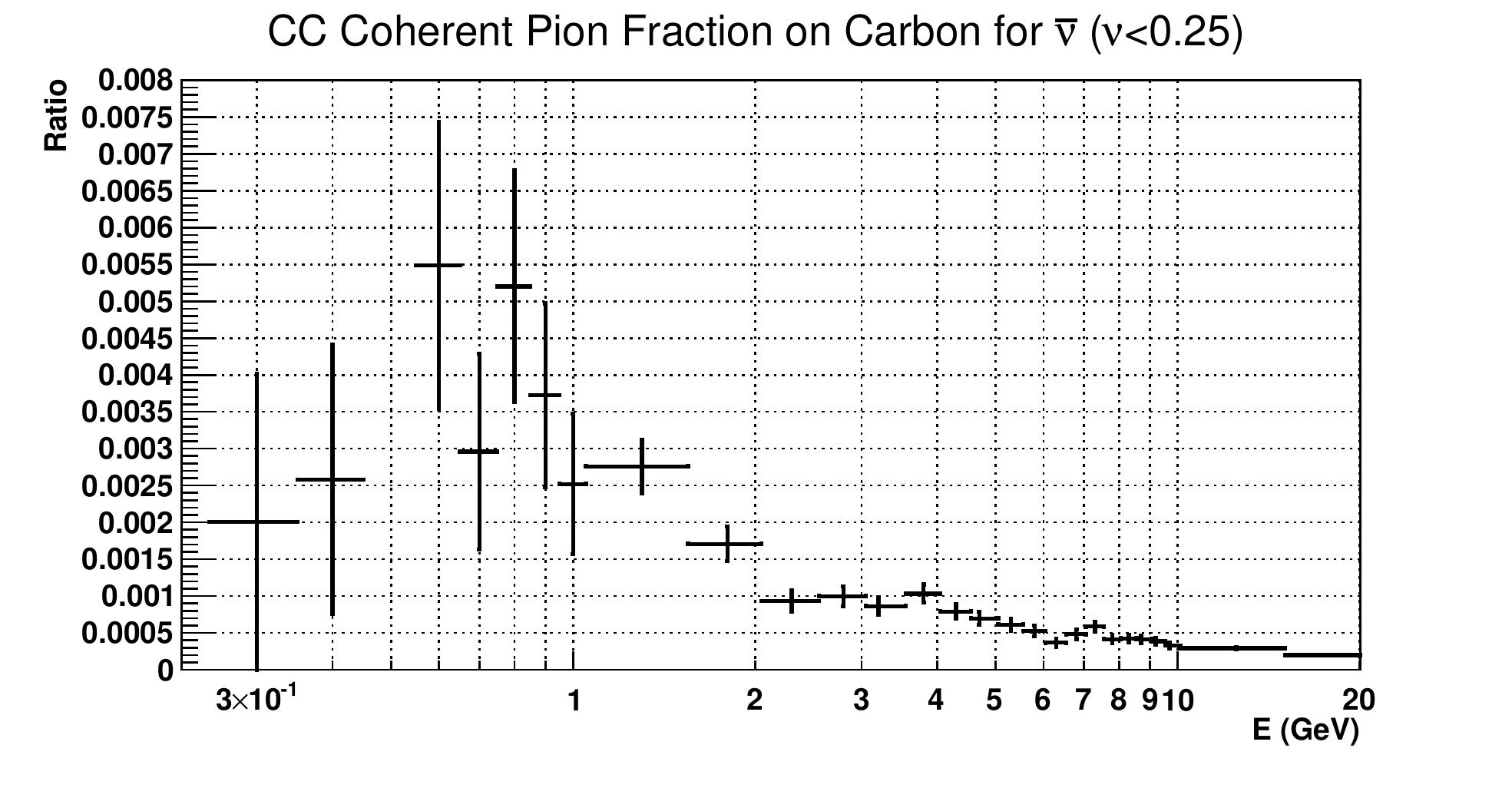}
\vspace{-0.3in}  
\caption{  The fraction of events from coherent pion production in the $\nu<0.25$ GeV event sample
(calculated with GENIE) as a function of neutrino energy:
Neutrinos (top panel) and antineutrinos (bottom panel).
}
\label{copi25}
\end{figure}

    \begin{figure}
\includegraphics[width=3.7in,height=2.9in]{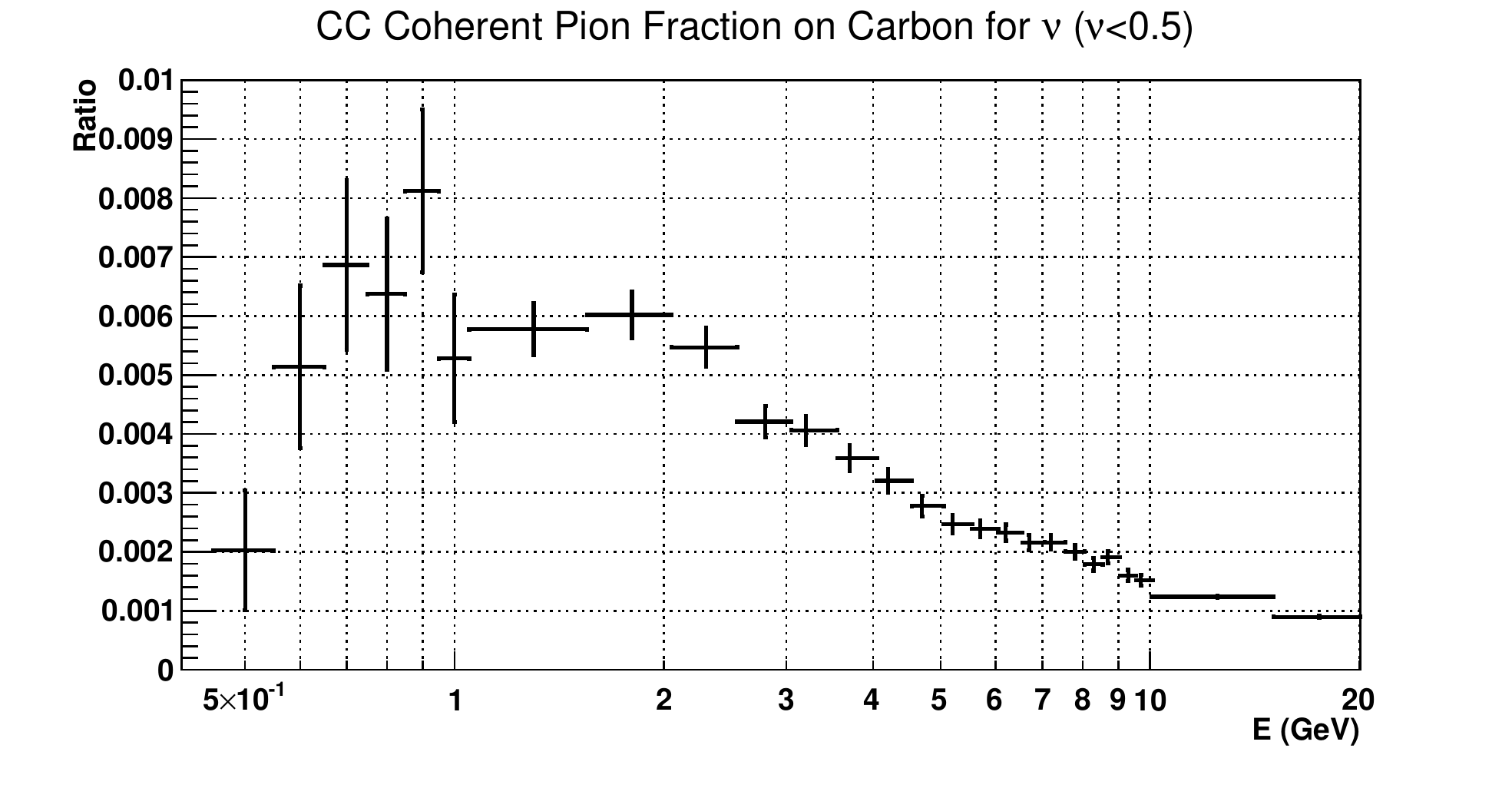}
     
     \vspace{-0.2in} 
     
\includegraphics[width=3.7in,height=2.9in]{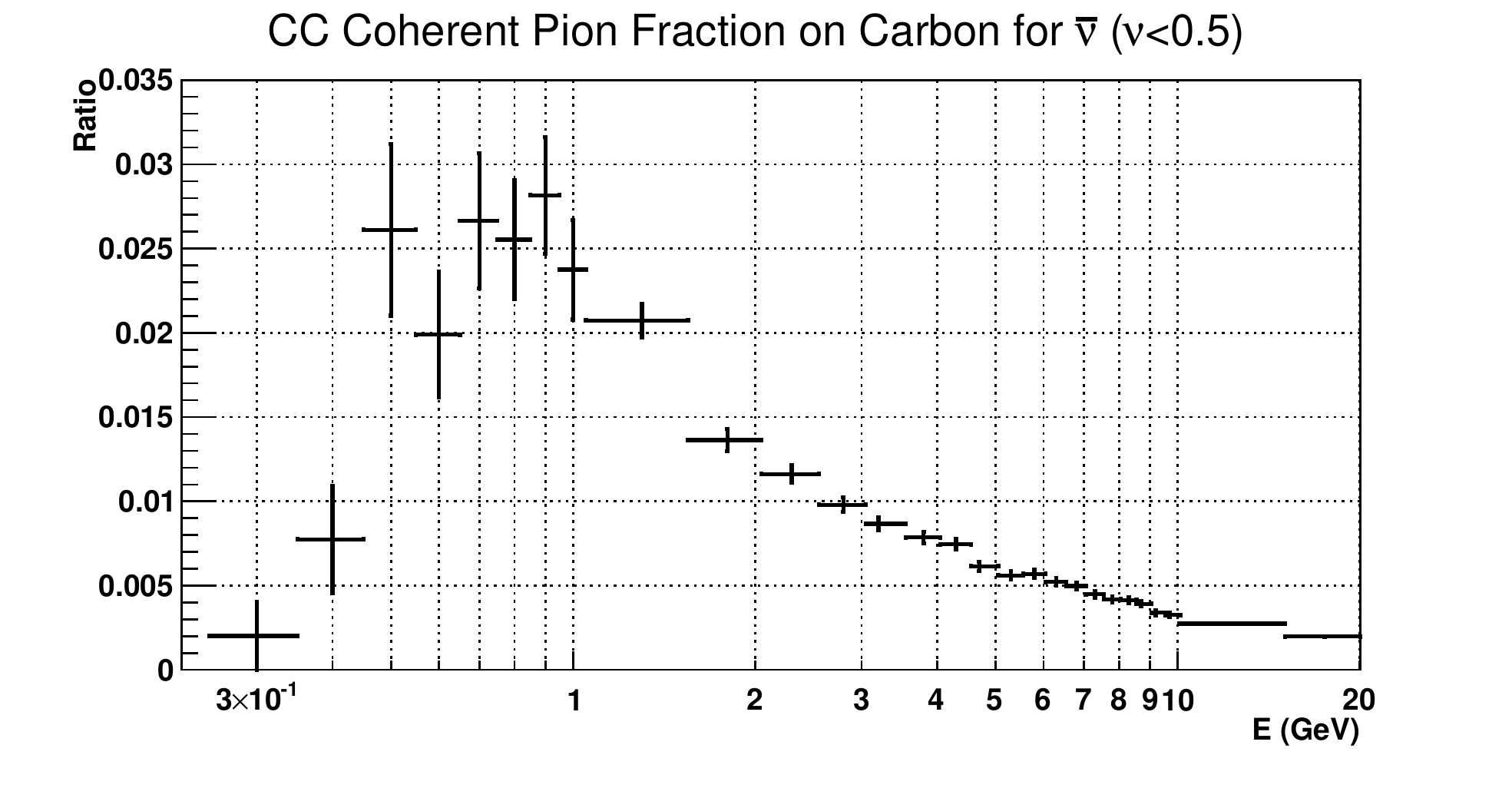}
\vspace{-0.3in}   
\caption{  Same as \ref{copi25} but for  the $\nu<0.50$ GeV sample.
}
\label{copi50}
\end{figure}

\subsection{Various parameterizations}
  
  The  low energy and high energy
  data for neutrino and antineutrino production of the $\Delta(1232)$ resonance
  are not entirely consistent.
 Therefore, we use range
  of parameterization to span the systematic error in our modeling of $\Delta$ production cross sections.
  
  The form factors for    $\nu_\mu P \to \mu^- \Delta^{++}$ and $\nub_\mu N \to \mu^+  \Delta^-$ should be
  the same. 
  The $d\sigma/dQ^2$  differential cross sections  ($W<1.4$ GeV) for
  $\nu_\mu P \to \mu^- \Delta^{++}$ measured at high energies are shown in the top panel
  of  Fig. \ref{dsdq2allasia} (Allasia et. al., BEBC90\cite{BEBC90} data on deuterium)   and also
  on the top panel of   Fig. \ref{dsda2allen} (Allen et. al. BEBC80\cite{BEBC80} data on hydrogen)
  The bottom panel of  Fig. \ref{dsdq2allasia} shows the $d\sigma/dQ^2$  cross sections  at high energies ($W<1.4$ GeV)   for  $\nub_\mu N \to \mu^+  \Delta^-$  measured by Allasia et. al. (BEBC90)  data on deuterium.   The black curve labeled Paschos-2011($M_A^{\Delta}$=1.05, $C_5^A$ = 1.2) is from fits to lower energy $\nu_\mu P \to \mu^- \Delta^{++}$ data (BNL and Argonne).  The   red curve
labeled  FIT-A1 ($M_A$=1.93, $C_5^A$ = 0.62) is a fit to the  BEBC90 $\nub_\mu N \to \mu^+  \Delta^-$  data .  The  blue curve labeled
FIT-A2 ($M_A$=1.75, $C_5^A$ = 0.49) is a fit to the BEBC90 $\nu_\mu P \to \mu^- \Delta^{++}$  data. The variation among the three curves is taken as a systematic error. 

The bottom panel of Fig.  \ref {dsda2allen}  shows values of $d\sigma/dQ^2$ 
differential cross sections for   $\nu_\mu N \to \mu^-  \Delta^+$   (for $W<1.4$ GeV) measured by BEBC90 
 on free nucleons on deuterium.  This reaction has different form factors then $\nu_\mu P \to \mu^- \Delta^{++}$.
  The green curve labeled  FIT-B ($M_A$=1.62, $C_5^A$ = 1.27) represents a fit to the BEBC90 $\nu_\mu N$ data.

We use the above models with the addition of Pauli suppression in order to model the differential cross sections 
on nuclear targets. 

Fig.  \ref {dsdq2Amodel} shows our three $\nu_\mu P/ \nub_\mu N$  $d\sigma/dQ^2$  cross sections  models  (for $W<1.4$ GeV) with Pauli suppression for nuclear
targets at an energy of 40.5 GeV.  The cross sections for for $\nu_\mu P \to \mu^- \Delta^{++}$ are shown on the top panel and the cross sections for $\nub_\mu N \to \mu^+  \Delta^-$  are shown on the bottom panel.  These two reactions should be described by the same form factors.  The black curve labeled Paschos-2011($M_A^{\Delta}$=1.05, $C_5^A$ = 1.2) is from fits to lower energy $\nu_\mu P$ free nucleon data (BNL and Argonne).  The   red curve
labeled  FIT-A1 ($M_A$=1.93, $C_5^A$ = 0.62) is from a  fit to the  BEBC90 $\nub_\mu N$  free nucleon data.  The  blue curve labeled
FIT-A2 ($M_A$=1.75, $C_5^A$ = 0.49) is from a  fit to the BEBC90 $\nu_\mu P$ free nucleon data.  The variation among the three curves is taken as a systematic error.

Fig.  \ref {dsdq2Bmodel} shows our $\nu_\mu N/ \nub_\mu P$  $d\sigma/dQ^2$   cross sections  model  (for $W<1.4$ GeV) with Pauli suppression for nuclear
targets at an energy of 40.5 GeV.  The cross sections for for $\nu_\mu N \to \mu^+ \Delta^{-}$ are shown on the top panel and the cross sections for $\nub_\mu P \to \mu^+  \Delta^0$  are shown on the bottom panel.
The green curve labeled 
FIT-B ($M_A$=1.62, $C_5^A$ = 1.27) is extracted  fit to the BEBC90 $\nu_\mu N$  free nucleon data.

 \section{Appendix II: Coherent Pion Production}  
 
 Fig. \ref{copi25} shows  the fraction of events from coherent pion production in the $\nu<0.25$ GeV event sample
(calculated with GENIE) as a function of neutrino energy. Neutrinos are shown on the top panel
and antineutrinos are shown on the bottom panel. For the $\nu<0.25$ GeV sample, the contribution from
coherent pion production is less than 0.1\% for neutrinos and less than  0.6\% for antineutrinos

Fig. \ref{copi50} shows the fraction of events from coherent pion production in the $\nu<0.50$ GeV event sample
(calculated with GENIE) as a function of neutrino energy. Neutrinos are shown on the top panel
and antineutrinos are shown on the bottom panel. For the $\nu<0.50$ GeV sample, the contribution from
coherent pion production is less than 0.7\% for neutrinos and less than  3\% for antineutrinos


\begin{thebibliography}{9}

\bibitem {T2K}   Y. Itow  et al., (T2K) arXiv:hep-ex/0106019;
\bibitem {MINOS}  D.G. Michael et al., (MINOS) Phys. Rev. Lett. {\bf 97}, 191801 (2006);  
http://www-numi.fnal.gov/Minos/
\bibitem {MINOS2} P. Adamson  et al., (MINOS) Phys. Rev. D {\bf 81}, 072002 (2010).
\bibitem {NOVA}  http://www-nova.fnal.gov/
\bibitem{minerva} http://minerva.fnal.gov/
\bibitem{mishra} S. R. Mishra, in {\it Proceedings of the Workshop on Hadron
Structure Functions and Parton Distributions} , edited by D. Geesaman {\it et al}
\bibitem{seligman} W. Seligman, Ph.D. thesis, Columbia University, 1997), Nevis 292.
\bibitem{neugen3}  H. Gallagher, Nucl. Phys. Proc. Suppl. 112 (2002).
\bibitem{bardin} D. Bardin and V. Dokuchaeva, Preprong JINR-E2-86-260 (1986).
\bibitem{paschos} O. Lalakulich and  E. A. Paschos,
Phys. Rev. D {\bf 71}, 074003 (2005), and Phys. Rev. D {\bf 74},014009 (2006); E. A. Paschos and
D. Schalla, Phys. Rev. D {\bf 84}, 013004 (2011))
\bibitem{pyu}  E. A. Paschos and J. Y. Yu,
Phys. Rev. D {\bf 65}:033002 (2002).
\bibitem{GENIE} C.Andreopoulos (GENIE),  Nucl. Instrum. Meth.A614, 87,2010;
 H. Gallagher, (NEUGEN) Nucl. Phys. Proc. Suppl. 112 (2002);Y. Hayato (NEUT), Nucl Phys. Proc. Suppl.. {\bf 112}, 171 (2002);D. Casper (NUANCE) , Nucl. Phys. Proc. Suppl. 112, 161 (2002);  http://nuint.ps.uci.edu/nuance/
\bibitem{Lle72} C. H. Llewellyn Smith, Phys. Rep. 3C (1972); E. A. Paschos, Electroweak Theory,
Cambridge University Press (2007). 
\bibitem{transverse} A. Bodek, H. Budd and E. Christy,  Eur.Phys. J. {\bf C}71, 172 (2011)
\bibitem{steffens} 	
F.M. Steffens and  K. Tsushima , Phys. Rev. D {\bf 70}, 094040 (2004)
\bibitem{MEC4}
J. Carlson, J. Jourdan, R. Schiavilla, and I. Sick, Phys. Rev. C {\bf 65}  024002 (2002)
\bibitem{quasi} 
A. Bodek, S. Avvakumov, R. Bradford, and  H. Budd, Eur. Phys. J. C{\bf
53}, 349 (2008).
\bibitem{JUPITER} JUPITER collaboration, Jefferson Lab experiment E04-001,  Arie Bodek, Cynthia Keppel and M. Eric Christy, spokespersons.
\bibitem{vahe-thesis} V. Mamyan, Ph.D. dissertation, University of Virginia, 2010.  
\bibitem{MEC5}
M. Martini, M. Ericson, G. Chanfray, and J. Marteau, Phys. Rev. C 80: 065501, 2009; ibid
 Phys. Rev. C 81: 045502, 2010. 
\bibitem {MiniBooNE}   A. A.   Aguilar-Arevalo et al., (MiniBooNE) Phys. Rev. Lett {\bf 98}, 231801(2007);  A.A. Aguilar-Arevalo {\em et al.} Measurement of the neutrino component of an anti-neutrino beam observed by a non-magnetized detector. e-Print: arXiv:1102.1964 [hep-ex] 
\bibitem{NOMAD}  
  V.~Lyubushkin  et al.  (NOMAD Collaboration),
  Eur.\ Phys.\ J.\  C {\bf 63}, 355 (2009);  Q. Wu {\it et al.}(NOMAD Collaboration), Phys. Lett. 
{\bf B60}, 19 (2008).
\bibitem{Serp96} V.B. Anikeev, et al. (Serpukhov) Z. Phys. C {\bf 70}, 39 (1996)
\bibitem{BNL82} N. J. Baker et al. (BNL)  Phys. Rev. D {\bf 25}, 617 (1982).	
 \bibitem{ANL73}    J. Campbell et. al. (ANL), Phys. Rev. Lett. {\bf 30}, 335 (1973) 
  \bibitem{ANL79}    S. J.  Barish et al. (ANL)  Phys. Rev. D {\bf 19}, 2521 (1979) 
  \bibitem{ANL82}  G.M. Radecky et al. ( ANL)   Phys. Rev. D {\bf 25}, 1161 (1982)
 \bibitem{BNL86}   T. Kitagaki et al.  (BNL) Phys. Rev. D {\bf 34}, 2554 (1986)  
 \bibitem{FNAL78}    J.  Bell 1978  et al. (FNAL) Phys. Rev. Lett. {\bf 41}, 1008  (1978); ibid 1012  (1978)
\bibitem{FNAL81} V.I. Efremenko et al.  (FNAL) ITEP-83-1981 (unpublished).
  \bibitem{BEBC80} P. Allen et al. (BEBC) , Nucl. Phys. B176, 269(1980); ibid B{\bf264} ,221 (1986)
\bibitem{BEBC90}  D. Allasia et al. (BEBC), Nucl. Phys. B {\bf 343}, 285 (1990). 
  \bibitem{GGM78}  W. Lerche et al. (GGM)  Phys.Lett.  B{\bf 78}, 510  (1978)
\bibitem{SKAT88}   V.V. Ammosov et al ( SKAT)   Soviet J. Nucl. Phys. {\bf  50} ,57  (1988) 
\bibitem{SKAT89} H. J. Grabosch et. al. (SKAT)  Z. Phys. C {\bf 41}, 527 (1989)
\bibitem{GGM79}   T. Bolognese et al. (GGM)  Phys. Lett.  B {\bf 81}, 393  (1979)
\bibitem{rs} D. Rein and L. M. Sehgal, Annals Phys. {\bf 133}
79 (1981); R. Belusevic and D. Rein,
Phys. Rev. D {\bf 46}, 3747 (1992)
\bibitem{E180} 
V.V. Ammosov et al. (FNAL E180) JETP Lett.43 716,1986, Pisma Zh.Eksp.Teor.Fiz.43 554,1986.
    \bibitem{radcor} J. Kiskis,
 Phys. Rev. {\bf D8} 2129 (1973); Roger J. Barlow and Stephen Wolfram,
 Phys.~Rev. {\bf D20} 2198 (1979); A.B. Arbuzov , D.Yu. Bardin, L.V.
Kalinovskaya
 hep-ph/0407203) JHEP 0506 (2005).
\bibitem{rujula} A.~De R\'{u}jula, R.~Petronzio, and
A.~Savoy-Navarro, Nucl.~Phys.~B {\bf 154}, 394 (1979); Gunther Sigl,  Phys.~Rev. {\bf D57} 3786 (1998);  A. Bodek,  " Muon internal bremsstrahlung: A Conventional explanation for the excess nu(e) events in MiniBoone" , arXiv:0709.4004 [hep-ex]

\bibitem{Hernandez:2007qq}
E.~Hernandez, J.~Nieves, and M.~Valverde,
\newblock Phys. Rev. {\bf D76}, 033005 (2007), hep-ph/0701149.

\bibitem{Lalakulich:2010ss}
O.~Lalakulich, T.~Leitner, O.~Buss, and U.~Mosel,
\newblock Phys. Rev. {\bf D82}, 093001 (2010), 1007.0925.
\bibitem{Leitner:2008ue}
T.~Leitner, O.~Buss, L.~Alvarez-Ruso, and U.~Mosel,
\newblock Phys. Rev. {\bf C79}, 034601 (2009), 0812.0587.

\bibitem{Graczyk:2009qm}
K.~M. Graczyk, D.~Kielczewska, P.~Przewlocki, and J.~T. Sobczyk,
\newblock Phys. Rev. {\bf D80}, 093001 (2009), 0908.2175.

\bibitem{Hernandez:2010bx}
E.~Hernandez, J.~Nieves, M.~Valverde, and M.~J. Vicente~Vacas,
\newblock Phys. Rev. {\bf D81}, 085046 (2010), 1001.4416.

\bibitem{AlvarezRuso:1998hi}
L.~Alvarez-Ruso, S.~K. Singh, and M.~J. Vicente~Vacas,
\newblock Phys. Rev. {\bf C59}, 3386 (1999), nucl-th/9804007.

\bibitem{SajjadAthar:2009rc}
M.~Sajjad~Athar, S.~Chauhan, and S.~K. Singh,
\newblock J. Phys. {\bf G37}, 015005 (2010), 0908.1442.

\bibitem{Radecky:1981fn}
G.~M. Radecky {\em et~al.},
\newblock Phys. Rev. {\bf D25}, 1161 (1982).

%
\end{thebibliography}
\end{document}